%% file: main.tex
\documentclass[preprint,number]{elsarticle}
\usepackage[utf8]{inputenc}
\usepackage{multirow}
\usepackage{amsthm}
\usepackage{algorithmic}
\usepackage{algorithm}
\usepackage{graphicx}
\usepackage{wrapfig}
\usepackage{appendix}
\usepackage{subcaption}
\usepackage{siunitx}

\usepackage{pdflscape}

\usepackage[draft]{todonotes}

\usepackage{booktabs}

\usepackage{float}
\floatname{algorithm}{Model}

\newtheorem{mydef}{Definition}

\begin{document}
	\begin{frontmatter}
		
		\author[ul]{Davide Nunes\corref{cor1}}
		\ead{davide.nunes@di.fc.ul.pt}
		\author[ul]{Luis Antunes}
		\ead{xarax@ciencias.ulisboa.pt}
		
		\cortext[cor1]{Corresponding author}

		\address[ul]{GUESS/BioISI--Instituto de Biosistemas e Ciências Integrativas,
                  Faculdade de Ciências, Universidade de Lisboa, 1749-016 Lisboa, Portugal.}
                \title{Modelling Structured Societies: a Multi-relational Approach to Context
                  Permeability}
		
		\begin{abstract}
                  The structure of social relations is fundamental for the construction of plausible
                  simulation scenarios. It shapes the way actors interact and create their identity
                  within overlapping social contexts. Each actor interacts in multiple contexts
                  within different types of social relations that constitute their social space. In
                  this article, we present an approach to model structured agent societies with
                  multiple coexisting social networks. We study the notion of \textit{context
                    permeability}, using a game in which agents try to achieve global consensus. We
                  design and analyse two different models of permeability. In the first model,
                  agents interact concurrently in multiple social networks. In the second, we
                  introduce a \textit{context switching} mechanism which adds a dynamic temporal
                  component to agent interaction in the model. Agents switch between the different
                  networks spending more or less time in each one. We compare these models and
                  analyse the influence of different social networks regarding the speed of
                  convergence to consensus. We conduct a series of experiments that show the impact
                  of different configurations for coexisting social networks. This approach unveils
                  both the limitations of the current modelling approaches and possible research
                  directions for complex social space simulations.
		\end{abstract} 
		

\end{frontmatter}
`	\newpage
	\section{Introduction}
	\label{sec:introduction}
	
        \noindent In most simulations of complex social phenomena, agents are considered to inhabit
        a space in which structure is very simple. This space has little resemblance with the social
        world it is supposed to depict and for which conclusions are supposed to be
        extrapolated. This simplicity does not come by chance, rather, it is necessary and desired
        by the researchers: the problems to be approached are themselves so complex that whichever
        factors of complexity can be reduced (or at least postponed), the reduction is always
        welcome. So, it is common practice that geographical space is reduced to a two-dimensional
        grid, and all social relations between agents are condensed into one more or less structured
        abstract relation. Most social simulation and modelling approaches disregard the fact that
        we engage in multiple social relations. Moreover, each kind of social relation can possess
        distinctive characteristics that include: rich information such as degree of connectivity,
        centrality, trust, interactions frequency, asymmetry, and so on.
	
        To explore the addition of multiple relations and its consequences for the dissemination of
        phenomena in social simulations, we put forward to \textit{reduce} the emphasis given on
        agent individual interactions. We accomplished this by choosing a simple -- and especially
        neutral -- game to model those interactions. Our main concern was that the game itself would
        not provide biases or trends in the collective phenomenon being studied, so we chose the
        \textit{consensus formation game}, with a straightforward majority rule to decide the
        outcome of each individual move.

        While most models are simplified descriptions -- reductions -- of real-world phenomena, many
        constitute complex systems themselves and thus, we need techniques such as simulation to
        explore their properties. We can make a distinction between two types of complexity: the
        complexity of real systems (\textit{ontological complexity}) and the complexity of models
        (\textit{descriptive complexity})~\cite{Emmeche1997}. The levels of abstraction in a
        simulation model can then range from data-driven paradigms to more abstract descriptions
        that allow us to create \textit{what-if-scenarios}. The study of this abstract
        \textit{descriptive complexity} in simulation models is as valuable as its
        \textit{data-driven} counterpart. We may not be able to make predictions based on the direct
        application of such models; but, their study can inspire the engineering of artificial
        complex systems and reveal properties with applications beyond the explanation of observable
        phenomena. Studying consensus formation can thus advance our understanding of related
        real-world social phenomena.

        Examples of thematic real-world phenomena in social simulation models include for instance
        the joint assessments of policies or, in the context of economics and politics, \textit{the
          voting problem}. Herbert Simon's investigations on this problem were one of the first
        stepping stones to the field of social simulation \cite{Simon1954}; which then inspired
        models (like the one we present in this article) related to evolution and dissemination of
        opinions, which we call \textit{opinion dynamics}. In agent-based \textit{opinion dynamics},
        agent interactions are guided by \textit{social space} abstractions. In some models, the
        dimensionality and structure of this space is irrelevant (any agent can interact with any
        other agent). Other models use an underlying artefact that structures agent
        neighbourhoods. Axelrod for instance, represents agent neighbourhoods as a bi-dimensional
        grid in its model of \textit{dissemination of culture} \cite{Axelrod1997}. In an attempt to
        mimic real social systems, one can also make use of \textit{complex network models} to
        create the infrastructure that guides agent interaction (see \cite{Weisbuch2004} for an
        example).
	
	In real-world scenarios, actors engage in a multitude of social relations different in kind
        and quality. Most simulation models don't explore \textit{social space} designs that take
        into account the differentiation between coexisting social networks. Modelling multiple
        coexisting relations was an idea pioneered by Peter Albin in \cite{Albin1975} but without
        further development. The process of interacting in these different complex social dimensions
        can be seen as the basis for the formation of our social identity
        \cite{Roccas2002,Ellemers2002}.
	
	In this article, we explore the of modelling opinion dynamics with multiple social
        networks. We look at how the properties of different network models influence the
        convergence to opinion consensus. We present a series of models that use these networks in
        different ways: which create distinct emergent dynamics. We want to study the consequences
        of using multiple social networks at the same time while maintaining the interaction model
        as simple and abstract as possible (following the methodology in most opinion dynamics
        literature). We present two models where agents can: interact at the same time in the
        multiple networks (choosing partners from any network); or switch between networks (choosing
        only partners from their current network).
	
	\subsection{Article Structure}
        \noindent This article is organised as follows. In the next section, we present the work
        related to \textit{opinion dynamics}, \textit{social space} modelling and \textit{complex
          network models}. In the following section, we describe our game of consensus and introduce
        our multiple model variations. These are designed to study the notion of \textit{context
          permeability} and consensus formation in multiple social networks. Section
        \ref{sec:experimental-setup} describes the experimental setup; the set of tool and
        methodologies followed to conduct our investigations. In section
        \ref{sec:results-discussion}, we present and discuss our results and compare the different
        simulation models. Finally, we summarise what we learned from our experiments and point out
        future research directions.
	
	\section{Related Work}
	\label{sec:relatedwork}
        \noindent In this section, we present work related to \textit{opinion dynamics},
        \textit{social space} modelling, and \textit{complex network models}.
	\subsection{Opinion Dynamics and Consensus Formation}
        \noindent Formal opinion dynamics models provide an understanding, if not an analysis, of
        opinion formation processes. An early formulation of such models was created as a way to
        understand complex phenomena found empirically in groups \cite{French1956}. The work on
        consensus building (in the context of decision-making) was first studied by DeGroot
        \cite{Degroot74} and Lehrer \cite{Lehrer1975}. Empirical studies of opinion formation in
        large populations have methodological limitations. Computational sociology arises with a set
        of tools -- simulation models in particular -- to cope with such limitations. We use
        \textit{multi-agent simulation (MAS)} as a methodological framework to study such social
        phenomena in a larger scale. Most opinion dynamics models make use of binary opinion values
        \cite{Galam1997,Antunes2009}, or continuous values \cite{Deffuant2000,Deffuant2002}. For a
        detailed analysis over some opinion dynamics models, refer to \cite{Hegselmann2002}.
	
	Opinion dynamics models allow us to discover under which circumstances a population of
        agents reaches consensus or polarisation. Agent-based models can have broader application
        outside social simulation though. In computational distributed systems, consensus is a means
        by which processes agree on some data~value needed during computation. Typically, this
        agreement is the result of a negotiation process (often with the aid of a mediator). In
        human societies, social conventions emerge to deal with coordination and subsequently with
        cooperation problems \cite{Lewis1969}. These conventions are regularities of behaviour that
        can turn normative if they come to be persistent solutions to recurrent problems. MAS are
        also capable of producing emergent conventions in coordination or cooperation problems
        \cite{Delgado2002}. The decentralised nature of these computational models of consensus is
        highly desirable for dynamic control problems. In these scenarios, creating conventions
        before hand (off-line), or developing a central control mechanism for generating them, can
        be a difficult and intractable task: either due to the uncertainty and complexity associated
        with the environment, or due to the system scale and heterogeneity (which makes it very
        difficult to handcraft each component).
	
	Consensus models are also investigated as means to create conventions in MAS. Shoham and
        Tennenholtz create a bridge between economic literature and machine learning by studying a
        series of models in which multiple agents are engaged in learning a particular convention
        \cite{Shoham1994}. They call this process \textit{co-learning}. The complexity of these
        systems comes from its concurrent nature: one agent adapts to the behaviour of another
        agents it has encountered, these other agents update their behaviour in a similar fashion,
        which results in a highly non-linear system dynamics. In their work, Shoham and Tennenholtz
        define the notion of \textit{stochastic social games} and compare different rules with the
        objective of establishing conventions in a decentralised fashion. In their stochastic games,
        they present one particular rule that we use in our models:
	
	\begin{mydef}
		\label{def:external-majority}
		\textit{External Majority (EM)} update rule: adopt action $i$ if so far it was
                observed in other agents more often than other action and remain with your current
                action in the case of equality.
	\end{mydef}
	
	This EM rule was shown to coincide with \textit{Highest Cumulative Reward (HCR)}, which is a
        simple rule that states that ``an agent should adopt an action that has yielded the highest
        cumulative reward to date.'' We use the EM rule in our simulation models for its simplicity
        and success in the evolution of conventions.
	
	\subsection{Social Structure in Simulation Models}
        \noindent The usage of a bi-dimensional grid to represent abstract social spaces in
        simulation models, is one of the most widely used approaches in the agent-based simulation
        and modelling literature. This has its origins in the \textit{``checkerboard''} model
        introduced in computational social sciences by Schelling and Sakoda
        \cite{Schelling1971,Sakoda1971}. Another famous example that uses bi-dimensional grids as
        social structure is the social simulation model of dissemination of culture from Axelrod
        \cite{Axelrod1997}.
	
	Cellular automata (CA) are an example from the area of \textit{artificial life} that also
        makes use of bi-dimensional structures to model neighbourhood interactions. While these
        models are idealised frameworks, their exploration can bring us deeper insights than models
        with high level of descriptive complexity (which would render their analysis very
        difficult). The work of Flache and Hegselmann relaxes the standard CA assumptions about the
        regular bi-dimensional grids by considering irregular grid structures (Voronoi diagrams)
        \cite{FlacheHegselmann2001}. They present results on the robustness of some important
        general properties in relation to variations in the typical grid structure. There are also
        contributions that incorporate both the complexity of network models and the behaviour of
        dynamic processes. One example is the work in \cite{Sullivan2000}, which explores a
        graph-based cellular automaton to study the relationship between spatial forms of urban
        systems and the robustness of different process dynamics under spatial change. Moreover,
        this shows how we can use real geographical information to construct graph-based cellular
        automata (thus making a connection between purely abstract models and data-driven models).
	
	Finally, another type of social space models are the \textit{random graph/network
          models}. Each complex network -- or \textit{class} of complex networks -- captures
        specific topological properties. These properties are found in real-world network
        structures. One of the first complex network models is the random graph from Erd\H{o}s and
        Rényi \cite{Erdos1959}. Other examples include the model of preferential attachment
        \cite{Barabasi1999}, and the small-world networks \cite{Watts1998}. One common issue of
        these models is the fact that once generated, the network structure is static. The models
        are not really suitable for the description of highly dynamic groups or communities. An
        example of a model that takes such dynamics into account is the model of team
        formation in~\cite{Gaston2005}. A more in-depth review of network models (and their properties)
        is given in~\cite{Nunes2012}.

	\section{Multi-context Models}
	\label{sec:multi-context-models} 
\begin{figure}[h]
          \centering
          \includegraphics[width=0.7\linewidth]{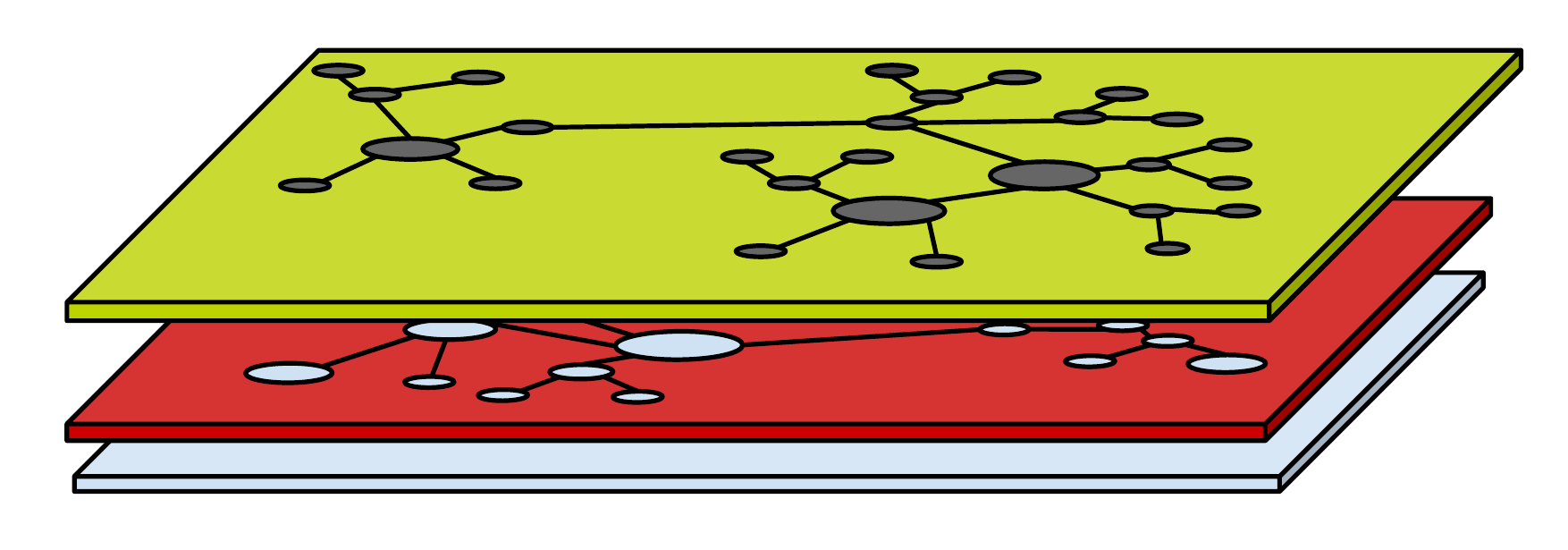}
          \begin{minipage}{0.9\linewidth}
            \caption{Multiple social network structure that shapes the social space in our
              simulation models.}
            \label{img:multiple-relations}
          \end{minipage}
	\end{figure}

        \noindent In this section, we present our modelling approach to explore the concept of
        permeability between contexts. We use multiple social networks to represent the complex
        social space in which an agent is inserted. In a simulation model, this setting can be seen
        as a n-dimensional scenario where each dimension contains a network that represents a
        different social relation (see figure \ref{img:multiple-relations}). The word ``context'' is
        used in many different senses and it is by itself subjected to many analysis
        \cite{Hayes1997}. Here, we use the term \textit{social context} in a simpler and more strict
        way. Agents belong to distinct social contexts which are their neighbourhoods in these
        multiple networks. The \textit{social context} of an agent is thus a set of neighbours
        currently available for interaction at a particular network or set of
        networks. \textit{Context permeability} is the ability of a particular norm, strategy,
        opinion, or trend, to permeate from one social relation to another. Agents belonging to
        disjoint neighbourhoods in different networks serve as bridges between subsets of the
        population that wouldn't otherwise be influenced by a consensus formed in a distant
        coalition.	
	
	We study the notion of context permeability in different simulation models in which agents
        interact using a simple consensus game. The society of agents has to adopt a binary opinion
        value according to a majority rule. We use the speed of consensus as a measure for
        self-organisation and explore the relationship between different network topologies and this
        measure.
	
	We partitioned the design space of our experiments using different simulation models. Each
        model is designed to analyse different aspects of context permeability. In a first model
        \cite{Antunes2007, Antunes2010}, we study the notion of permeability by overlapping social
        networks. In a second model \cite{Antunes2009}, we analyse the dynamics introduced by
        switching between \textit{social contexts}(neighbourhoods in different networks). This adds
        a temporal component that changes the dynamics of the game: agents can only perform
        encounters with neighbours that are active in their current context. (It also adds the
        chance for some agents to be isolated during the simulation.)
	
	\subsection{Simulation and Consensus Game}
	\label{sec:sim_consensus}
        \noindent Our agent-based models are designed as \textit{discrete-event} models. On each
        simulation cycle, every agent executes a simulation step. The agents are selected in a
        random uniform fashion for execution. This is common practice in agent-based simulation and
        guarantees that there is no bias caused by the order in which the agents are executed. For
        each simulation model, we present a description of the individual agent behaviour. Every
        model we present can be seen as a binary opinion dynamics model, or a consensus game. In
        this game, the society of agents tries to reach an arbitrary global consensus about two
        possible choices or opinion values. During a simulation run, each agent keeps track of the
        number of each opinion value of their interaction partners. In each iteration, each agent
        selects an available neighbour to interact with and observes its current opinion value. The
        agent decides to switch its current opinion choice if the observed opinion value becomes the
        majority of the two possible choices. The goal of the game is to reach a global consensus,
        but the particular choice that gets collectively selected is irrelevant. What is important
        is that overall agreement is achieved. In the consensus game we consider, each agent uses
        the previously presented external majority (EM) rule to update its opinion value (see
        definition \ref{def:external-majority} in section \ref{sec:relatedwork}).
	
	The reason we are not interested in which exact option gets selected is the same reason why
        we chose such a content-neutral game for the individual interactions. From the very
        beginning, our research questions were focused on the collective properties, the structure
        of the networks involved, and the dynamical outcomes of the simulations: thus leaving out
        issues related with individual motivations and desired results (either for any individual
        agent or the society). Granted, most applications of this research will often include such
        goals, for inquisition into complex social issues, including possibly policy design, in
        which the individual rationality must be considered, and will have a very strong influence
        on the collective outcomes.
	
	\subsection{Context Permeability}
        \noindent The first model of context permeability \cite{Antunes2007,Antunes2010} is designed
        to study the setting where an agent is immersed in a complex social world: where it engages
        in a multitude of relations with other agents. In this model, we consider that the
        \textit{context permeability} is created by agents that can belong \textit{simultaneously}
        to several relations. As an example, two agents can be simultaneously family and
        co-workers. While some links might not connect two agents directly, others can contain
        neighbourhoods in which they are related~(either directly or through a common
        neighbour). Model \ref{model:context_permeability} describes the behaviour of each agent for
        each simulated step.
	\begin{algorithm}[H]
		\caption{Context Permeability}
		\label{model:context_permeability}
		\algsetup{indent=2em}
		\begin{algorithmic}
			\vspace{0.5em}
			\STATE $M$ \COMMENT{Simulation Model / Environment}
			\STATE $A$ \COMMENT{Current Agent in Execution}
			\\ \hrulefill 
			
			\STATE\COMMENT{\textbf{Randomly select a network}}
			\STATE $r$ $\leftarrow$ randomUniform(0,$ M.networks.length() $)
			\STATE network $\leftarrow$ $M.networks[r]$
			
			\STATE
			\STATE \COMMENT{\textbf{Randomly select a neighbour}}
			\STATE neighbours $\leftarrow$ network.neighboursOf($A$) 
			\STATE $r$ $\leftarrow$ randomUniform(0,neighbours.length())
			\STATE partner $\leftarrow$ neighbours[r]
			\STATE
			\STATE \COMMENT{\textbf{Update opinion and memory}}
			\STATE \COMMENT{opinions take the values \{0,1\}--so they can be used as an index}
			\STATE A.memory[partner.opinion]++
			\IF {(A.memory[partner.opinion] $>$ A.memory[A.opinion])}
			\STATE A.opinion $\leftarrow$ partner.opinion
			\ENDIF
		\end{algorithmic}
	\end{algorithm}
	In this model, on each time step, each agent select a random neighbour from any 
	social network available and updates its opinion based on the opinion of the selected partner using the \textit{external majority} rule.

	\subsection{Context Switching}
	\label{sec:models_cs}
        \noindent In the context switching model \cite{Antunes2009}, agents are embedded in multiple
        relations represented as static social networks and they switch \textit{contexts} with a
        probability $\zeta_{C_i}$ associated with context $C_i$. The switching probability is the
        same for every agent in a given network.  Each agent switches from its current neighbourhood
        to another network. The agents are only active in one context at a time, and can only
        perform encounters with neighbours in the same context. We can think of context switching as
        a temporary deployment in another place, such as what happens with temporary
        immigration. The network topology is static; when an agent switches from one network to
        another, they become inactive in one network and active in their destination.

	\begin{algorithm}[H]
		\caption{Context Switching}
		\label{model:context_switching}
		\algsetup{indent=2em}
		\begin{algorithmic}
			\vspace{0.5em}
			\STATE $M$ \COMMENT{Simulation Model / Environment}
			\STATE $A$ \COMMENT{Current Agent in Execution}
			\\ \hrulefill 
			
			\STATE\COMMENT{\textbf{Get the current network (\textit{context} is a network index)}}
			\STATE cNetwork $\leftarrow$ M.networks[A.context]
			\STATE 
			\STATE \COMMENT{\textbf{Get the neighbours for the current network}}
			\STATE neighbours $\leftarrow$ cNetwork.neighboursOf($A$) 
			\STATE \COMMENT {filter by agents active in the same network}
			\STATE neighbours $\leftarrow$ \{n $|$ n $\in$ neighbours $\wedge$ (n.context = A.context)\}
			\STATE
			\STATE\COMMENT{\textbf{Randomly select a neighbour}} 
			\STATE $r$ $\leftarrow$ randomUniform(0,neighbours.length())
			\STATE partner $\leftarrow$ neighbours[r]
			\STATE
			\STATE \COMMENT{\textbf{Update opinion and memory}}
			\STATE \COMMENT{opinions take the values \{0,1\}--so they can be used as an index}
			\STATE A.memory[partner.opinion]++
			\IF {(A.memory[partner.opinion] $>$ A.memory[A.opinion])}
			\STATE A.opinion $\leftarrow$ partner.opinion
			\ENDIF
			\STATE
			\STATE \COMMENT{\textbf{Switch to another network}}
			\STATE switchingProb $\leftarrow$ M.params.switchingProb(A.context)
			\STATE $r \leftarrow $ randomUniform(0,1)
			\IF{($r < $ switchingProb)}
			\STATE numNets $\leftarrow$ M.networks.length()
			\STATE nextNetworks $\leftarrow$ \{i $|$ (i $\in$ [0,numNets[) $\wedge$ (M.networks[i] != A.context)\}
			\STATE $r$ $\leftarrow$ randomUniform(0,nextNetworks.length())
			\STATE A.context $\leftarrow$ nextNetworks[r]
			\ENDIF
		\end{algorithmic}
	\end{algorithm}
	\noindent In this model, agents select an (active) partner from their current
        neighbourhood. If a partner is available, the agents update their opinion value based on
        EM. At the end of the interaction, the agent switches \textit{from} the current context to a
        different one with a probability associated with its current context.
	
	This model describes an abstract way to represent the time spent on each network using the
        switching probability $\zeta_{C_i}$. Here, the permeability between contexts is achieved
        using this temporal component. Context switching introduces a notion that has not been
        explored in the literature so far: the fact that, although some social contexts can be
        relatively stable, our social peers are not always available for interaction and spend
        different amounts of time in distinct social contexts.
	
	\section{Experimental Setup}
	\label{sec:experimental-setup}
        \noindent In this section, we present all the tools and processes necessary to produce the
        current research output. The experiments were developed using the MASON~\cite{Luke2005}
        simulation framework written in Java. All the code for the simulation models can be found
        here \cite{Nunes:Software:11067}. Moreover, all the results presented in this article were
        made reproducible by using the \textit{statistical computing language R} \cite{R2008} and
        the R package \textit{knitr} \cite{knitr2014}. Knitr is used for generating reports that
        contain the code, the results of its execution (plots and tables), and the textual
        description for such results. All the R code used to produce the data analysis and the
        configuration files used to reproduce the data can be found in
        \cite{NunesAntunes2014:Analysis:11898}.

        For each configuration in our experiments, we performed 100 independent simulation runs. The
        results are analysed in terms of value distribution, average and standard deviation, or
        variance over the 100 runs. Like we described in section \ref{sec:sim_consensus}, our models
        are discrete-event models. In each cycle agents execute a step in a random order: this is so
        that this dynamical system --our simulation model-- is not influenced by the order in which
        agents are executed. Simulations end when total consensus is achieved or 2000 steps have
        passed.

        \subsection{Social Network Models}
        \label{sec:exp-setup_network_models}
        \noindent The social networks used in our models are \textit{k-regular} and
        \textit{scale-free}. A k-regular network is a network where all the nodes have the same
        number of connections. These networks are constructed by arranging the nodes in a ring and
        connecting each node to their next $k$ neighbours. Each network has $2k$ edges per vertex.

        Scale-free networks are networks in which the degree distribution follows a power law, at
        least asymptotically. That is, the fraction $P(k)$ of nodes in the network having $k$
        connections to other nodes goes for large values of $k$ as: $ P(k) \sim k^{-\gamma}$ where
        $\gamma$ is a parameter whose value is typically in the range $2 < \gamma < 3$.

        We use the method proposed in \cite{Barabasi1999} by Barab\'{a}si and Albert to construct
        the scale-free network instances using a preferential attachment. This model builds upon the
        perception of a common property of many large networks: a scale-free-power-law distribution
        of node connectivity. This feature was found to be a consequence of two generic mechanisms:
        networks expand continuously by the addition of new vertices, and new vertices attach
        preferentially to sites that are already well connected (more commonly known as
        ``preferential attachment''). In these networks, the probability $P(k)$ of two nodes being
        connected to each other decays as a power law, following $P(k) \sim k^{- \gamma}$.

        The network instances were generated using the \textit{b-have network library}
        \cite{Nunes:Software:11069}. This is a Java library that allows the creation and
        manipulation of network/graph data structures. It also includes the more commonly used
        random network models from the literature.

        \subsection{Measuring Network Properties}
        \noindent We measure two structural properties of our networks: the \textit{average path
          length} and the \textit{clustering coefficient}. The \textit{average path length} measures
        the typical separation between two vertices in the graph. The \textit{clustering
          coefficient} measures the average cliquishness of the graph neighbourhood. This measure
        quantifies how close the neighbours of a node are to forming a clique. Duncan J. Watts and
        Steven Strogatz \cite{Watts1998} introduced the measure to characterise a class of complex
        networks called small-world networks.

        These properties characterise structures that can be found in the real scenarios. The
        small-world networks generated by the models of Watts and Strogatz for instance, model
        real-world social networks with short average path length, local clustering structures, and
        triadic closures between the nodes \cite{Watts1998}. One shortcoming of this model, is the
        fact that these networks do not possess node hubs found in many real-world social
        networks. (Nevertheless, these models are designed to create networks with specific
        properties, not as general models for all kinds of real social networks)

        Since we construct scenarios where we overlap highly clustered k-regular networks and
        scale-free networks with node hubs, we are interested in the kind of properties that emerge
        from merging these structures -- and if/how they are comparable to the existing models
        --. The networks were produced by the \textit{b-have network library}
        \cite{Nunes:Software:11069}, exported to files, and analysed using the \textit{igraph R
          package}~\cite{igraph2006}.

        \section{Results and Discussion}
        \label{sec:results-discussion}

        \noindent In this section, we analyse and discuss the results for different experiments with
        both the context permeability and context switching models. We start by analysing the
        structural properties that arise from combining different networks. We then observe the
        impact of different model configurations in the convergence to consensus. We correlate the
        outcomes of our context permeability model simulations with the structural properties of the
        multi-network structures. We also explore how the new dynamics introduced by the context
        switching mechanism affects the consensus building process.

        \subsection{Overlapping Network Properties}
        \label{sec:network_properties}
        \noindent In this first analysis, we investigate the properties of different network
        topologies used in our models. One of the parameters in some experiments is the number of
        networks in which the agents interact. Adding more networks implies the addition of more
        connections. We show that what is important is \textit{not just} the the number of
        connections, but the properties of the resulting structure (when we merge the networks). We
        analyse the networks as follows.

        Each network is generated in such a way that the node indexes are randomised. This means
        that we can have multiple networks with the same topology and each node can have different
        neighbourhoods in different networks. Also, neighbourhoods do not necessarily overlap due to
        the node shuffling. Adding more networks to the social space is not the same as creating a
        single network with twice the number of connections.

\begin{figure}
          \centering
          \includegraphics[width=1\linewidth]{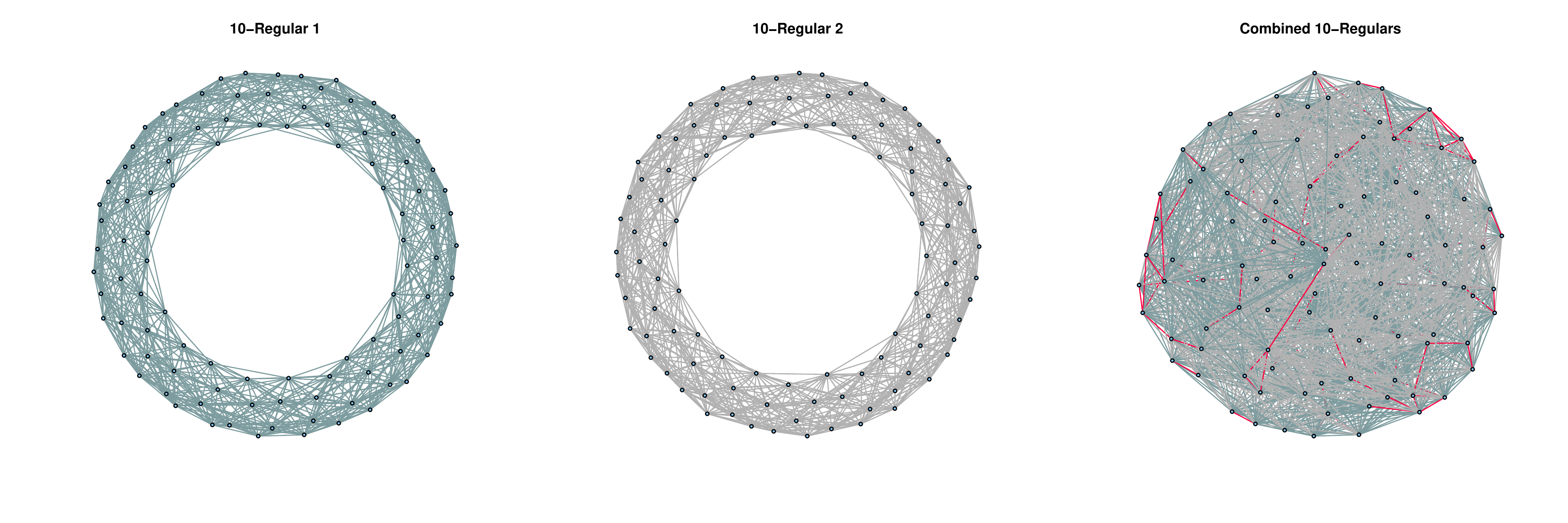} 
          \begin{minipage}{0.9\textwidth}
            \caption{Overlapping of two k-regular networks, each with k=10.}
            \label{fig:network_properties_merge_2_10regular}
          \end{minipage}
        \end{figure}

        \noindent Consider figure \ref{fig:network_properties_merge_2_10regular}. We created two
        random k-regular networks with k=10 and merged them ignoring edges from common neighbours.
        (In the simulation models this edge agglutination is not performed, we use it here for the
        purposes of network analysis.) Common edges are highlighted with a different colour.

        \begin{table}
          \centering
          \begin{minipage}{0.9\textwidth}
            \caption{Properties for overlapping of two k-regular networks, each with k=10.}
            \label{tab:network_properties_merge_2_10regular}
          \end{minipage}
          \begin{tabular}{lcccc}
            & Nodes &  Edges & Clustering Coef.	  &  Avg. Path Length \\ 
            \hline  10-regular 1 & 100 &  1,000  &  0.711 &  2.980 \\ 
            \hline  10-regular 2 & 100 & 1,000 & 0.711 &  2.980 \\ 
            \hline  Combined & 100 & 1,790  & 0.512 &  1.638 \\ 
            \hline 20-regular & 100 & 2,000	& 0.731	& 1.788 \\
            \hline 
          \end{tabular} 
        \end{table}

        \begin{wrapfigure}[13]{R}{0.46\textwidth}
          \includegraphics[width=1\linewidth]{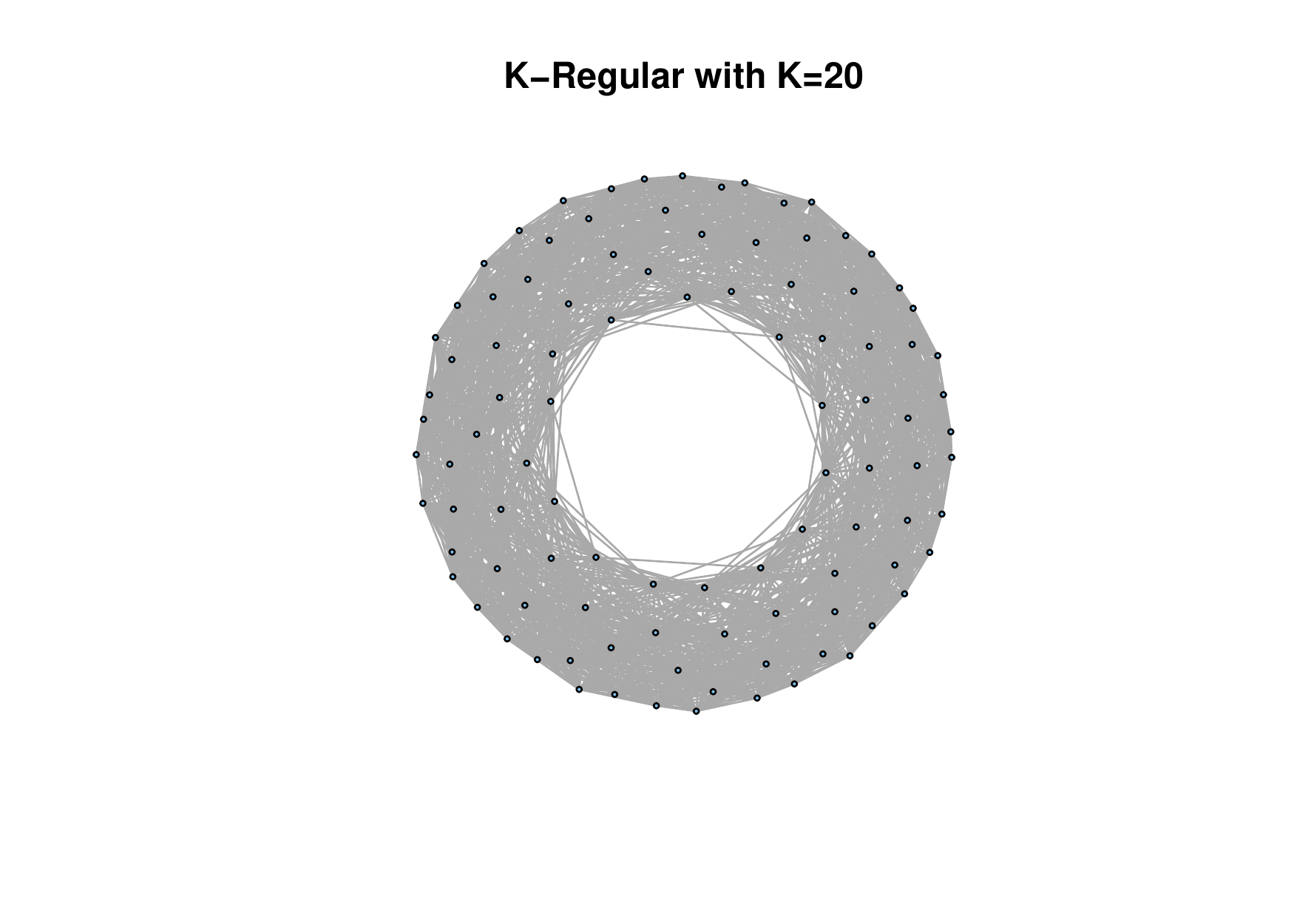}
          \begin{minipage}{0.7\linewidth}
            \caption{Single k-regular network with k=20.}
            \label{fig:network_properties_1_20regular}
          \end{minipage}
        \end{wrapfigure}

        The networks are drawn using the \textit{Kamada-Kawai} Layout \cite{Kamada19897} which
        treats the ``geometric'' (Euclidean) distance between two vertices in the drawing as the
        ``graph theoretic'' distance between them in the corresponding graph. We can see by the
        results of the layout in figure \ref{fig:network_properties_merge_2_10regular}, that the
        distance between nodes in the network has decreased, this is also confirmed by the reduced
        average path length of the combined networks (see table
        \ref{tab:network_properties_merge_2_10regular}). To illustrate our point, we analysed one
        k-regular network with k = 20 (see figure \ref{fig:network_properties_1_20regular} and the
        corresponding entry on table \ref{tab:network_properties_merge_2_10regular}). The clustering
        coefficient of a 20-regular network is approximately the same as the one of a 10-regular
        network (higher than the combination of two 10-regular networks). The average path length
        also decreases as there are more connections between previously distant nodes. As we will
        show, these properties may vary due to the node indexes being subjected to random
        permutations.

        Since we model the agent social space as a multitude of networks, we need to investigate the
        properties resulting from the merging of these networks. We do this empirically by taking
        the network instances used in the simulations, and analysing the distribution of clustering
        coefficient and average path length for the different configurations.

        \subsubsection{Properties of Overlapping K-Regular Networks}
        \label{sec:overlapping_kreg}
        \noindent We will now investigate what kind of properties we can get from merging k-regular
        networks. First, we analysed the distribution for the average path length and the clustering
        coefficient values over 100 network instances (each with 100 nodes). From a box plot
        preliminary analysis (see figures \ref{append_fig:network_properties_cc_kreg} and
        \ref{append_fig:network_properties_apl_kreg}), we can see that the average path length does
        not vary much for the 100 instances, as such, the average makes a good descriptor for these
        properties.

\begin{figure}[H] \centering
          \begin{subfigure}{.5\linewidth} \centering
            \includegraphics[width=1\linewidth]{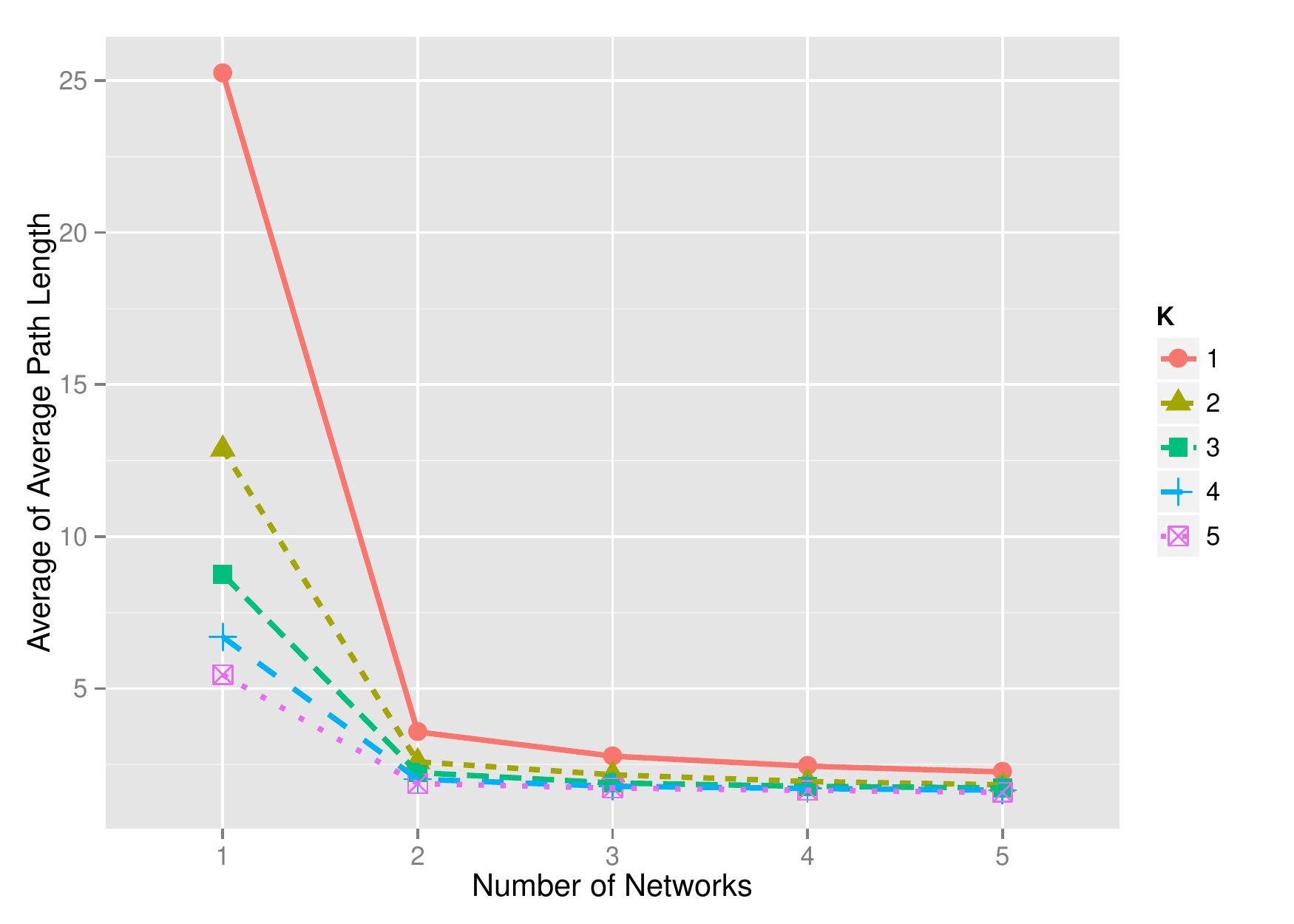}
            \caption{}
            \label{fig:network_properties_apl_line_kreg_12345}
          \end{subfigure}%
          \begin{subfigure}{.5\linewidth} \centering
            \includegraphics[width=1\linewidth]{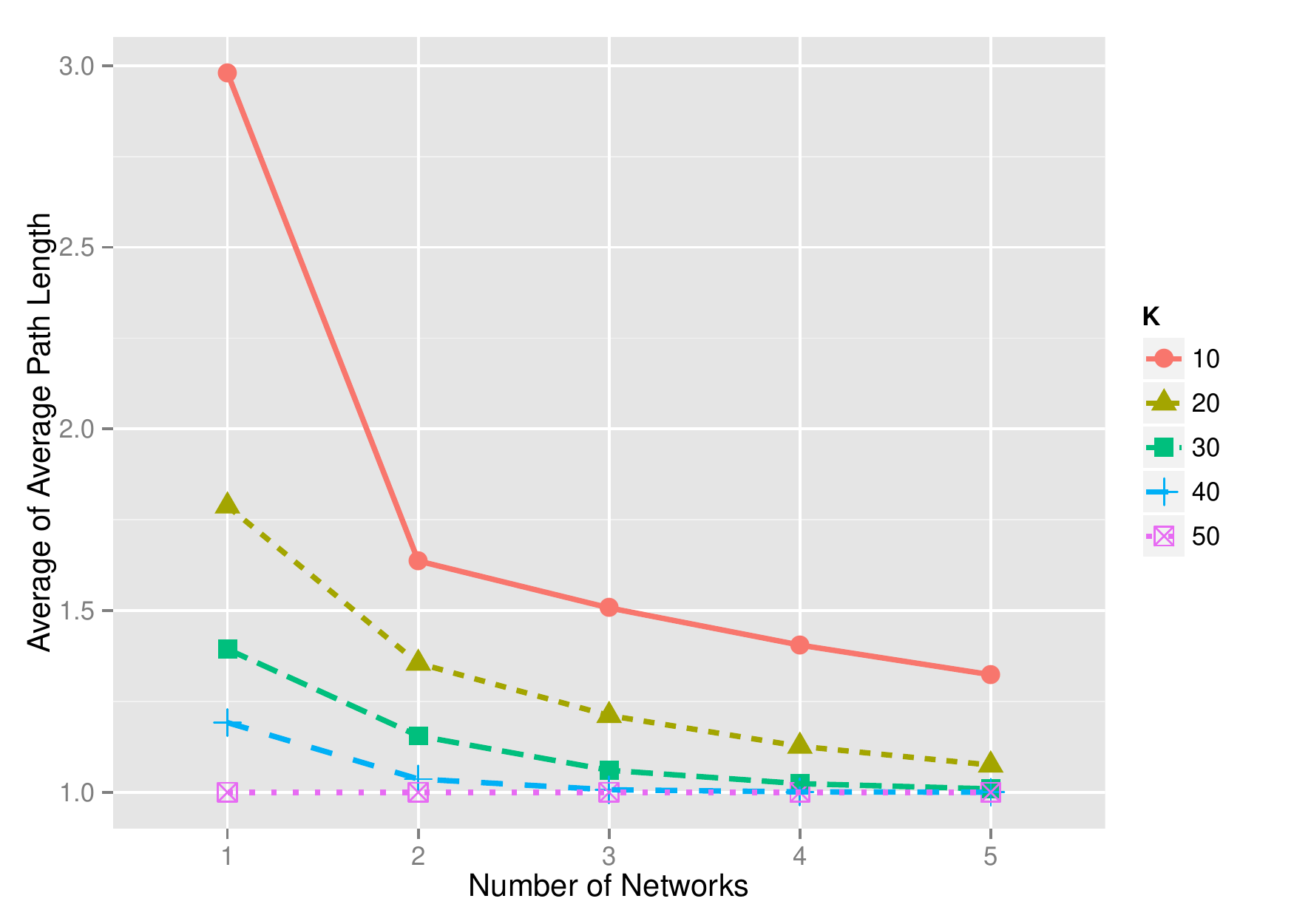}
            \caption{}
            \label{fig:network_properties_apl_line_kreg_1020304050}
          \end{subfigure}
     
     \begin{minipage}{0.9\textwidth} \vspace{0.2cm}
            \caption{Average value for the \textit{average path length} of 100 instances of
              overlapping k-regular networks (containing 100 nodes each) with
              k=\{1,2,3,4,5\}~(\ref{fig:network_properties_apl_line_kreg_12345}), and k=
              ~\{10,20,30,40,50\} (\ref{fig:network_properties_apl_line_kreg_1020304050}).}
		\label{fig:network_properties_apl_line_kreg}
              \end{minipage}

        \end{figure}

        Figure~\ref{fig:network_properties_apl_line_kreg} shows the average value for the
        \textit{average path length} of 100 k-regular network instances (each with 100 nodes). We
        can see that the \textit{average path length} changes more drastically when we go from 1 to
        2 networks. This is precisely the effect we can see in figure
        \ref{fig:network_properties_merge_2_10regular}. Merging these networks at random effectively
        creates multiple shortcuts between points that were not connected in the original k-regular
        topology. Beyond this point, adding more networks does not modify this structural property
        in a significant way. Note that, since each network has 100 nodes, with $k=50$ the network
        is fully connected (hence the average path length being 1).

\begin{figure}[H]
          \centering
          \begin{subfigure}{.5\linewidth}
		\centering
		\includegraphics[width=1\linewidth]{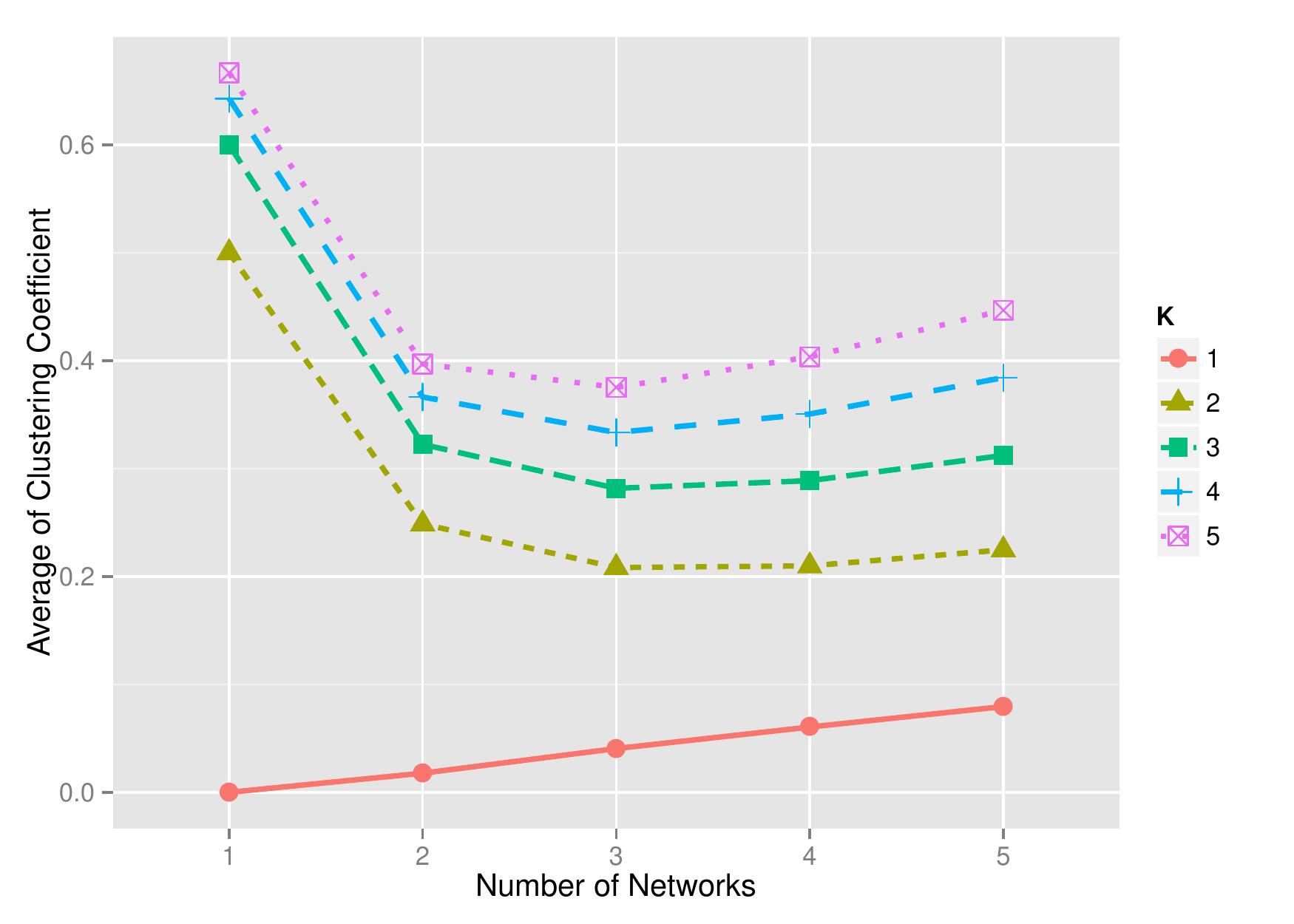}
		\caption{}
		\label{fig:network_properties_cc_line_kreg_12345}
              \end{subfigure}%
              \begin{subfigure}{.5\linewidth}
		\centering
		\includegraphics[width=1\linewidth]{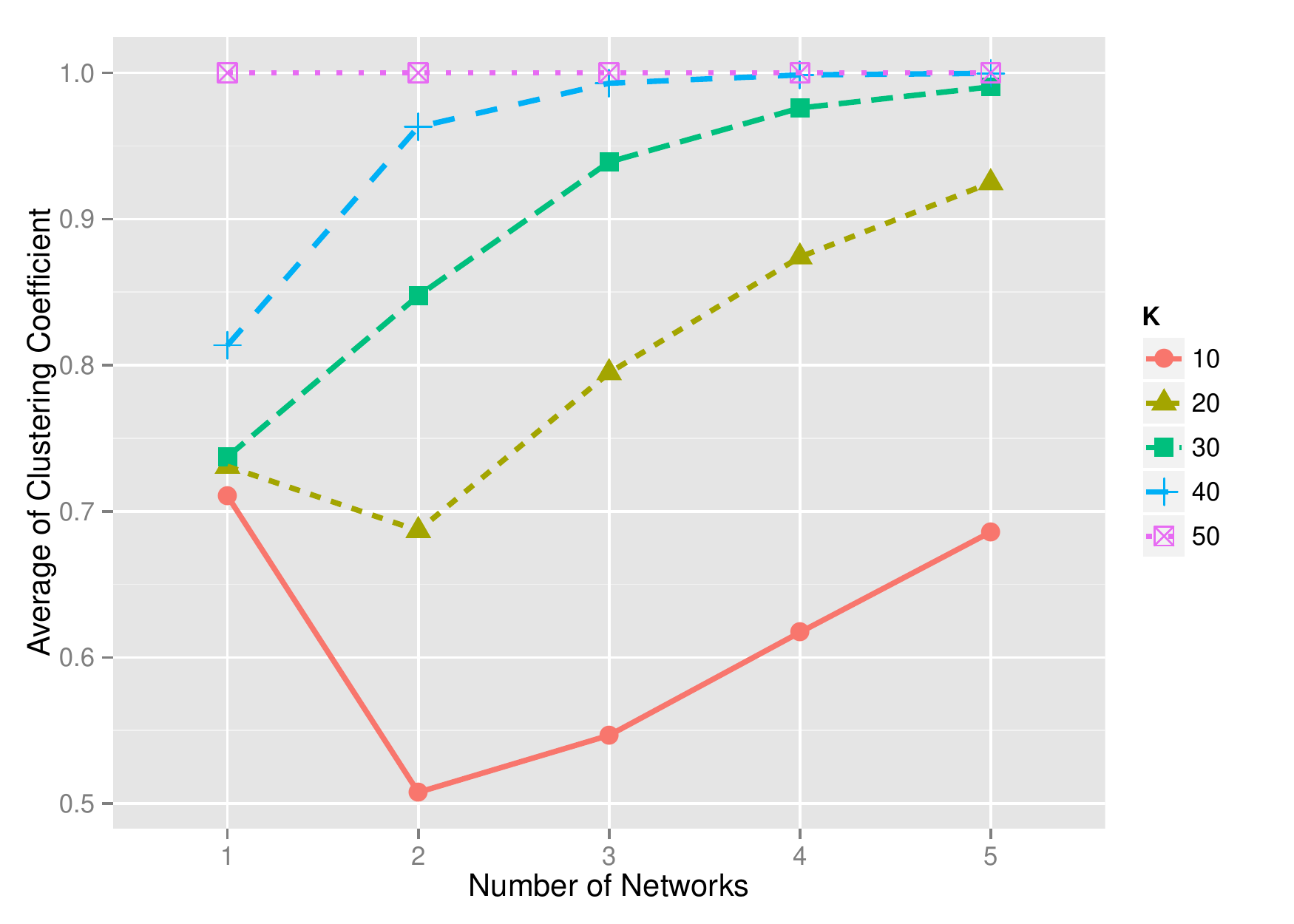}
		\caption{}
		\label{fig:network_properties_cc_line_kreg_1020304050}
              \end{subfigure}
              \begin{minipage}{0.9\textwidth}
		\vspace{0.2cm}
		\caption{Average value for the \textit{clustering coefficient} of 100 instances of overlapping k-regular networks (containing 100 nodes each) with ~k=\{1,2,3,4,5\}~(\ref{fig:network_properties_cc_line_kreg_12345}), and  k = \{10,20,30,40,50\} (\ref{fig:network_properties_cc_line_kreg_1020304050}).}
		\label{fig:network_properties_cc_line_kreg}
              \end{minipage}
        \end{figure}

        \noindent Figure~\ref{fig:network_properties_cc_line_kreg} shows the results for the average
        \textit{clustering coefficient} of 100 instances of overlapping k-regular networks. Adding
        multiple networks changes the clustering coefficient of the resulting structure. For
        $k \le 20$, the clustering coefficient drops when we add a second network. This happens
        because when we merge these networks, their connectivity is not high enough to generate
        shortcuts with tightly clustered neighbourhoods. The clustering coefficient is computed by
        taking the average local clustering coefficient of each node in the network. An initial
        highly clustered k-regular network (with all the nodes having the same structure) is
        affected when you modify the node neighbourhoods in such a way that some are more clustered
        than others. Networks with $k=1$ have a ring-like structure, so there are no nodes forming a
        clique\footnote{A clique is a group of nodes such that every two nodes are connected by an
          edge.} with their neighbours.

        \subsubsection{Properties of Overlapping Scale-free Networks}
\begin{figure}[H]
          \centering
          \begin{subfigure}{.5\linewidth}
		\centering
		\includegraphics[width=1\linewidth]{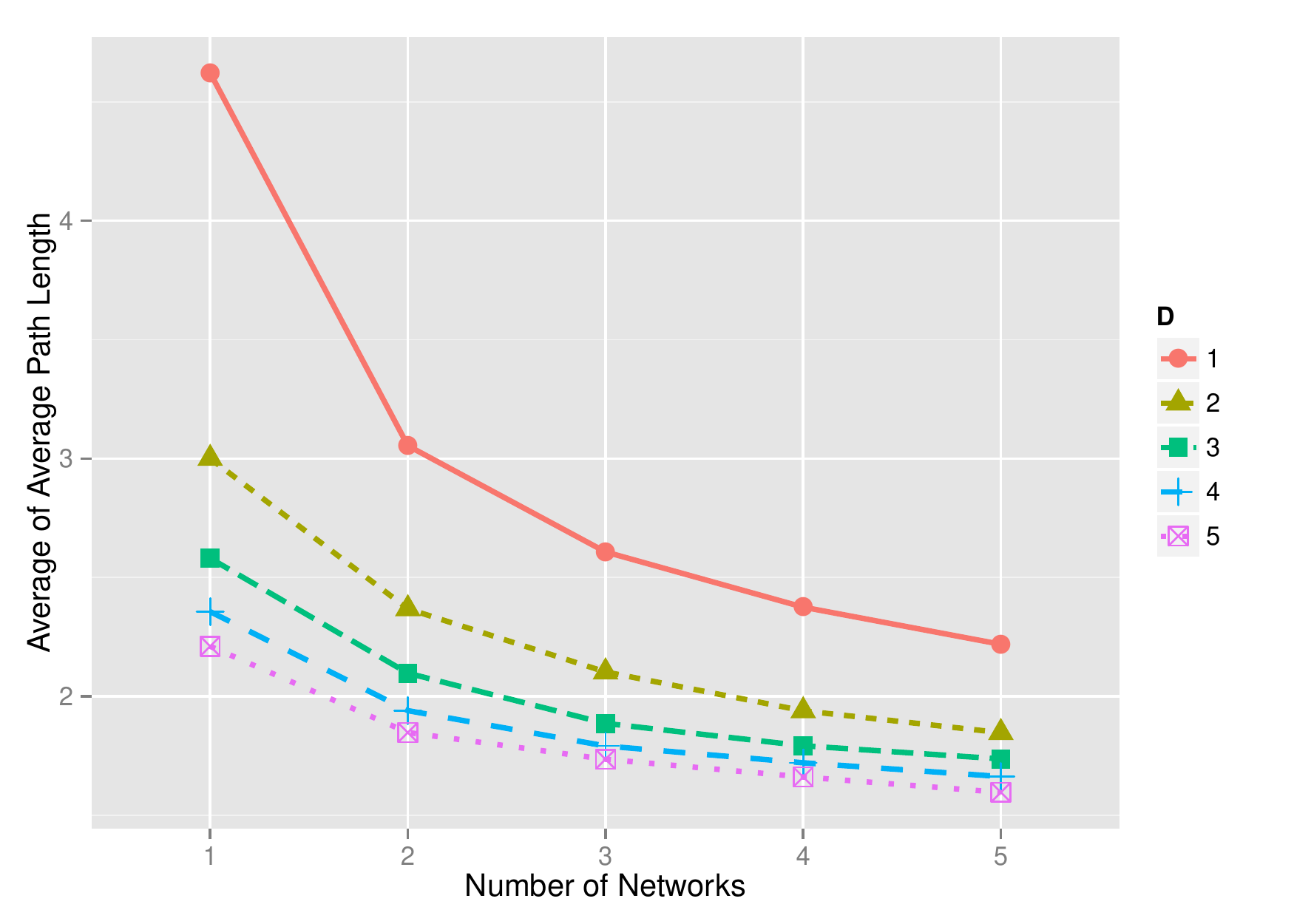}
		\caption{}
		\label{fig:network_properties_apl_line_sf_12345}
	\end{subfigure}%
	\begin{subfigure}{.5\linewidth}
		\centering
		\includegraphics[width=1\linewidth]{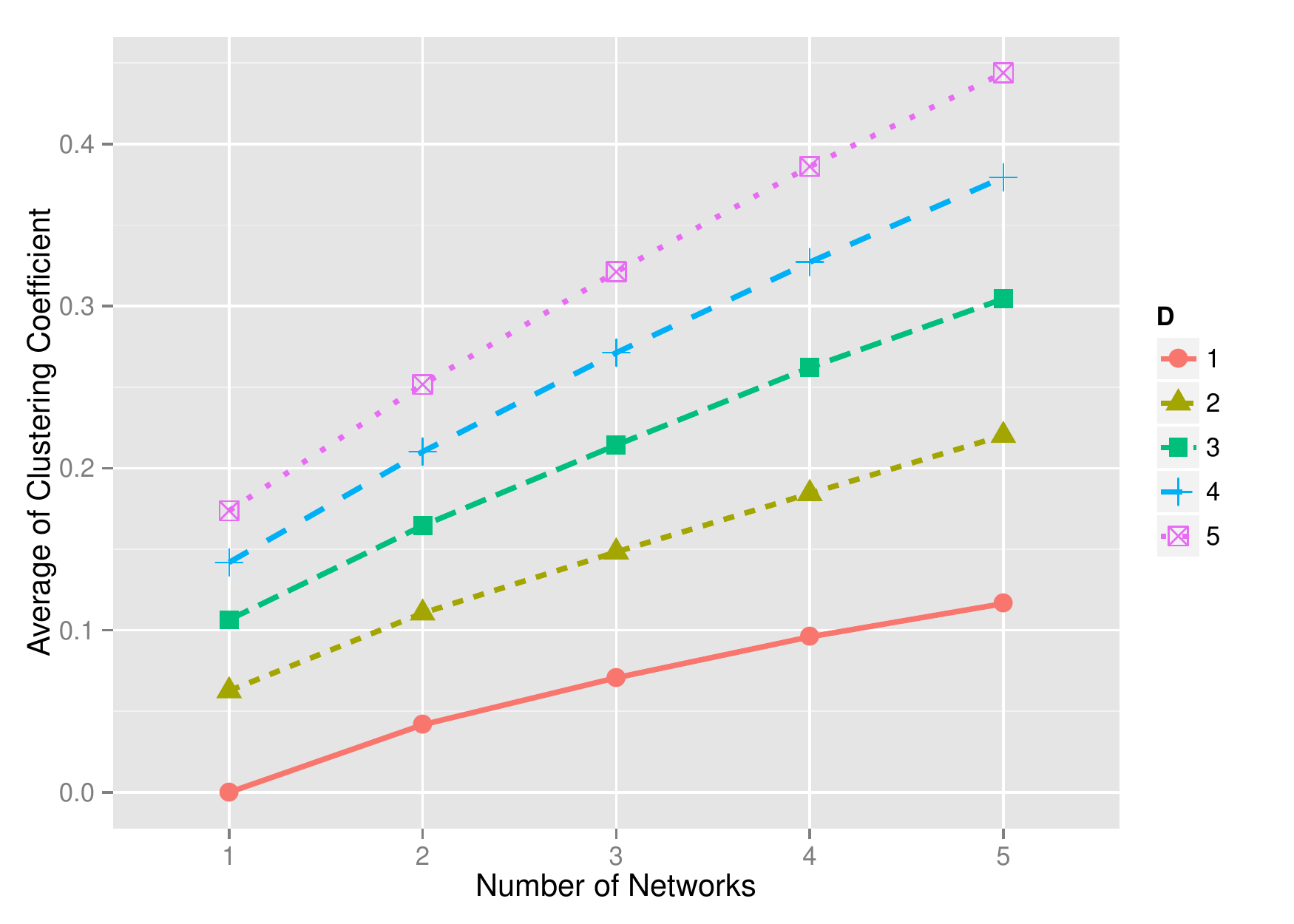}
		\caption{}
		\label{fig:network_properties_cc_line_sf_12345}
	\end{subfigure}
	\begin{minipage}{0.9\textwidth}
		\vspace{0.2cm}
		\caption{Average value for the \textit{average path length} (\ref{append_fig:network_properties_apl_sf_12345}) and \textit{clustering coefficient} (\ref{append_fig:network_properties_cc_sf_12345}) for 100 instances of overlapping \textit{scale-free} networks (containing 100 nodes each).}
		\label{fig:network_properties_line_sf}
	\end{minipage}
        \end{figure}

        \noindent We also looked at the properties that result from merging scale-free networks (see
        figure \ref{append_fig:network_properties_sf}). Like the previous results, the properties
        don't vary much between the 100 different instances. We can then say that specific
        configurations display a specific \textit{average path length} and \textit{clustering
          coefficient} both for \textit{k-regular} networks and \textit{scale-free} networks. We
        plotted the average of the \textit{average path length} (figure
        \ref{fig:network_properties_apl_line_sf_12345}) and \textit{clustering coefficient} (figure
        \ref{append_fig:network_properties_cc_sf_12345}) for each overlapping network configuration.

        The parameter $d$ (figure \ref{fig:network_properties_line_sf}) dictates how many edges are
        added by preferential attachment (see section\ref{sec:exp-setup_network_models}). For $d=1$,
        the resulting network is a \textit{forest}: a network composed of disjoint tree
        graphs. Figure~\ref{fig:network_properties_apl_line_sf_12345} shows us that the
        \textit{average path length} still decreases consistently when we add more networks. One of
        the characteristics of \textit{scale-free} networks is its short average path length, thus,
        for $d \ge 2$, adding more edges does not make a significant difference. Scale-free networks
        also have a reduced \textit{clustering coefficient} (see figure
        \ref{fig:network_properties_cc_line_sf_12345}). When we add more networks the
        \textit{clustering coefficient} increases.

        \subsubsection{Merging K-Regular with Scale-Free Networks}
\begin{figure}[H]
          \centering
          \includegraphics[width=1\linewidth]{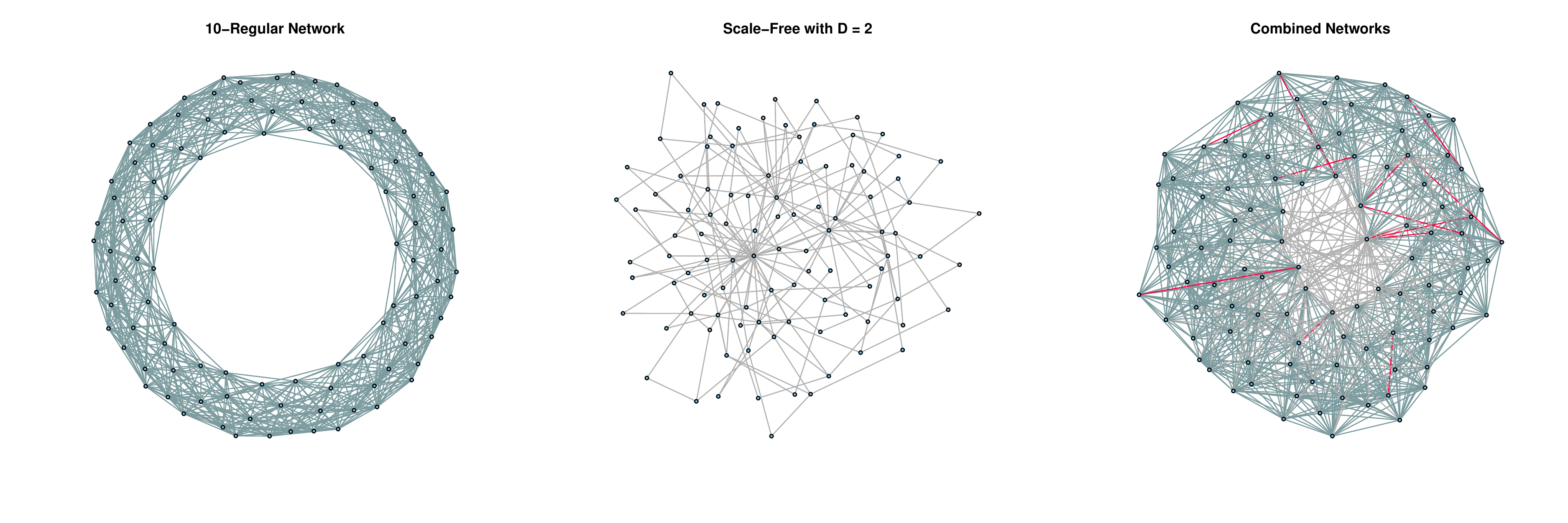}
          \begin{minipage}{0.9\textwidth}
            \caption{Overlapping of a 10-regular with a \textit{scale-free} network with D=2.}
            \label{fig:network_properties_merge_regular_scale-free}
          \end{minipage}
        \end{figure}

        \noindent Finally, we look at what happens when we merge k-regular networks with
        \textit{scale-free} networks. We didn't make an exhaustive analysis for this configuration,
        but we show what is the resulting structure of such merging. You can see in figure
        \ref{fig:network_properties_merge_regular_scale-free} that this results in a network with a
        lower \textit{average path length} due to the shortcuts created by the \textit{scale-free}
        network. It is also less dense than the structure with two 10-regular networks (figure
        \ref{fig:network_properties_merge_2_10regular}). We consider only the merging of two
        networks.

        \begin{figure}[H]
          \centering
          \begin{subfigure}{.5\linewidth}
            \centering
            \includegraphics[width=1\linewidth]{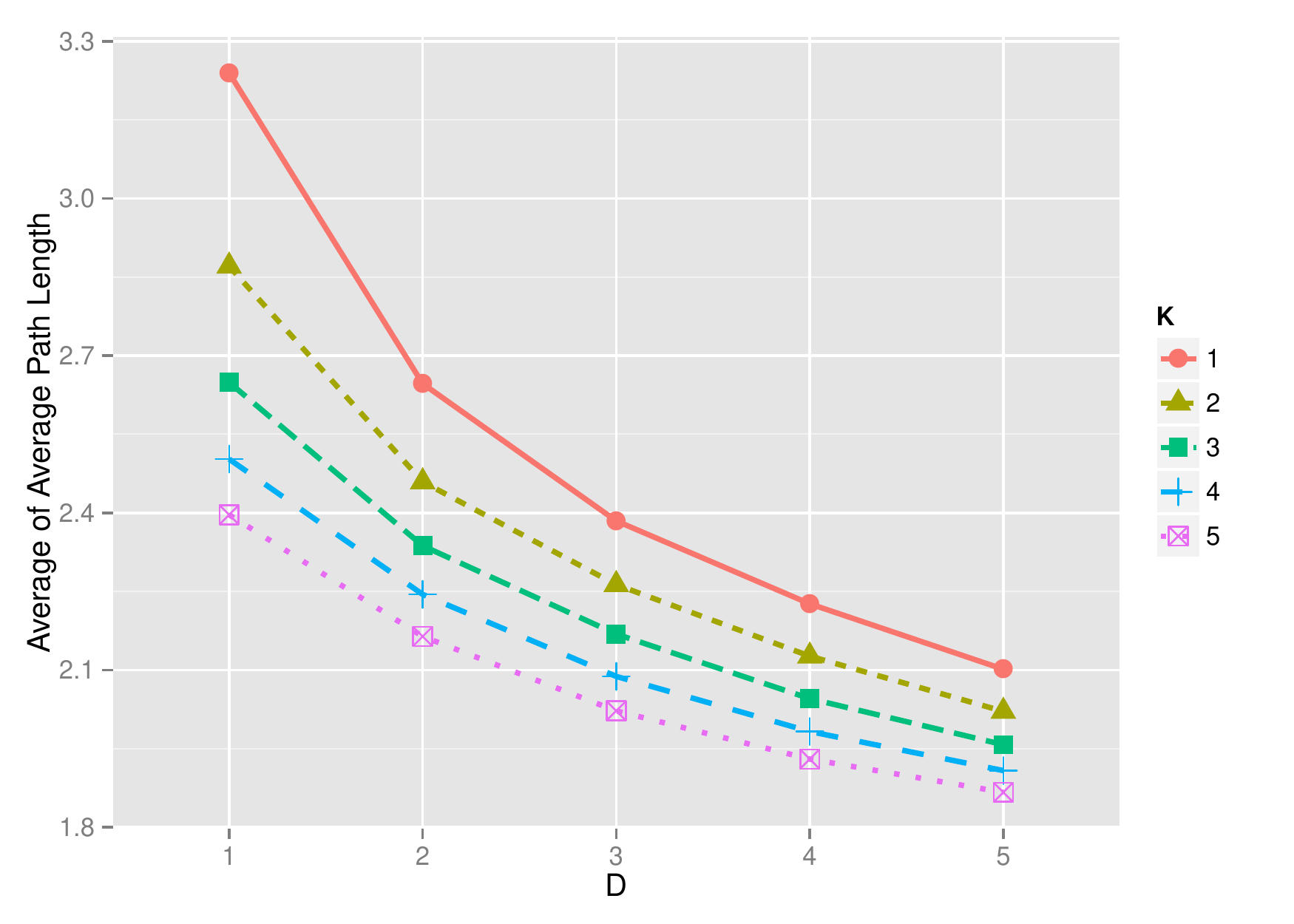}
            \caption{}
            \label{fig:network_properties_apl_line_sf_kreg_12345}
          \end{subfigure}%
          \begin{subfigure}{.5\linewidth}
            \centering
            \includegraphics[width=1\linewidth]{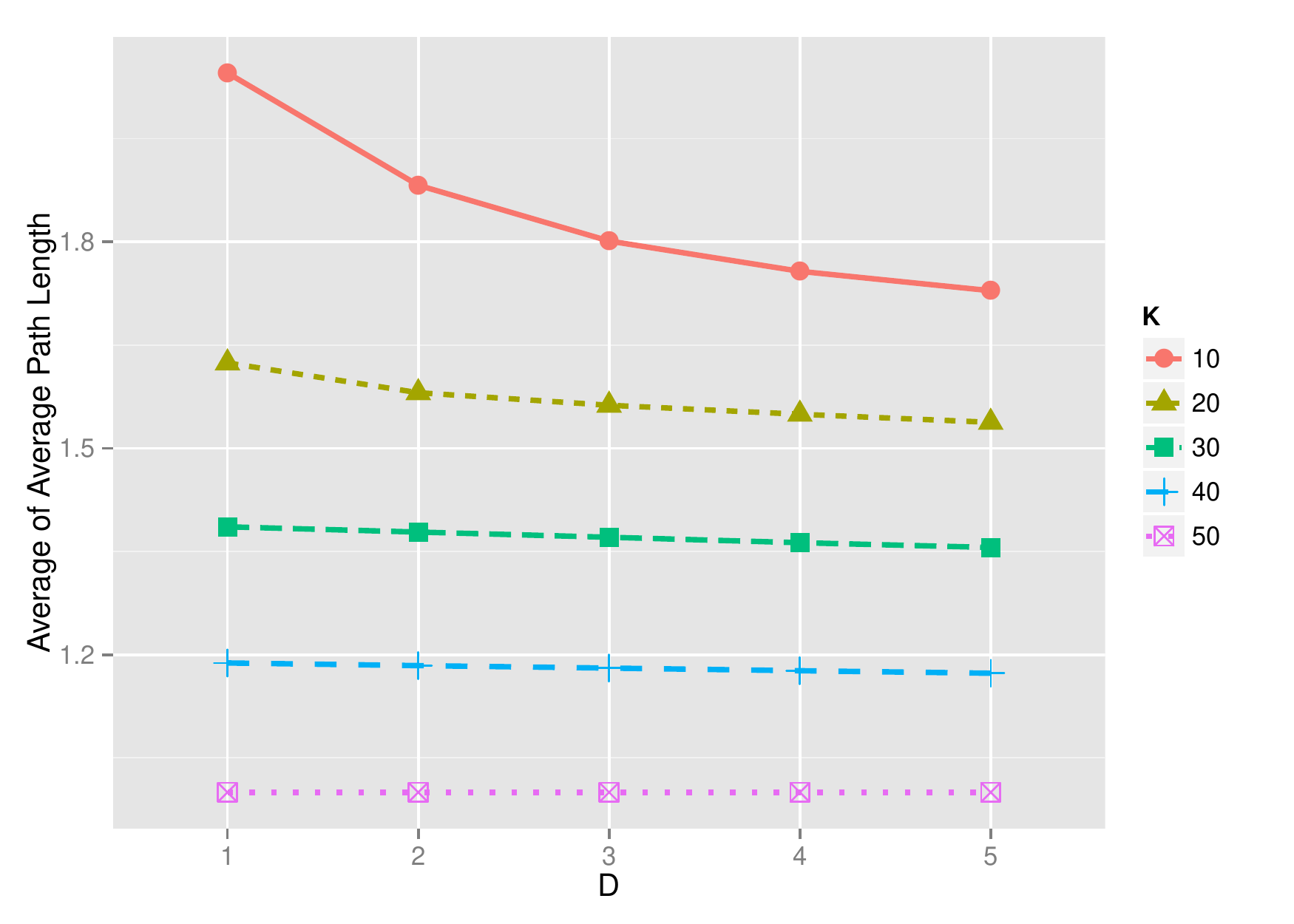}
            \caption{}
            \label{fig:network_properties_apl_line_sf_kreg_1020304050}
          \end{subfigure}
          \begin{minipage}{0.9\textwidth}
            \vspace{0.2cm}
            \caption{Average \textit{average path length} for 100 instances of heterogeneous network
              configurations: one \textit{scale-free} with $d=\{1,2,3,4,5\}$ and one
              \textit{k-regular} network with $k=\{1,2,3,4,5\}$
              (\ref{fig:network_properties_apl_line_sf_kreg_12345}) and $k=\{10,20,30,40,50\}$
              (\ref{fig:network_properties_apl_line_sf_kreg_1020304050}). (See the respective box
              plots in the appendix: figure \ref{append_fig:network_properties_apl_sf_kreg}).}
            \label{fig:network_properties_line_apl_sf_kreg}
          \end{minipage}
        \end{figure}

        Figure \ref{fig:network_properties_line_apl_sf_kreg} shows the \textit{average path length}
        of 100 independently generated instances of combinations of one \textit{k-regular} network
        and one \textit{scale-free} network with different values $k$ and $d$ respectively. For
        configurations where the \textit{k-regular} networks have low values of $k$ (figure
        \ref{fig:network_properties_apl_line_sf_kreg_12345}), adding a \textit{scale-free} network
        always decreases the \textit{average path length}. Moreover, the higher the $d$, the lower
        the path length (as more connections are formed between different nodes). With higher values
        of $k$ (figure \ref{fig:network_properties_apl_line_sf_kreg_1020304050}) however, increasing
        the value of $d$ does not decrease the \textit{average path length}. This is not surprising
        given the level of connectivity of this structure. The first network is already highly dense
        and connected, so adding a few extra connections doesn't matter; what does matter is that we
        have a scale-free network creating shortcuts between nodes that were not previously
        connected.

        \begin{figure}[H]
          \centering
          \begin{subfigure}{.5\linewidth}
            \centering
            \includegraphics[width=1\linewidth]{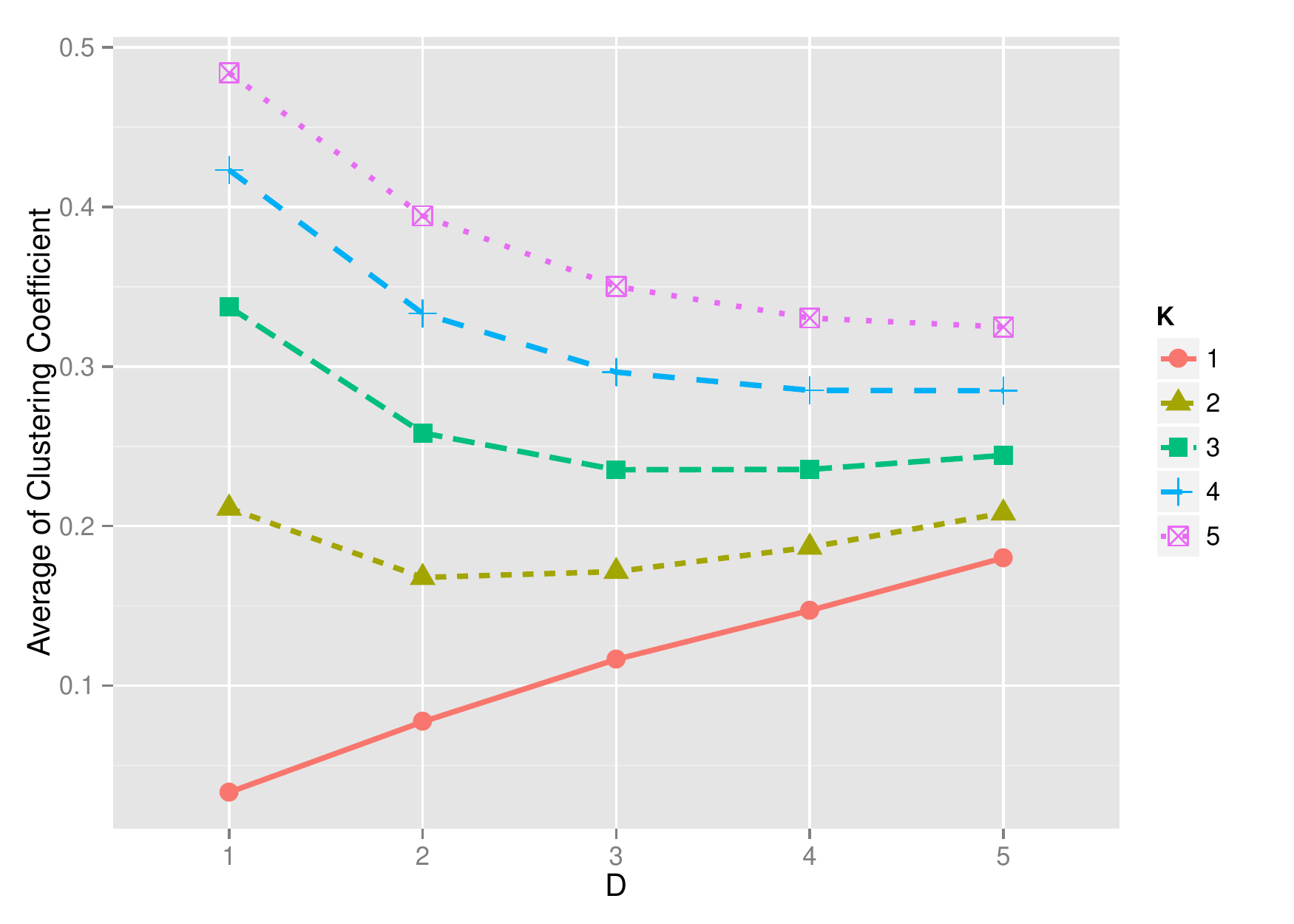}
            \caption{}
            \label{fig:network_properties_cc_line_sf_kreg_12345}
          \end{subfigure}%
          \begin{subfigure}{.5\linewidth}
            \centering
            \includegraphics[width=1\linewidth]{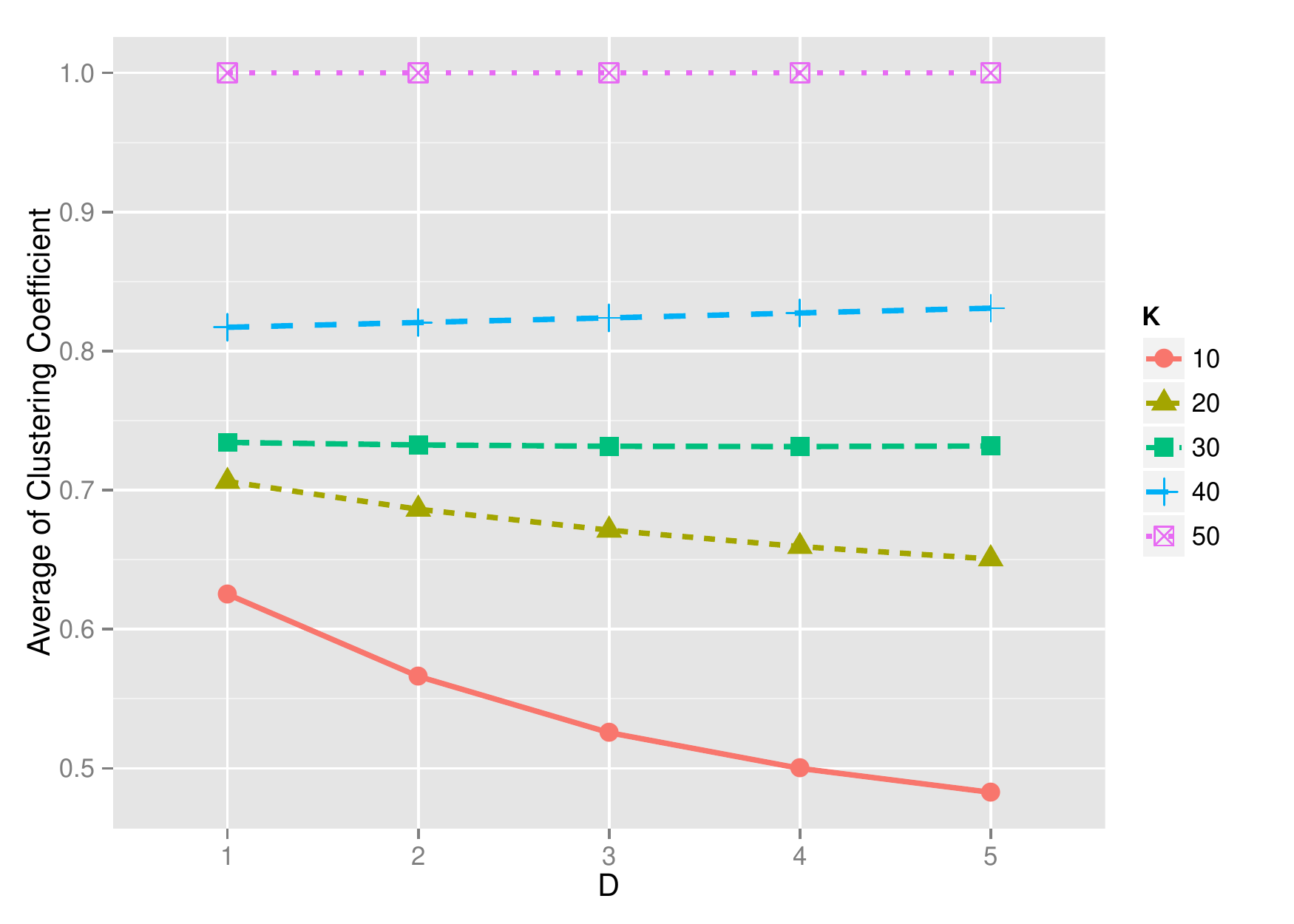}
            \caption{}
            \label{fig:network_properties_cc_line_sf_kreg_1020304050}
          \end{subfigure}
          \begin{minipage}{0.9\textwidth}
            \vspace{0.2cm}
            \caption{Average \textit{clustering coefficient} for 100 instances of heterogeneous
              network configurations: one \textit{scale-free} with $d=\{1,2,3,4,5\}$ and one
              \textit{k-regular} network with $k=\{1,2,3,4,5\}$
              (\ref{append_fig:network_properties_cc_sf_12345}) and $k=\{10,20,30,40,50\} $ (See the
              respective box plots in the appending:
              figure\ref{append_fig:network_properties_cc_sf_kreg}).}
		\label{fig:network_properties_line_cc_sf_kreg}
              \end{minipage}
        \end{figure}

        \noindent Figure \ref{fig:network_properties_line_cc_sf_kreg} shows the results for the
        \textit{clustering coefficient}. These are similar to the results we get for
        \textit{k-regular} networks (see figure \ref{fig:network_properties_cc_line_kreg}). The
        difference is that we are not adding more networks but rather increasing the parameter $d$
        of the scale-free (number of connections added by preferential attachment for each
        note). In this case, the effects are qualitatively similar with the ones of merged
        \textit{k-regular} network with low value of $k$. In contrast, for high values of $k$
        (figure \ref{fig:network_properties_cc_line_sf_kreg_1020304050}), a higher value of $d$ is
        not enough to make the network more clustered. This happens because scale-free networks are
        more sparse and have a very low number of triangles between nodes.

        \subsection{Context Permeability}
        \label{sec:results:context_permeability}

        \noindent In this section, we discuss the results for various sets of experiments on the
        \textit{context permeability model} \cite{Antunes2007,Antunes2010}. First, we used
        \textit{k-regular} and \textit{scale-free} networks with our model. Each network layer was
        configured with the same topology. We analyse the context permeability model in terms of
        convergence to consensus: the ratio of simulations that converged to total consensus over
        3000 runs; and how many agent encounters were needed for this to happen.

        \subsubsection{Convergence Ratio}
        \label{sec:ctx_perm_convergence}
        \noindent To analyse the ratio of convergence to consensus for different network
        configurations, we correlate this ratio with the properties that the combined network
        structure exhibits. As we have seen in section \ref{sec:network_properties}, the properties
        don't vary much for different random instances, so we use the average as the descriptor for
        these properties.

        The following tables show the results of our simulations. Table
        ~\ref{tab:regular_convergence} shows the percentage of total consensus achieved with 2000
        cycles for 3000 runs. For k-regular networks with a small $k$, consensus is rarely achieved
        with a single network. However, as soon as we add more networks, consensus is achieved in a
        significantly greater number of occasions. These results are especially interesting for the
        \textit{scale-free} networks (table \ref{tab:scale-free_convergence}): as soon as we add
        more networks, for a \textit{scale-free} with $d\ge2$, we can achieve consensus in a lot of
        occasions. This is to show that achieving consensus is not just a matter of
        connectivity. With \textit{scale-free} networks, we can achieve consensus more often with
        less clustered networks. This convergence is achieved not because of the clustering
        coefficient but rather due to lower \textit{average path length}. This is difficult to
        visualise using only the tables and the previous plots so, to confirm this hypothesis, we
        measured the \textit{Spearman's correlation}\footnote{The Spearman's rank correlation
          coefficient is a non-parametric measure of statistical dependence between two variables. It
          assesses how well the relationship between two variables can be described using a
          monotonic function.} between the ratio of convergences to consensus, the average
        \textit{average path length}, and the average \textit{clustering coefficient} for the
        corresponding network configurations (see table \ref{tab:convergence_correlation}).

        \begin{table}[H]
          \centering		
          \begin{minipage}{0.9\textwidth}
            \caption{Ratio of convergence to total consensus in 3000 independent runs with with a
              number of concomitant networks ($\#$ nets.) equal in kind: k-regular networks with
              $k=\{1,2,3,4,5,10,20,30,40,50\}$.}
		\label{tab:regular_convergence}
              \end{minipage}

              \setlength{\tabcolsep}{.30000em}
              \begin{tabular}{c|SSSSSSSSSS}
		
		\toprule
		& \multicolumn{10}{c}{value for k} \\ 
		\# nets.  & 1 & 2 & 3 & 4 & 5 & 10 & 20 & 30 & 40 & 50\\
		\midrule
		1  & 0.0000 & 0.0000 & 0.0006667 & 0.0050  & 0.01633 & 0.1830 & 0.6713 & 0.9673 & 1.0000 & 1.0000 \\
		2  & 0.0770 & 0.5887 & 0.8136667 & 0.8967 & 0.93467 & 0.9817 & 0.9977 & 1.0000 & 1.0000 & 1.0000 \\
		3  & 0.6560 & 0.9417 & 0.9780000 & 0.9920 & 0.99433 & 0.9997 & 1.0000 & 1.0000 & 1.0000 & 1.0000 \\
		4  & 0.8780 & 0.9867 & 0.9966667 & 0.9993 & 1.00000 & 1.0000 & 1.0000 & 1.0000 & 1.0000 & 1.0000 \\
		5  & 0.9497 & 0.9937 & 0.9993333 & 1.0000 & 1.00000 & 1.0000 & 1.0000 & 1.0000 & 1.0000 & 1.0000 \\
		\bottomrule
              \end{tabular}
            \end{table}

            \begin{table}[H]
              \centering
              \begin{minipage}{0.9\textwidth}
		\caption{Ratio of convergence to total consensus in 3000 independent runs with a
                  number of concomitant networks ($\#$ nets.) equal in kind: \textit{scale-free}
                  networks with $d=\{1,2,3,4,5\}$.}
		\label{tab:scale-free_convergence}
              \end{minipage}
	
	\setlength{\tabcolsep}{.30000em}
	\begin{tabular}{c|SSSSS}
          \toprule
          & \multicolumn{5}{c}{value for d} \\ 
          $\#$ nets.  & 1 & 2 & 3 & 4 & \multicolumn{1}{c}{5} \\ 
          \midrule
          1  &  0.0000  &  0.2903  &  0.8590  &  0.9573  &  0.9873  \\
          2  &  0.2840  &  0.9287  &  0.9877  &  0.9973  &  0.9980  \\
          3  &  0.7793  &  0.9850  &  0.9970  &  1.0000  &  1.0000  \\
          4  &  0.9287  &  0.9977  &  0.9997  &  1.0000  &  1.0000  \\
          5  &  0.9737  &  0.9997  &  1.0000  &  1.0000  &  1.0000  \\
          \hline 
	\end{tabular}
      \end{table}

      \begin{table}[H]
	\caption{Spearman's correlation between convergence ratio (CR), average \textit{average path length} (APL) and \textit{clustering coefficient} (CC) along with the respective p-value and confidence interval, for a confidence level of 95\%.}
	\label{tab:convergence_correlation}

	\centering	
	\begin{tabular}{r|cc|cll}
		
          \toprule
          network model & X & Y & Correlation & CI(95\%) & p-value \\
          \midrule
          k-regular & CR & APL & -0.726 & [-0.836, -0.561] & \num[scientific-notation = true]{2.437e-09} \\
          k-regular & CR & CC & 0.264   &   [-0.015,  0.505]       &    \num[scientific-notation = true]{0.064}     \\
          \midrule
          scale-free & CR & APL &    -0.912     &  [-0.961, -0.808]       &   \num[scientific-notation = true]{2.28e-10} \\
          scale-free & CR & CC & 0.633   &   [0.318, 0.823]  &   \num[scientific-notation = true]{0.0006776} \\
          \hline
		
          \hline
	\end{tabular}
      \end{table}

      \subsubsection{Number of Encounters to Achieve Consensus}
      \label{sec:ctx_perm_encounters}
      \noindent We now analyse the convergence in terms of average number of meetings during a
      simulation run. Since the maximum number of simulation cycles is 2000, the maximum number of
      encounters is 2~000~000 (we have 100 agents and each one performs one encounter per cycle). We
      show that for some configurations, the average of encounters is not a good descriptor since it
      varies greatly from run to run. Rather than considering the average, it is better to look at
      the distributions of these measures. Nevertheless, we included the data with all the average
      number of encounters, as well as the respective standard deviations for different network
      configurations. You can find this data in appendix \ref{append_ctx_permeability}.

\begin{figure}[H]
        \centering
	\begin{subfigure}{.49\linewidth}
		\centering
		\includegraphics[width=1\linewidth]{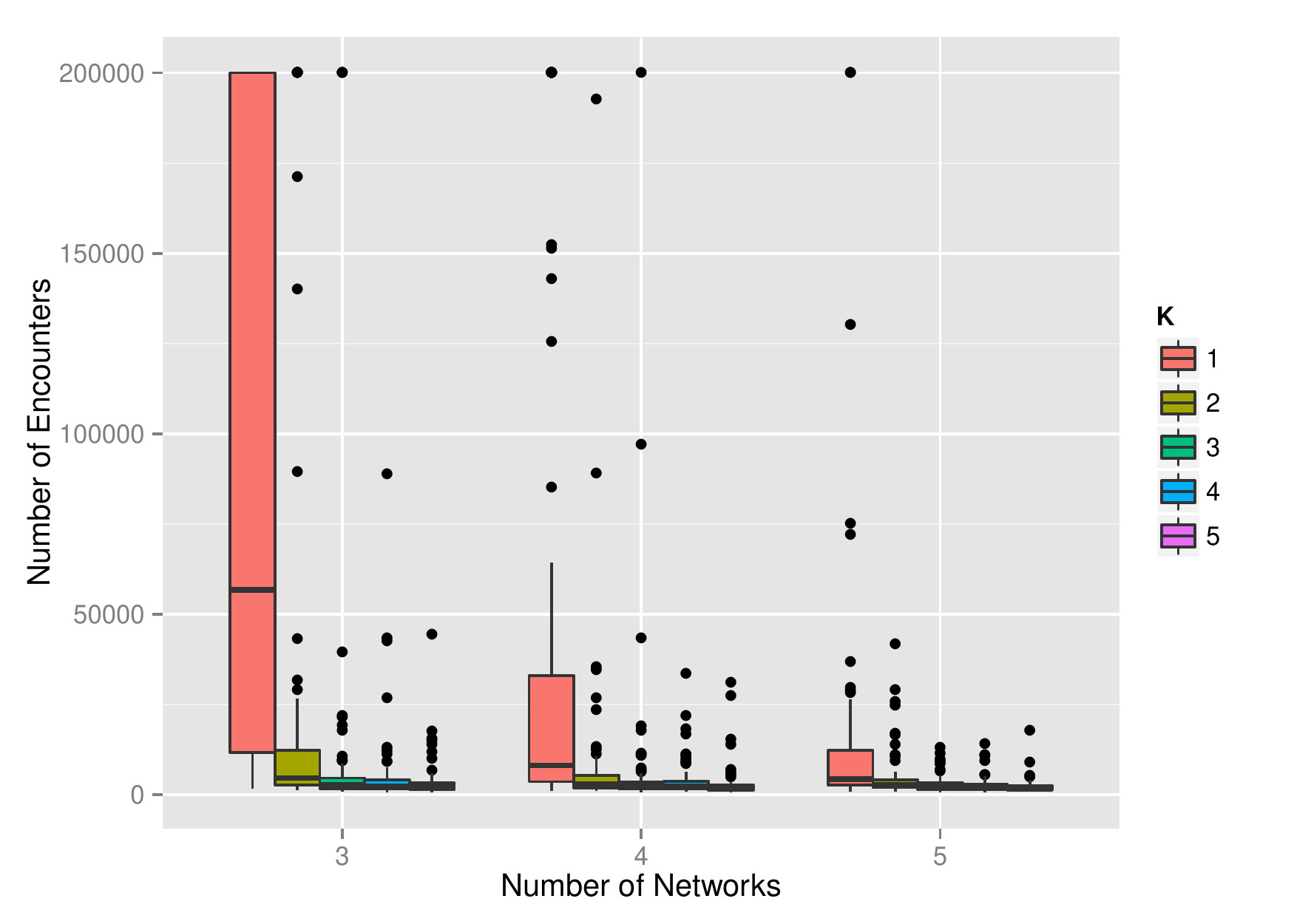}
		\caption{}
		\label{fig:ctx_perm_kreg_12345}
              \end{subfigure}%
              \begin{subfigure}{.49\linewidth}
		\centering
		\includegraphics[width=1\linewidth]{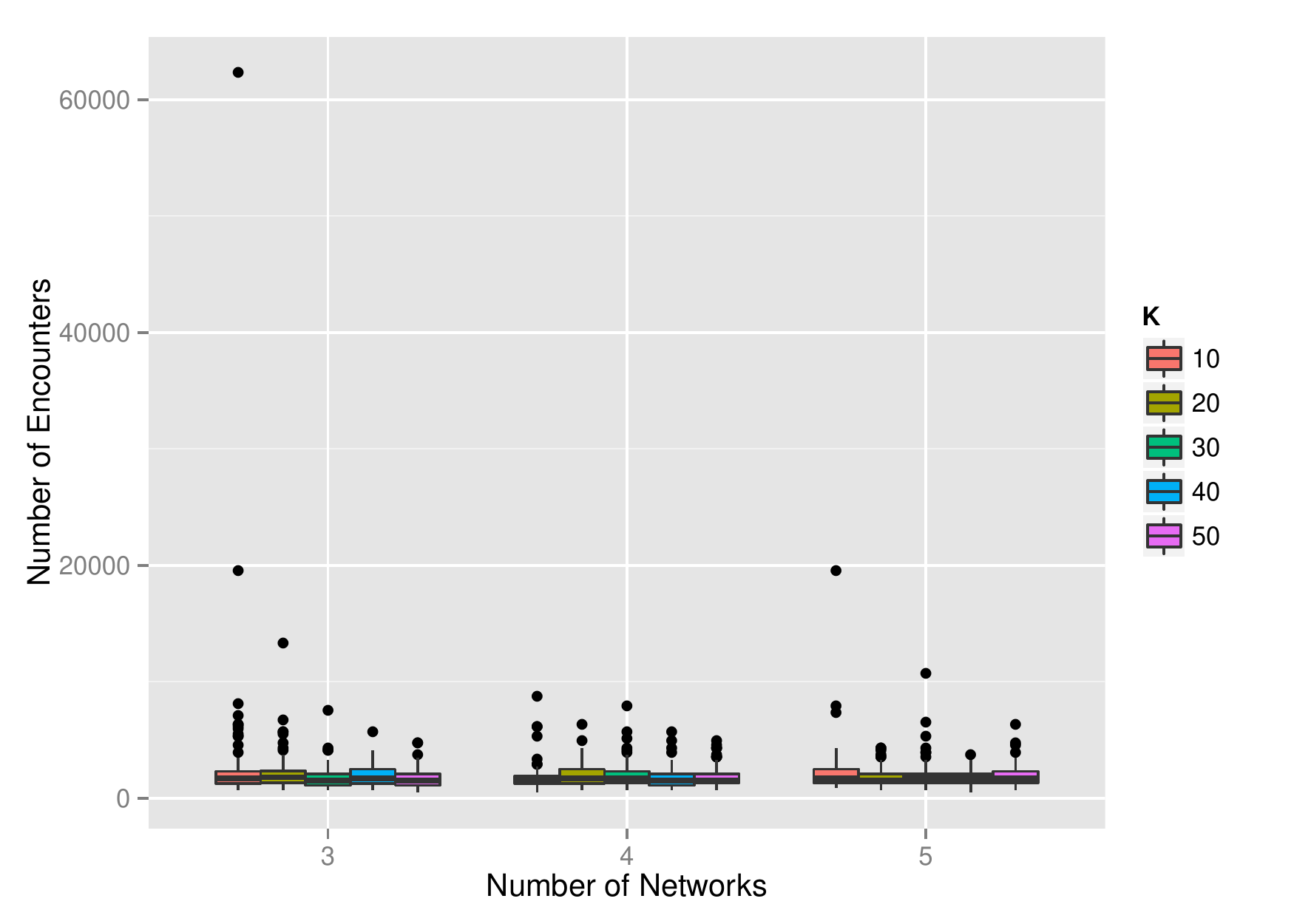}
		\caption{}
		\label{fig:ctx_perm_kreg_1020304050}
              \end{subfigure}
              \begin{minipage}{0.9\linewidth}
		\vspace{0.2cm}
		\caption{Number of meetings over 100 independent simulation runs for multiple
                  k-regular networks: $k=\{1,2,3,4,5\}$~(\ref{fig:ctx_perm_kreg_12345}) and
                  $k=\{10,20,30,40,50\}$~(\ref{fig:ctx_perm_kreg_1020304050})).}
		\label{fig:ctx_perm_kreg}
              \end{minipage}
      \end{figure}

      First, we present the results for k-regular networks. To observe the distribution of the
      number of encounters, consider the box plots for $k = \{1,2,3,4,5\}$ (figure
      \ref{fig:ctx_perm_kreg_12345}) and $k =\{10,20,30,40,50\}$ (figure
      \ref{fig:ctx_perm_kreg_1020304050}). Figure \ref{fig:ctx_perm_kreg} shows the results for
      configurations with a number of networks $\ge 3$ (since less networks lead to much worse
      results in terms of convergence speed). These plots reveal two types of situations: the
      typical runs, in which the number of encounters doesn't vary much, and some ``outlier runs,"
      which present some interesting dynamics, as we will see later on.

      \noindent Figure~\ref{fig:ctx_perm_kreg} shows us results qualitatively similar to what we
      previously observed for convergence ratios (see section
      \ref{sec:ctx_perm_convergence}). Networks with lesser connectivity and a greater average path
      length lead to total convergence less often. The simulation outcomes also vary much more in
      comparison with the other configurations ($k>1$ and $networks \ge 3$).

      Notice that for $k=3$ and $networks=3$ (\ref{fig:ctx_perm_kreg_1020304050}), we still have
      runs that deviate from what we consider the typical behaviour. Instead of disregarding this as
      an outlier, we explore this behaviour to better understand the dynamics of our
      \textit{External Majority} consensus game.

\begin{figure}[H]
        \centering
	\begin{subfigure}{.49\linewidth}
          \centering
          \includegraphics[width=1\linewidth]{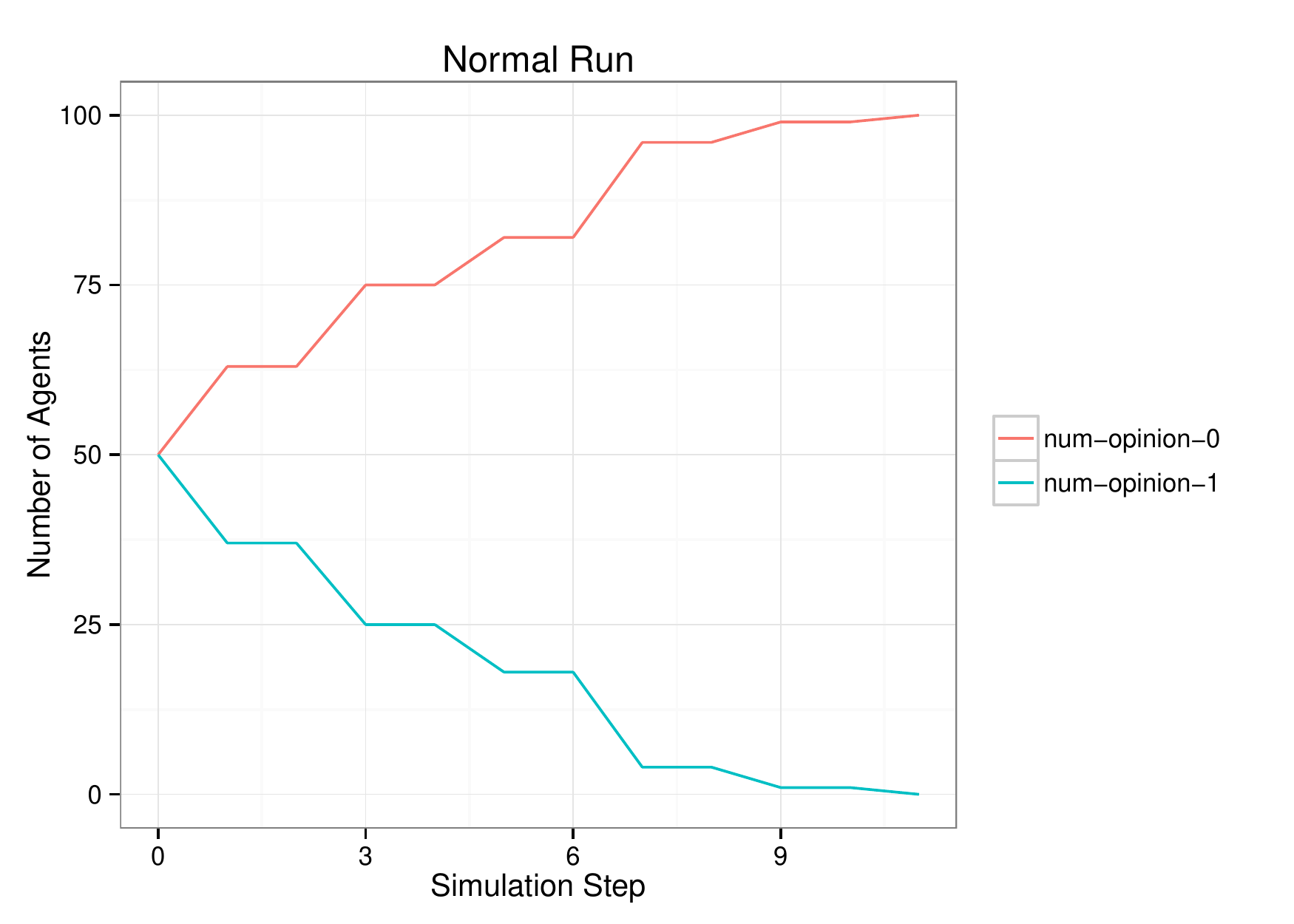}
          \caption{Typical Run}
          \label{fig:ctx_perm_kreg_run_normal}
	\end{subfigure}%
	\begin{subfigure}{.49\linewidth}
          \centering
          \includegraphics[width=1\linewidth]{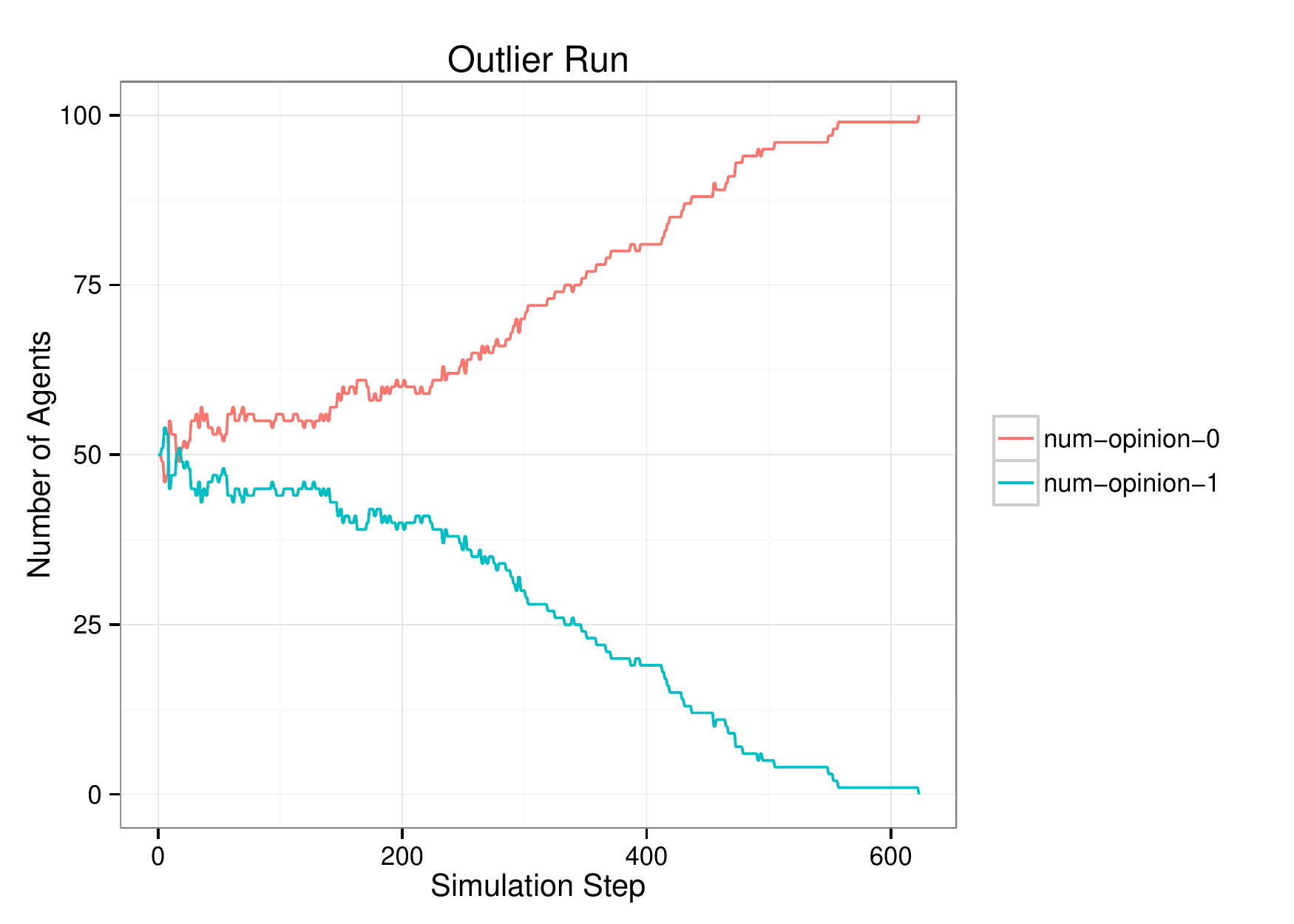}
          \caption{Outlier Run}
          \label{fig:ctx_perm_kreg_run_outlier}
	\end{subfigure}
	
	\begin{minipage}{0.9\linewidth}
          \vspace{0.2cm}
          \caption{Number of agents with each opinion during a simulation run: typical run in
            \ref{fig:ctx_perm_kreg_run_normal} and ``outlier run" in
            \ref{fig:ctx_perm_kreg_run_outlier}.}
		\label{fig:ctx_perm_kreg_runs}
              \end{minipage}
      \end{figure}

      We analysed the network structure from this run and observed that there were \textit{no
        distinct differences} between the network instances from the ``outlier run" and the ones of
      the ``typical" run. This pointed to the consensus game itself as a possible cause for the
      irregular behaviour -- not to the network structure --. What happens is that the agents reach
      a state from which converging towards consensus is significantly harder. Villatoro
      \cite{Villatoro2013} found that some complex network models lead consensus games to form
      metastable sub-conventions. These are very difficult to break and reaching 100\% agreement is
      not as straightforward as assumed by previous researchers. We are not concerned with this
      difficulty, instead, we want to explore how to describe what happens in similar
      cases. Figure~\ref{fig:ctx_perm_kreg_runs} shows the consensus progression for two simulation
      runs a typical case and the outlier.

\begin{figure}[H]
        \centering
	\begin{subfigure}{1\linewidth}
          \centering
          \includegraphics[width=1\linewidth]{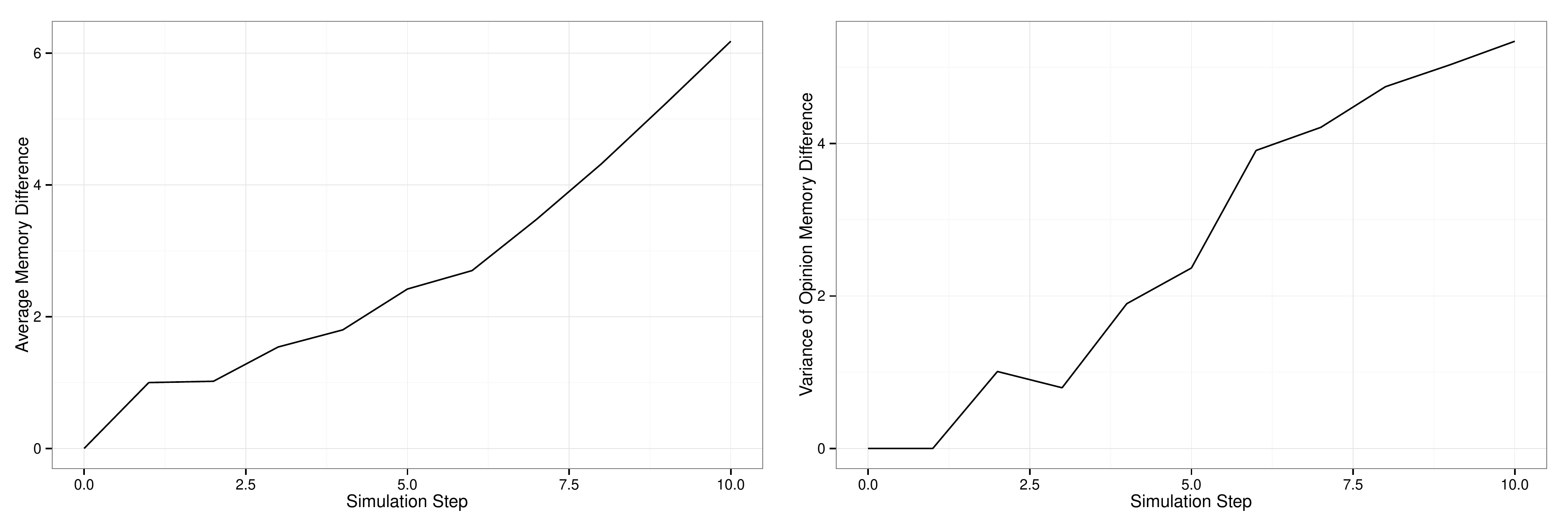}
          \caption{Typical Run}
          \label{fig:ctx_perm_kreg_run_normal_diff}
	\end{subfigure}%
	\\
	\begin{subfigure}{1\linewidth}
          \centering
          \includegraphics[width=1\linewidth]{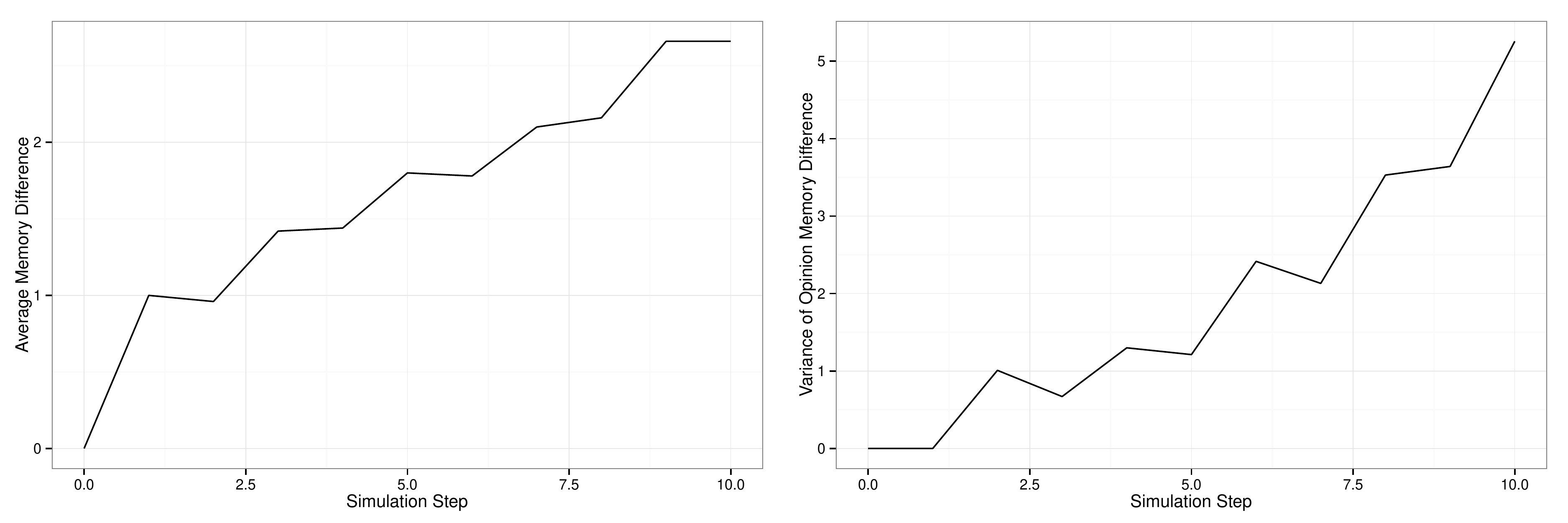}
          \caption{Outlier Run}
          \label{fig:ctx_perm_kreg_run_outlier_diff}
	\end{subfigure}
	
	\begin{minipage}{0.9\linewidth}
          \vspace{0.2cm}
          \caption{Average and variance of the difference between opinions observed for the 100
            agents: typical run in figure \ref{fig:ctx_perm_kreg_run_normal_diff} and first steps of
            the``outlier run" in figure \ref{fig:ctx_perm_kreg_run_outlier_diff}.}
		\label{fig:ctx_perm_kreg_runs_diff}
              \end{minipage}
      \end{figure}

      \noindent We take the outlier in the configuration with $k=3$ and $networks=3$, and look at
      the memory that agents use in the consensus game. As we described previously, agents record
      the number of individuals encountered with each opinion value (two possible values in this
      case). We look at the average difference between the number of values observed for all the
      agents. We also look at the variance to find how the opinion observations evolve throughout
      the simulation. Figure~\ref{fig:ctx_perm_kreg_runs_diff} shows the average difference between
      opinion memory for the first 10 steps (which was when the simulation converged for the normal
      run).

      We can see that both the average difference in the opinion memory and the variance grow faster
      in the normal run than in the outlier run. This is not enough to establish a significant
      difference between the two. Moreover, the memory differences are quite similar, but the number
      of agents for each opinion was quite even. This happens due to the position of agents in the
      network and their initial opinion values. These circumstances led the agents to a initial
      stability. Note that the opinion difference continues to rise throughout the simulation along
      with the variance. The variance starts to increase more rapidly after 200 steps. The
      exponential growth reveals that the convergence to consensus is not done evenly throughout the
      network, this is one of the reasons why consensus takes more time to be achieved. Also, the
      low variance in the first $200$ steps shows us that the opinion strength was evenly
      distributed: we have almost the same number of agents for each opinion but the memory
      differences are qualitatively the same for both opinions. As soon as this variance increases,
      one of the opinions breaks the stable state and convergence towards global consensus begins.

      \begin{figure}[H]
	\centering
	\includegraphics[width=1\linewidth]{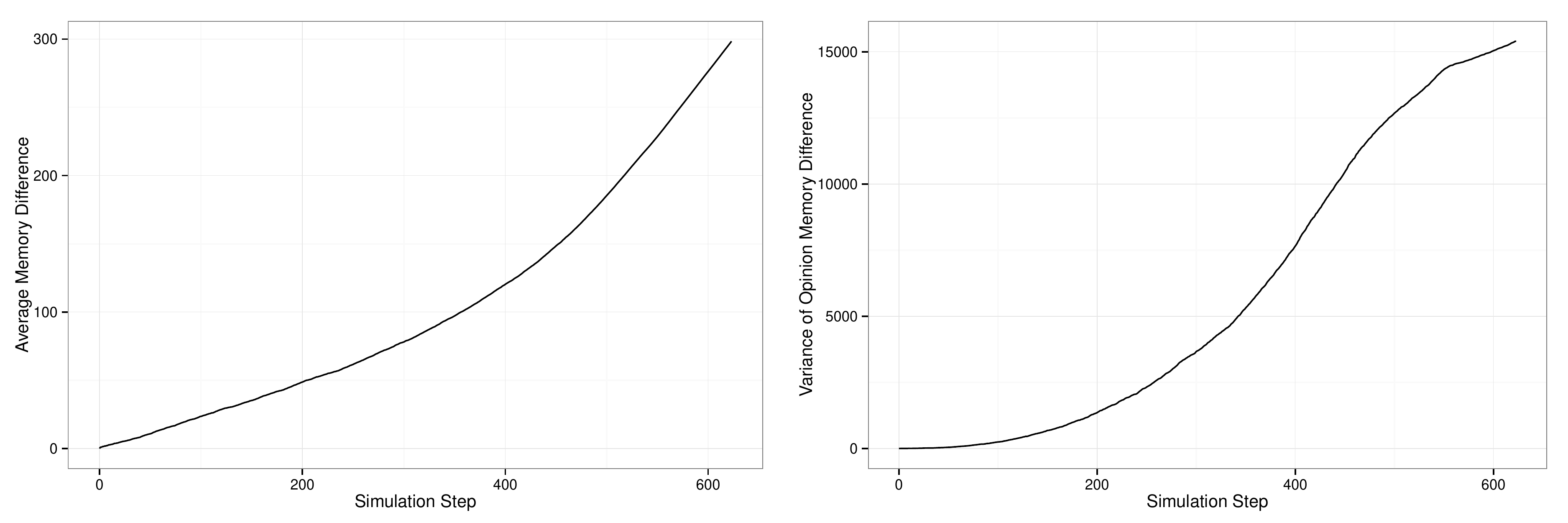}
	\begin{minipage}{0.9\textwidth}
          \caption{Average and variance of the difference between opinions observed for the 100
            agents throughout the outlier run.}
          \label{fig:ctx_perm_kreg_run_outlier_diff_full}
	\end{minipage}
      \end{figure}

      \noindent In other words, different neighbourhoods in the network lead to different opinion
      observation configurations (possibly even creating self-reinforcing structures). Thus, the
      influence of heterogeneous connectivity is reflected on the growth of opinion difference
      variance.

      \begin{figure}[H]
        \centering
	\begin{subfigure}{.49\linewidth}
          \centering
          \includegraphics[width=1\linewidth]{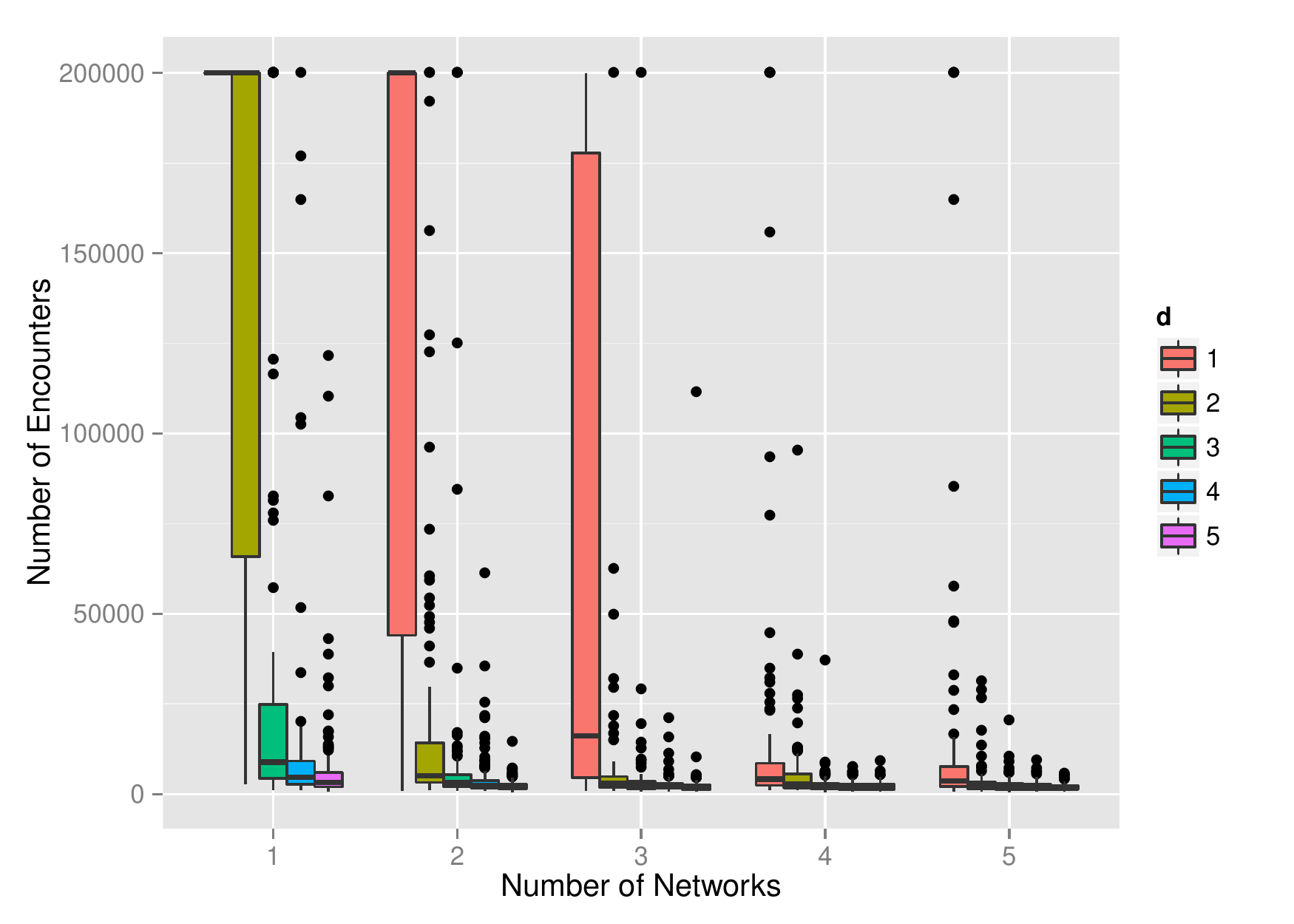}
          \caption{}
          \label{fig:ctx_perm_sf_full}
	\end{subfigure}%
	\begin{subfigure}{.49\linewidth}
          \centering
          \includegraphics[width=1\linewidth]{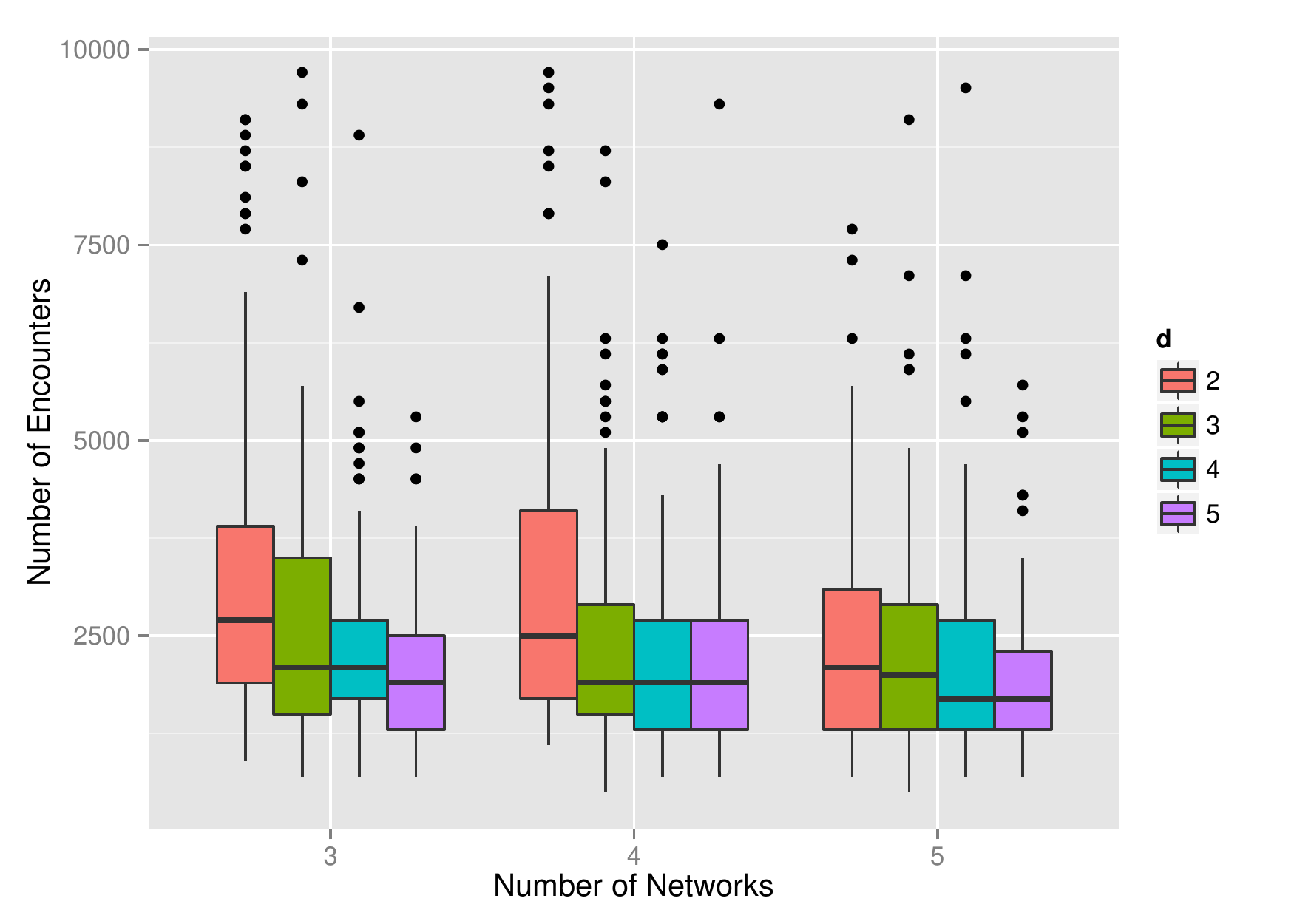}
          \caption{}
          \label{fig:ctx_perm_sf_best}
	\end{subfigure}
	\begin{minipage}{0.9\linewidth}
		\vspace{0.2cm}
		\caption{Number of meetings over 100 independent simulation runs for multiple
                  \textit{scale-free} networks: with d=\{1,2,3,4,5\} (\ref{fig:ctx_perm_sf_full})
                  with a zoom on the best configurations in \ref{fig:ctx_perm_sf_best}.}
		\label{fig:ctx_perm_sf}
              \end{minipage}
      \end{figure}

      \noindent Figure~\ref{fig:ctx_perm_sf} shows the average number of encounters during 100
      independent simulation runs for \textit{scale-free} networks. (We don't include the
      configuration for one network and $d = 1$ because these never converge.) The convergence ratio
      is very low for the configurations with one network $d = 2$ and 2 networks with $d = 1$; hence
      the high number of encounters~(see also table \ref{tab:scale-free_convergence}).

      These results are similar to those of \textit{k-regular} in that adding more networks speeds
      up the convergence to consensus. The difference is that more ``atypical" runs occur with
      multiple \textit{scale-free} networks. Again, this is the effect of these types of
      topologies. As we discussed previously, in these less connected topologies, agents are more
      prone to be arranged in such a way that progression towards consensus is more difficult to
      attain due to bottlenecks. Different regions in the network might converge towards different
      opinion values.

      \subsubsection{Heterogeneous Network Configuration}
      \noindent We also performed experiments with mixed network topologies: we used one
      \textit{scale-free} and one \textit{k-regular}
      networks. Table~\ref{tab:regular_scale-free_convergence} shows the convergence ratio for the
      heterogeneous network configuration with the multiple values for $d$ and $k$ for the
      \textit{scale-free} and \textit{k-regular} respectively.

      \begin{table}[H]
        \centering
	\begin{minipage}{0.9\textwidth}
          \caption{Ratio of convergence to total consensus in 3000 independent runs with two
            networks: one \textit{k-regular} and one \textit{scale-free} network with different $k$
            and $d$ values respectively.}
          \label{tab:regular_scale-free_convergence}
	\end{minipage}
	\setlength{\tabcolsep}{.30000em}
	\begin{tabular}{c|SSSSSSSSSS}
		
          \toprule
          & \multicolumn{10}{c}{k} \\ 
          d  & 1 & 2 & 3 & 4 & 5 & 10 & 20 & 30 & 40 & \multicolumn{1}{c}{50} \\ 
          \midrule
          1  & 0.18 & 0.47 & 0.55 & 0.67 & 0.61 & 0.76 & 0.77 & 0.81 & 0.75 & 0.82 \\
          2  & 0.63 & 0.72 & 0.88 & 0.89 & 0.90 & 0.97 & 1.00 & 1.00 & 0.99 & 1.00 \\
          3  & 0.77 & 0.86 & 0.87 & 0.97 & 0.98 & 0.99 & 0.98 & 0.99 & 1.00 & 1.00 \\
          4  & 0.73 & 0.92 & 0.92 & 0.97 & 0.96 & 0.96 & 0.99 & 1.00 & 1.00 & 1.00 \\
          5  & 0.87 & 0.85 & 0.94 & 0.94 & 0.98 & 0.99 & 1.00 & 1.00 & 1.00 & 1.00 \\
          \bottomrule
	\end{tabular}
      \end{table}

      \noindent One of the differences between this configuration and the previous ones can be
      observed in the simulations with $d=1$ and $k=1$. The convergence ratio is higher than with
      $2$ $1-regular$ networks (see table \ref{tab:regular_convergence}) -- but not better than two
      \textit{scale-free} networks. This happens because with $k=1$ the network is basically a ring
      and has the maximum possible average path length for a connected graph. Adding two rings
      improves the convergence ratio but the underlying structure is still very susceptible to
      self-reinforcing structures (a connected sub-graph is basically a line). Adding a
      \textit{scale-free} changes this drastically as we are mixing tree-like network components
      with a ring.

      \begin{figure}[H]
        \centering
	\begin{subfigure}{.49\linewidth}
          \centering
          \includegraphics[width=1\linewidth]{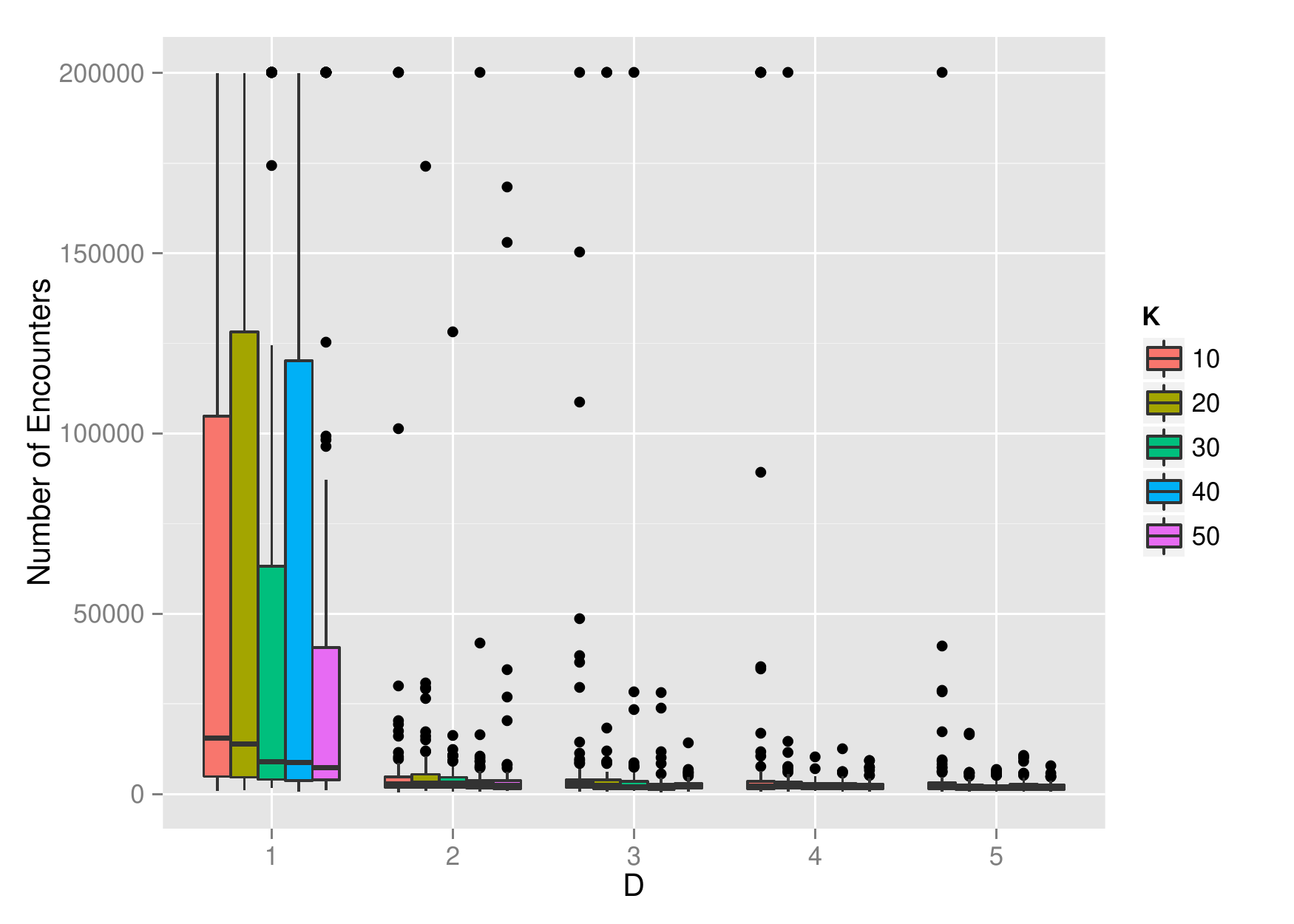}
          \caption{}
          \label{fig:ctx_perm_sfkreg_1020304050full}
	\end{subfigure}%
	\begin{subfigure}{.49\linewidth}
		\centering
		\includegraphics[width=1\linewidth]{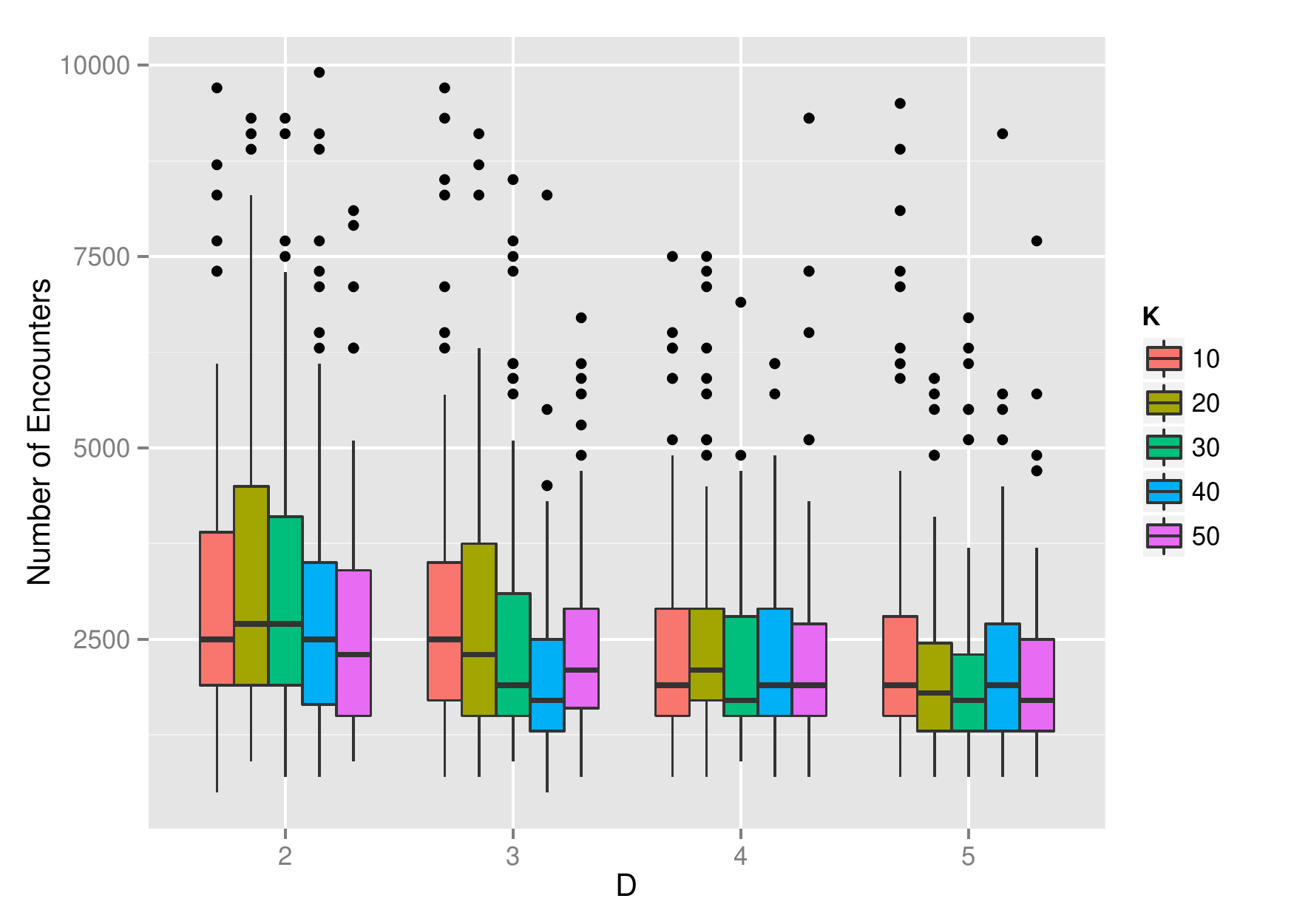}
		\caption{}
		\label{fig:ctx_perm_sfkreg_1020304050best}
              \end{subfigure}
              \begin{minipage}{0.9\linewidth}
		\vspace{0.2cm}
		\caption{Number of meetings over 100 independent simulation runs for heterogeneous
                  configuration: one \textit{k-regular} network with $k=\{1,2,3,4,5\}$ and one
                  \textit{scale-free} network with $d\{1,2,3,4,5\}$. Figure
                  \ref{fig:ctx_perm_sfkreg_1020304050best} shows a zoom on the best configurations.}
		\label{fig:ctx_perm_sfkreg}
              \end{minipage}
      \end{figure}

      Figure \ref{fig:ctx_perm_sfkreg} shows the results for the number of encounters necessary for
      the agents to achieve consensus. We focus on the \textit{k-regular} networks with higher
      values of $k$. We can see that with these two networks, the configurations with values of
      $d>=2$ for \textit{scale-free} networks produce drastically better results in terms of speed
      of convergence. The data for the average number of encounters, as well as the respective
      standard deviations referent to the results in figure \ref{fig:ctx_perm_sfkreg}, can be found
      in appendix \ref{append_ctx_permeability}, table \ref{append_tab:ctx_perm_kreg_sf_encounters}.

      
      \subsection{Context Switching}
      \noindent In this section, we present the results relative to the context switching model. The
      difference between this model and the previous one is that the agents no longer interact in
      multiple networks at the same time (being able to select any neighbour at a given simulation
      step). In this model, they become active in a single network at a time. Agents can switch to a
      different network at the end of each step. The switching mechanism uses a probability
      associated with each network. With this new idea of swapping contexts, we covered some space
      left undeveloped in Antunes and colleagues original work \cite{Antunes2007,Antunes2010}.

      As we discussed in section \ref{sec:models_cs} (see model \ref{model:context_switching}), the
      switching probability dictates the frequency with which an agent switches \textit{from} the
      current network \textit{after} an encounter has been performed. In an abstract manner, this
      allows us to model how much time agents spend on each network. A further development of this
      model will be to assign different preferences to different networks for each agent. This can
      help us model phenomena such as real-world agents that dedicate more or less time consuming
      content from different social networks. For now, we attribute this probability to the network
      to restrain the model complexity.

      \subsubsection{Exploring the Switching Probability}
\begin{figure}[H]
        \centering
	\begin{subfigure}{0.49\linewidth}
          \centering
          \includegraphics[width=1\linewidth]{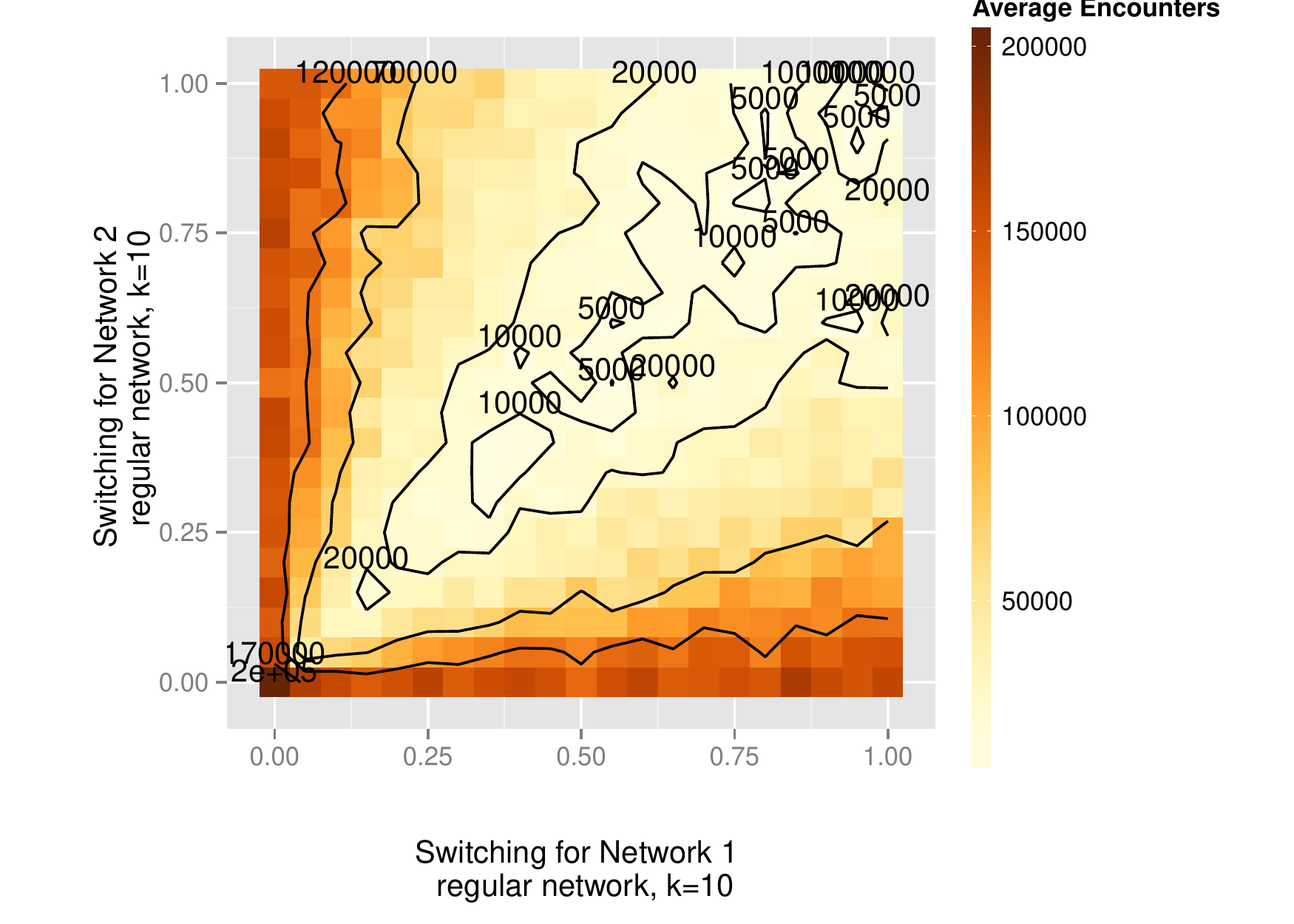}
          \caption{10-regular Networks}
          \label{fig:ctx_cs_2_10kreg}
	\end{subfigure}%
	\begin{subfigure}{0.49\linewidth}
          \centering
          \includegraphics[width=1\linewidth]{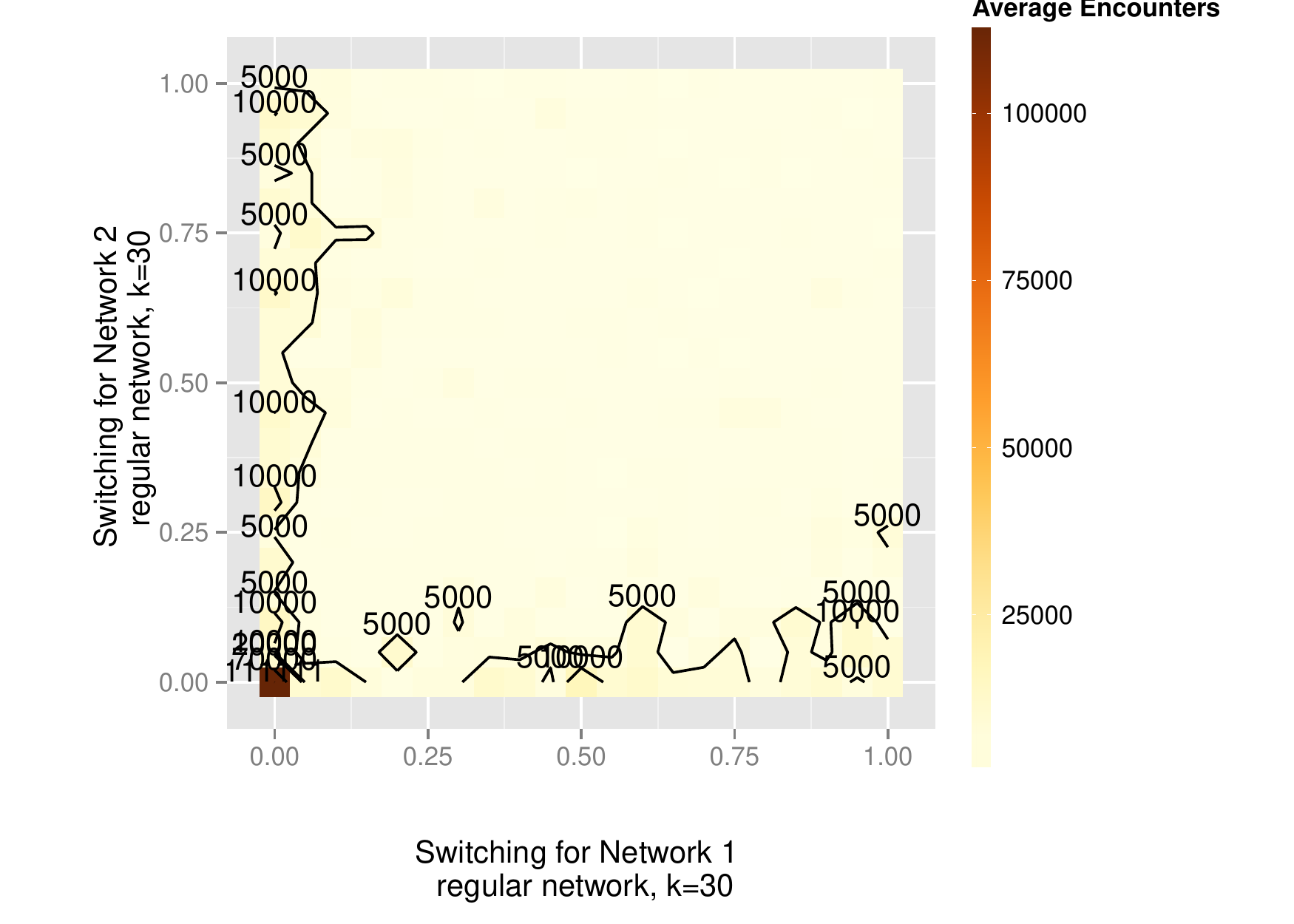}
          \caption{30-regular Networks}
          \label{fig:ctx_cs_2_30kreg}
	\end{subfigure}
	
	\begin{minipage}{0.9\linewidth}
          \vspace{0.2cm}
          \caption{Contour plot for the average number of meets during a simulation for 100
            independent runs: 2 \textit{10-regular} networks (\ref{fig:ctx_cs_2_10kreg}) and
            \textit{30-regular} networks (\ref{fig:ctx_cs_2_30kreg}). See the perspective plots in
            appendix \ref{append_ctx_switching}.}
          \label{fig:ctx_cs_2_kregular}
	\end{minipage}
      \end{figure}

      \noindent We first look at the influence of the new parameter (switching probability) in the
      convergence to consensus. We span the switching probability between $0$ and $1$ in intervals
      of $0.05$. We do this for 2 networks with \textit{k-regular} and \textit{scale-free}
      topologies. Figure \ref{fig:ctx_cs_2_kregular} shows the results for two \textit{k-regular}
      networks: one configuration with $k = 10$ and one configuration with $k = 30$. We use these
      two configurations to study the relationship between the number of encounters, the switching
      probability parameter, and the connectivity of the networks. Remember that with $k=30$ the
      average path length is also lower (section \ref{sec:overlapping_kreg}).

\begin{figure}[H]
        \centering
	\begin{subfigure}{0.49\linewidth}
          \centering
          \includegraphics[width=1\linewidth]{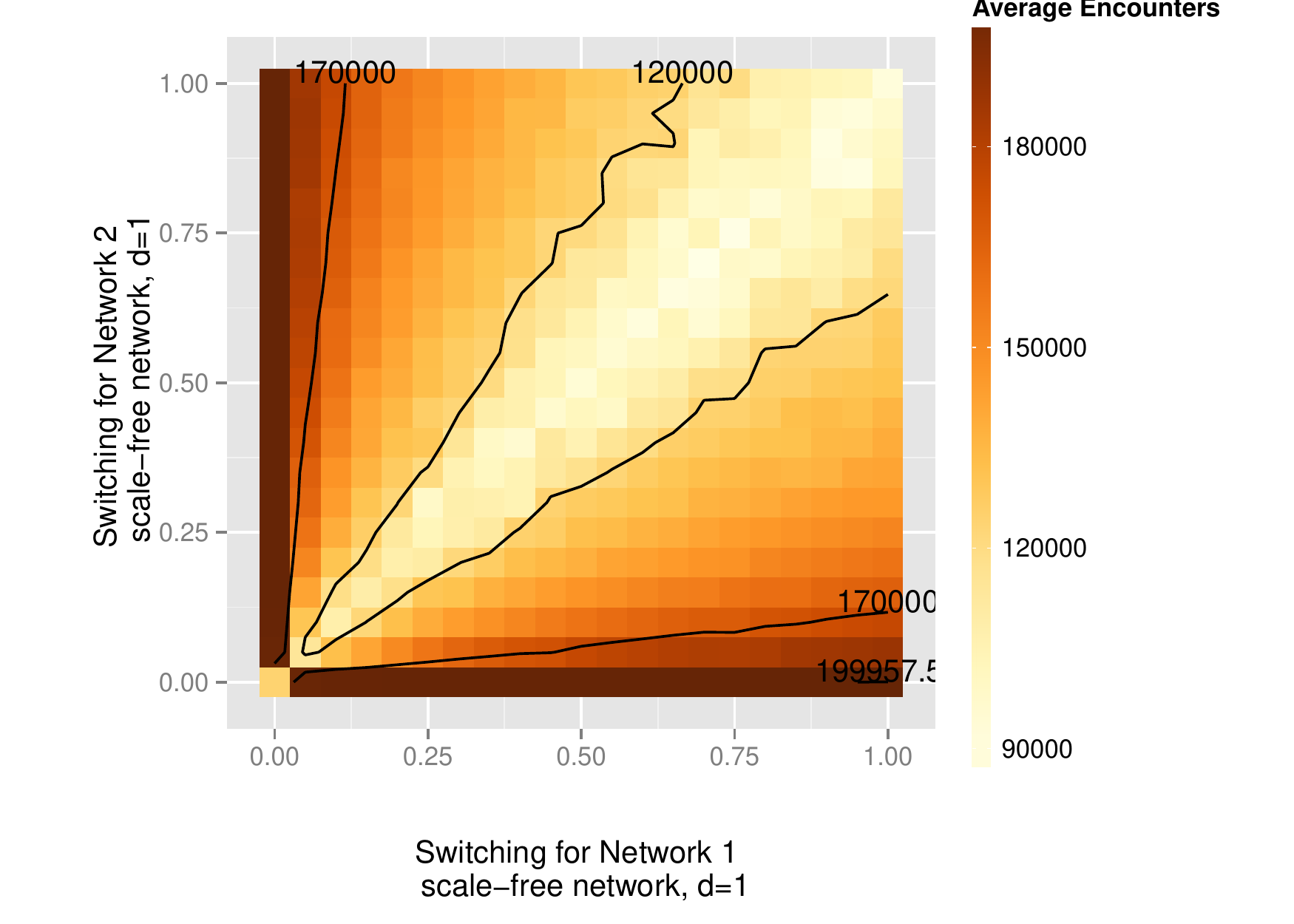}
          \caption{\textit{Scale-Free} $d=1$}
          \label{fig:ctx_cs_2_sf_d1}
	\end{subfigure}%
	\begin{subfigure}{0.49\linewidth}
          \centering
          \includegraphics[width=1\linewidth]{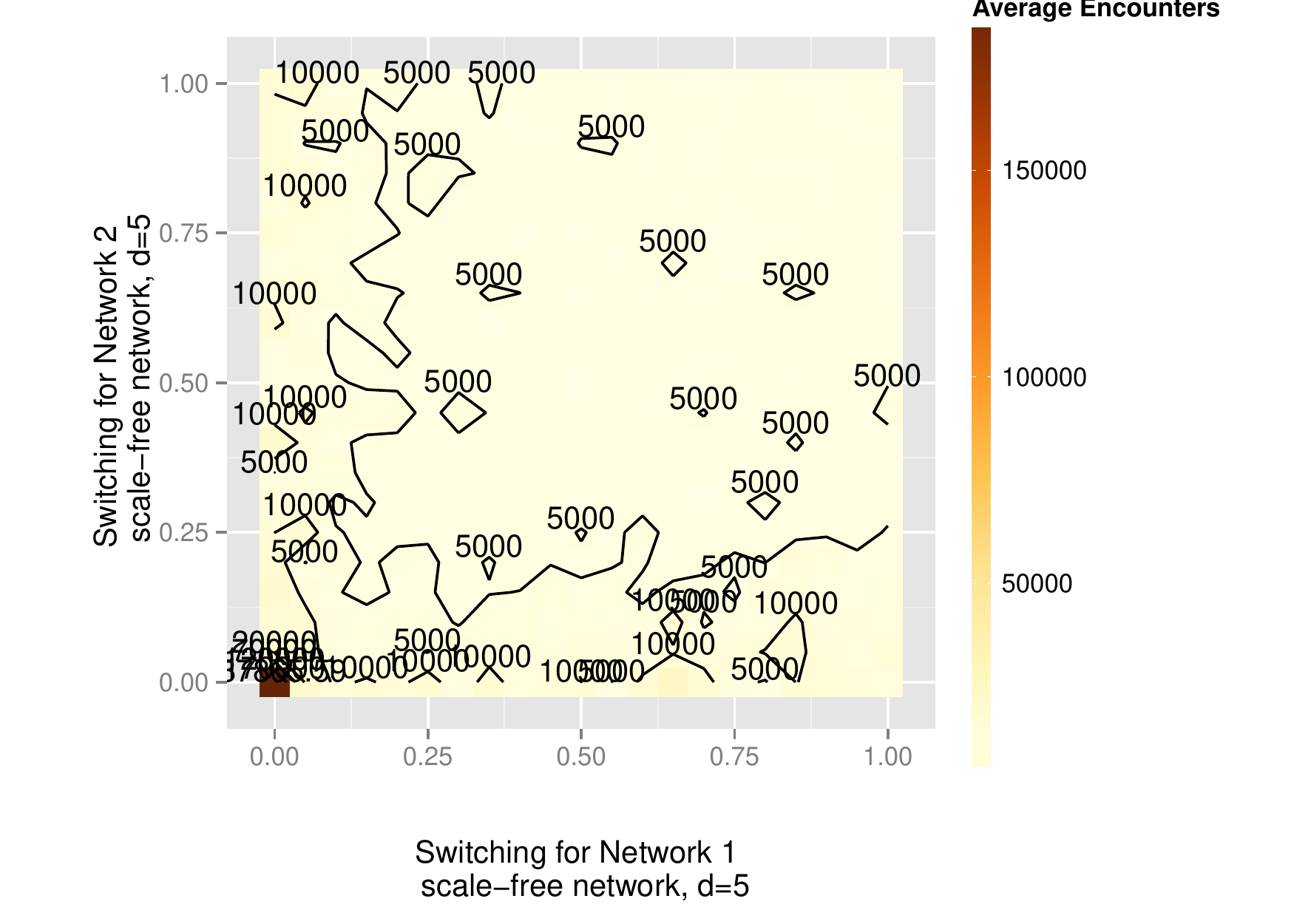}
          \caption{\textit{Scale-Free} $d=5$}
          \label{fig:ctx_cs_2_sf_d5}
	\end{subfigure}
	
	\begin{minipage}{0.9\linewidth}
          \vspace{0.2cm}
          \caption{Contour plot for the average number of meets during a simulation for 100
            independent runs: 2 \textit{scale-free} networks with $d=1$ (\ref{fig:ctx_cs_2_sf_d1})
            and $d=5$ (\ref{fig:ctx_cs_2_sf_d5}). See the perspective plots in appendix
            \ref{append_ctx_switching}.}
          \label{fig:ctx_cs_2_sf}
	\end{minipage}
      \end{figure}

      \noindent For smaller values of $k$, symmetry in the context switching probability (having the
      same probability in both networks) is more important if the agent switches less from one of
      the networks. Switching less from one network means spending more time in that network. This
      means that switching more from the other network can be disruptive to a neighbourhood that has
      already converged to a sub-convention. This can be observed in configurations with
      \textit{k-regular} networks (figure \ref{fig:ctx_cs_2_10kreg}) but it is especially apparent
      in \textit{scale-free} networks (figure \ref{fig:ctx_cs_2_sf_d1}).

      When we increase the connectivity (and lower the average path length), both in
      \textit{k-regular} and \textit{scale-free} networks, the switching probability becomes
      ``irrelevant'' --for any probability value, the speed of convergence will be the same--. The
      exception is obviously when considering the switching probability with value $0$. In this
      case, agents are isolated (they are initially equally distributed throughout the networks) and
      reaching consensus becomes very hard. (Especially because agents are active only in one network
      and this can create disconnected graphs.)

      One major difference from the previous model is that the agents no longer interact with a
      neighbour from any network at any time. The neighbour selection has to be made from the
      network they are currently active in. This causes the agents to spend different amounts of
      time in different networks. An agent can be very influential in one network -- being a very
      central node in that context -- and marginally important in another: in a sense its opinion in
      one network can be more important to overall convergence than in other network, due to its
      position in the network.

\begin{figure}[H]
	\centering
	\includegraphics[width=1\linewidth]{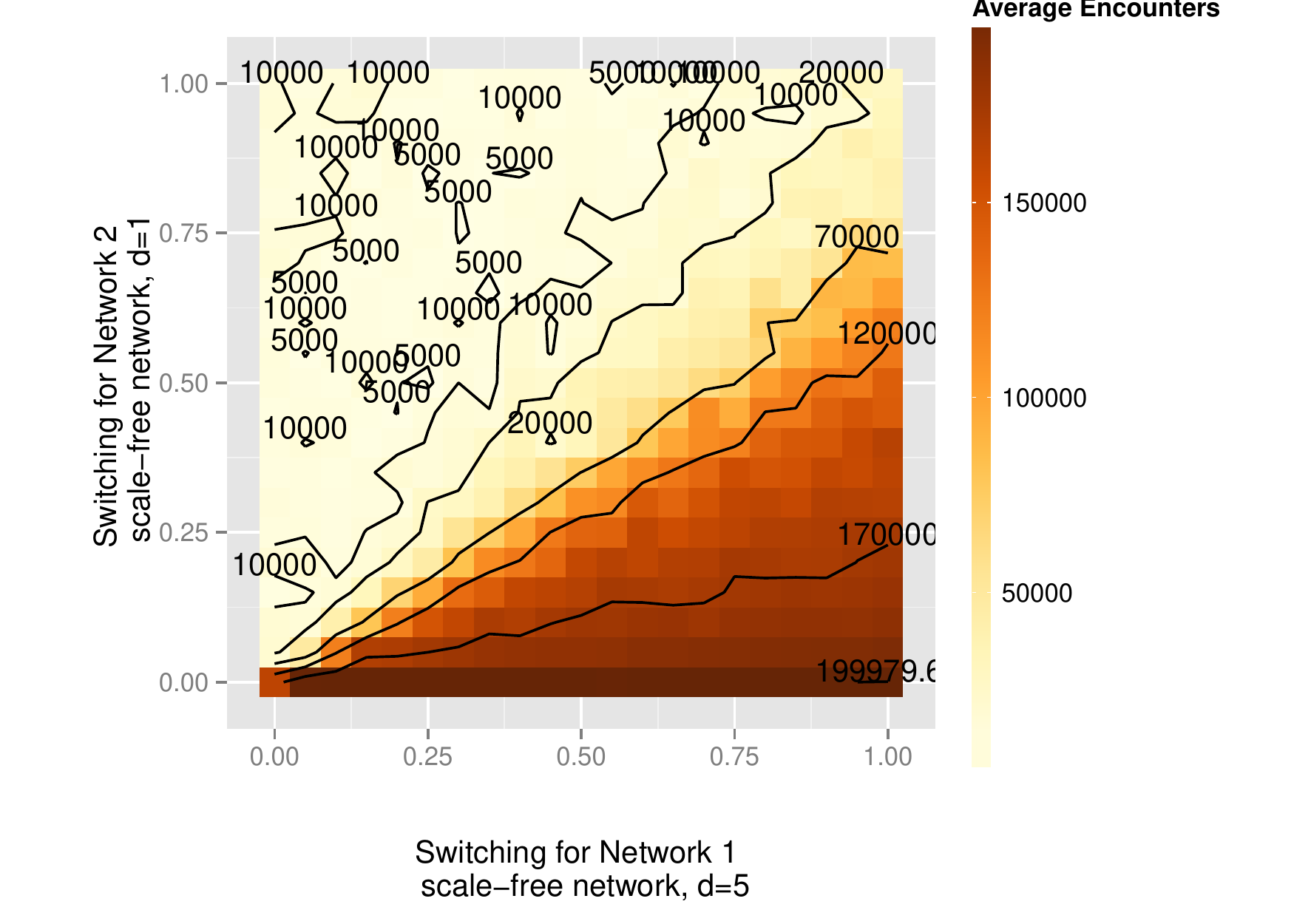}
	\begin{minipage}{0.9\textwidth}
          \caption{Contour plot for the average number of meets during a simulation for 100
            independent runs: 2 \textit{scale-free} networks, the first with $d=5$ and the second
            with $d=1$. See the perspective plots in appendix \ref{append_ctx_switching} (figure
            \ref{append_fig:ctx_cs_sf_d5_d1}).  }
		\label{fig:ctx_cs_sf_d5_d1}
              \end{minipage}
      \end{figure}

      To explore the contribution of connectivity to the new context switching dynamics, we setup
      two \textit{scale-free} networks; one with $d=1$ and the other with $d=5$. We then spanned the
      switching probability in increments of $0.05$. The results can be seen in figure
      \ref{fig:ctx_cs_sf_d5_d1}.

\begin{figure}[H]
        \centering
	\includegraphics[width=1\linewidth]{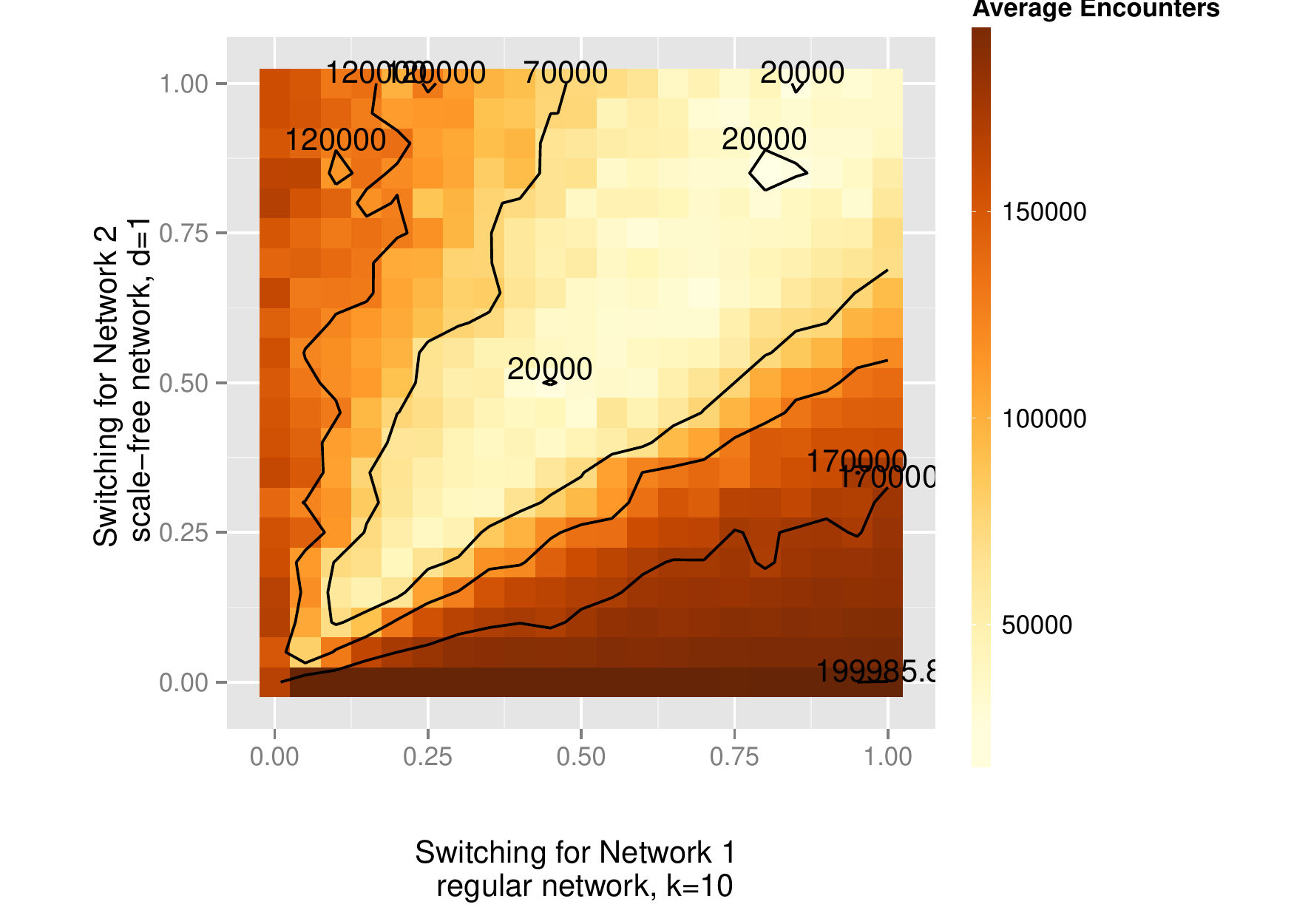}
	\begin{minipage}{0.9\textwidth}
          \caption{Contour plot for the average number of meets during a simulation for 100
            independent runs: one \textit{10-regular} network (k=10) and a \textit{scale-free}
            network with $d=1$. See the perspective plots in appendix \ref{append_ctx_switching}
            (figure \ref{append_fig:ctx_cs_kreg10_sfd1}).  }
          \label{fig:ctx_cs_kreg10_sfd1}
	\end{minipage}
      \end{figure}

      In this case it is more important (to convergence speed) to switch \textit{from} the network
      with lower average path length and a forest-like composition (and a probability of at least
      the same value or more than the one of the other network with $d=5$). The same can be observed
      when we mix \textit{k-regular} with \textit{scale-free} networks (see figure
      \ref{fig:ctx_cs_kreg10_sfd1}).

      Switching less from the less connected network is bad in both cases: it is possible that
      spending more time in the network with bigger neighbourhoods allows for a stabler
      convergence. Sub-conventions can emerge in \textit{scale-free} networks with $d=1$ because
      usually some network regions are isolated by a single node -- the root of the sub-tree they
      belong to.

      \subsubsection{Comparison with Context Permeability}
      \noindent Finally, to compare the context switching model with the previous model of context
      permeability, we froze the switching probabilities with values $\zeta={0.25,0.5,0.75}$ and
      varied the number of networks. (We considered these values because they conveniently
      characterise the probability in low, medium and high switching.) We compare the results with
      the context permeability in terms of average number of encounters to achieve consensus. Note
      that in some cases, due to the distribution of the number of encounters, the average is not an
      accurate descriptor for the convergence. Nonetheless, configurations with highly variable
      outcomes usually produce an average that is qualitatively distinct from the rest (see figure
      \ref{fig:ctx_perm_kreg} in section \ref{sec:ctx_perm_encounters}). We took the average number
      of encounters during the simulations to compare the context switching model with the context
      permeability model.

      \begin{figure}[H]
        \centering
	\begin{subfigure}{.49\linewidth}
          \centering
          \includegraphics[width=1\linewidth]{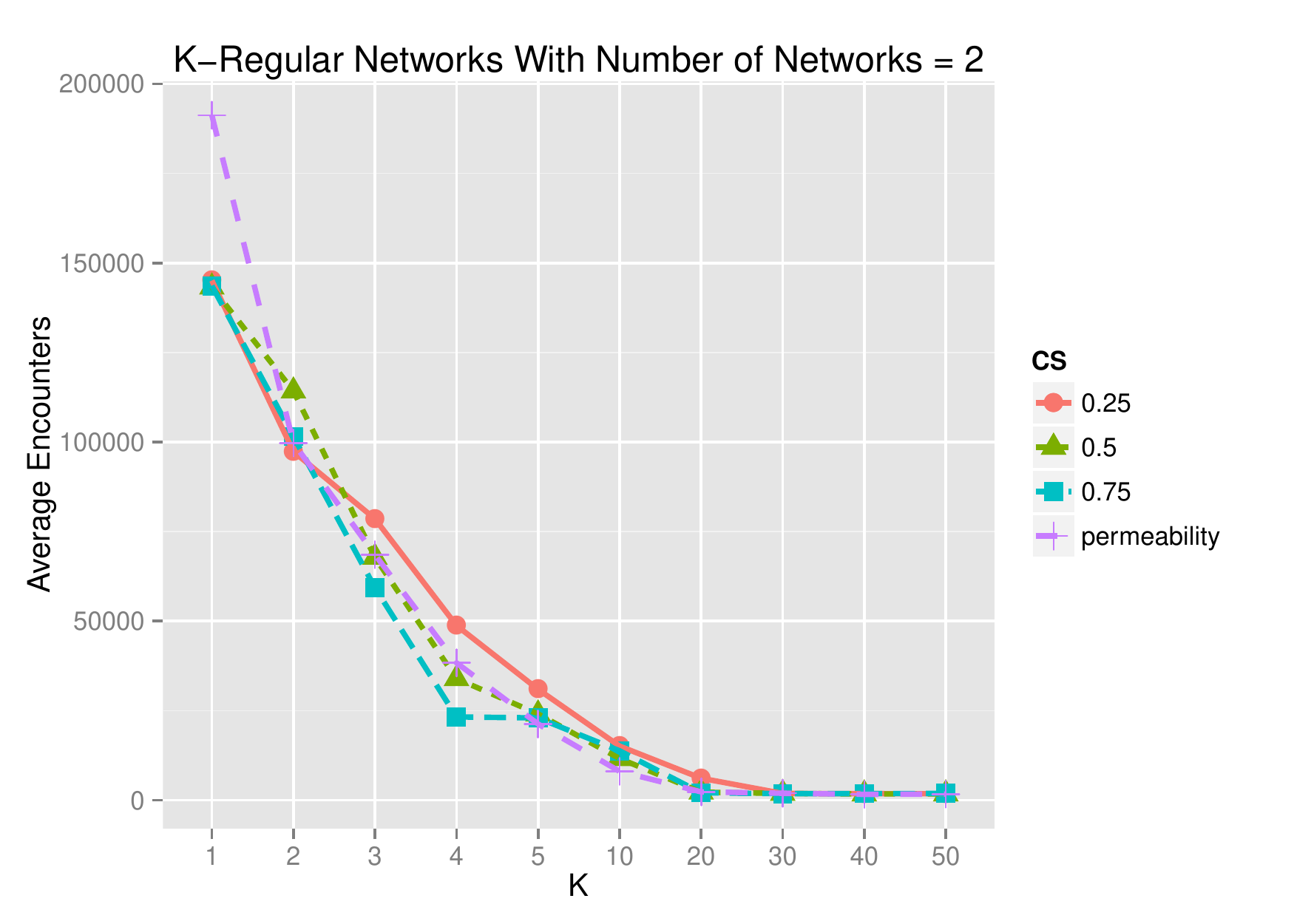}
          \caption{}
          \label{fig:ctx_switching_comp_kreg_2}
	\end{subfigure}%
	\begin{subfigure}{.5\linewidth}
          \centering
          \includegraphics[width=1\linewidth]{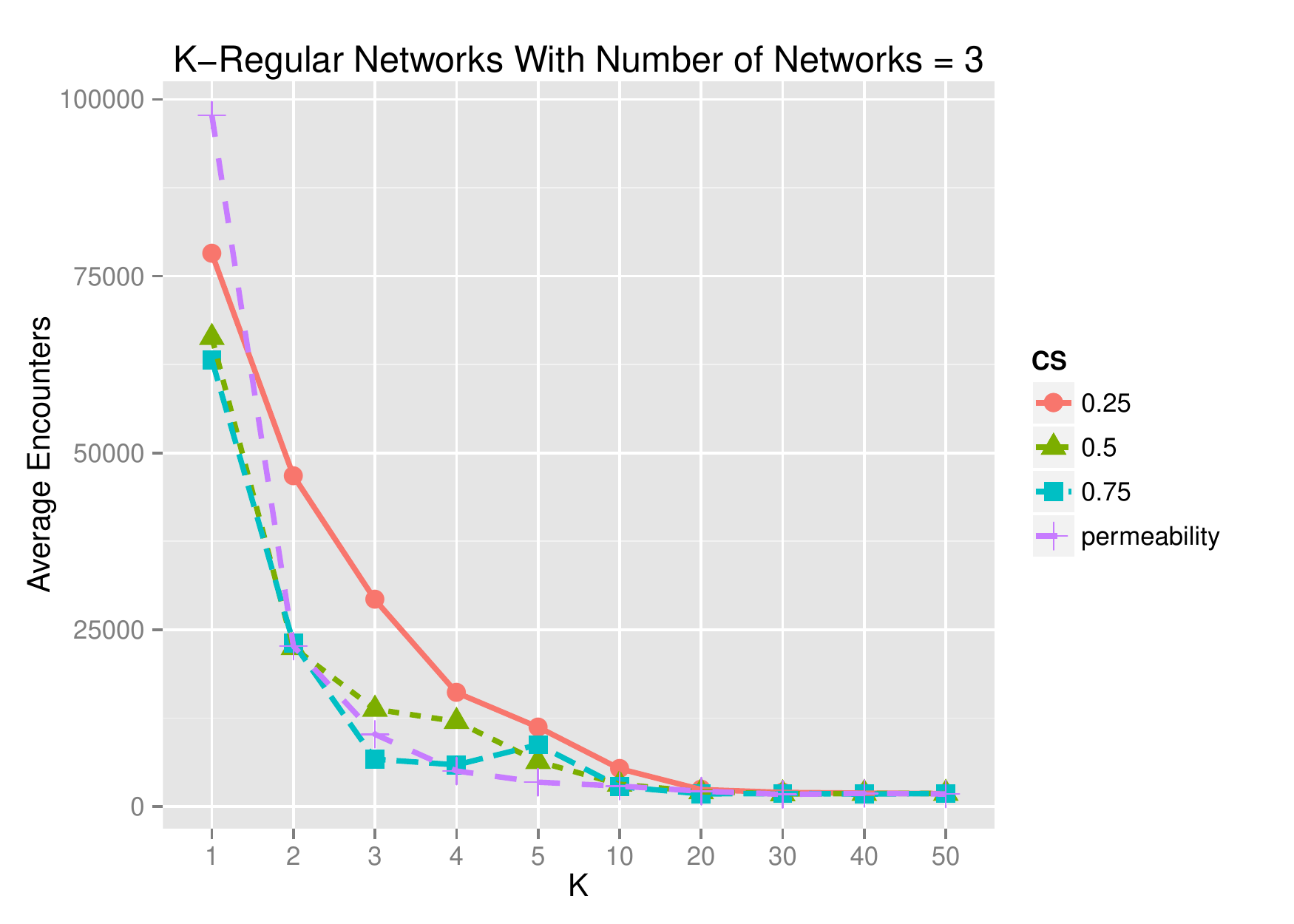}
          \caption{}
          \label{fig:ctx_switching_comp_kreg_3}
	\end{subfigure}\\
	\begin{subfigure}{.49\linewidth}
          \centering
          \includegraphics[width=1\linewidth]{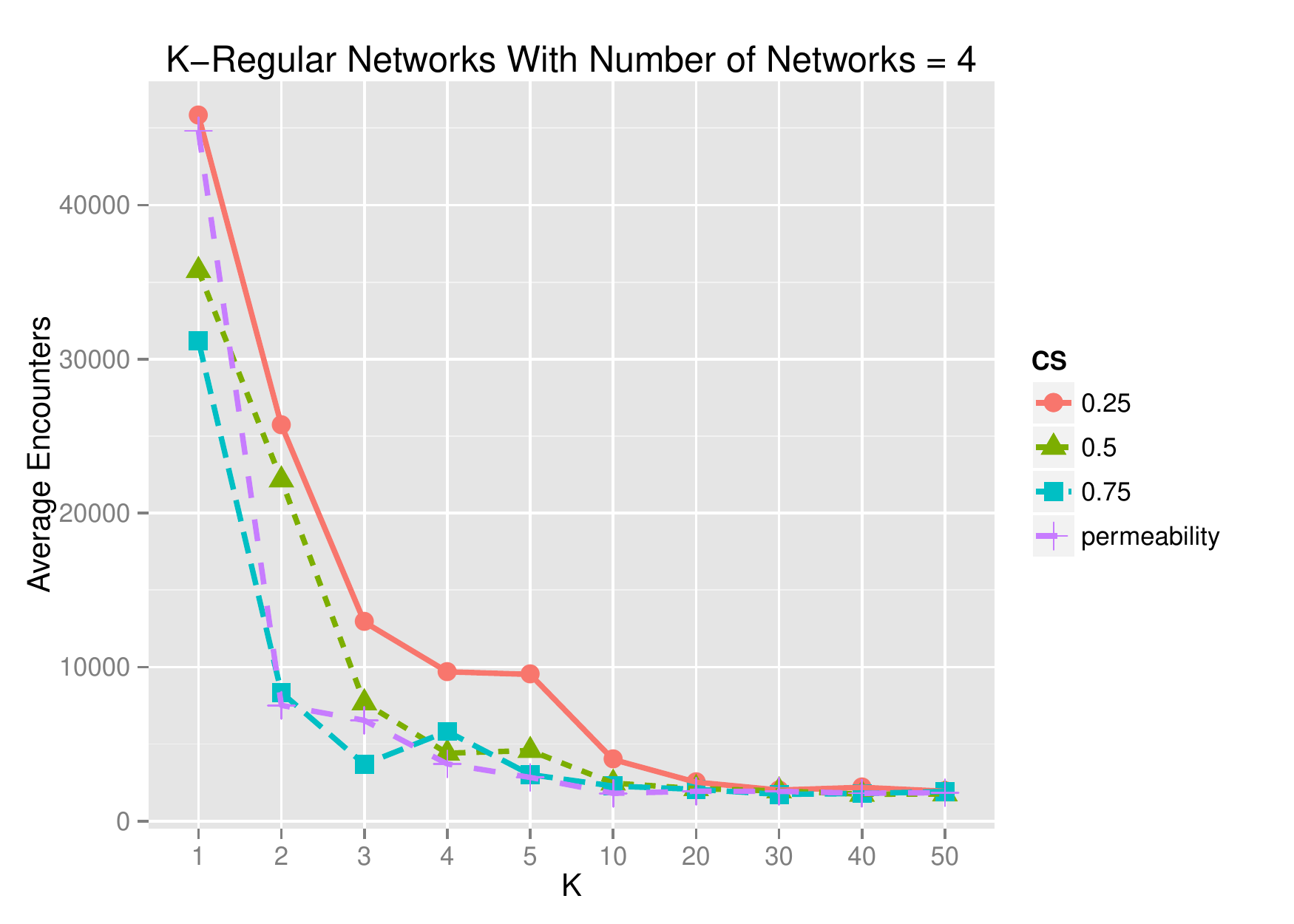}
          \caption{}
          \label{fig:ctx_switching_comp_kreg_4}
	\end{subfigure}
	\begin{subfigure}{.5\linewidth}
          \centering
          \includegraphics[width=1\linewidth]{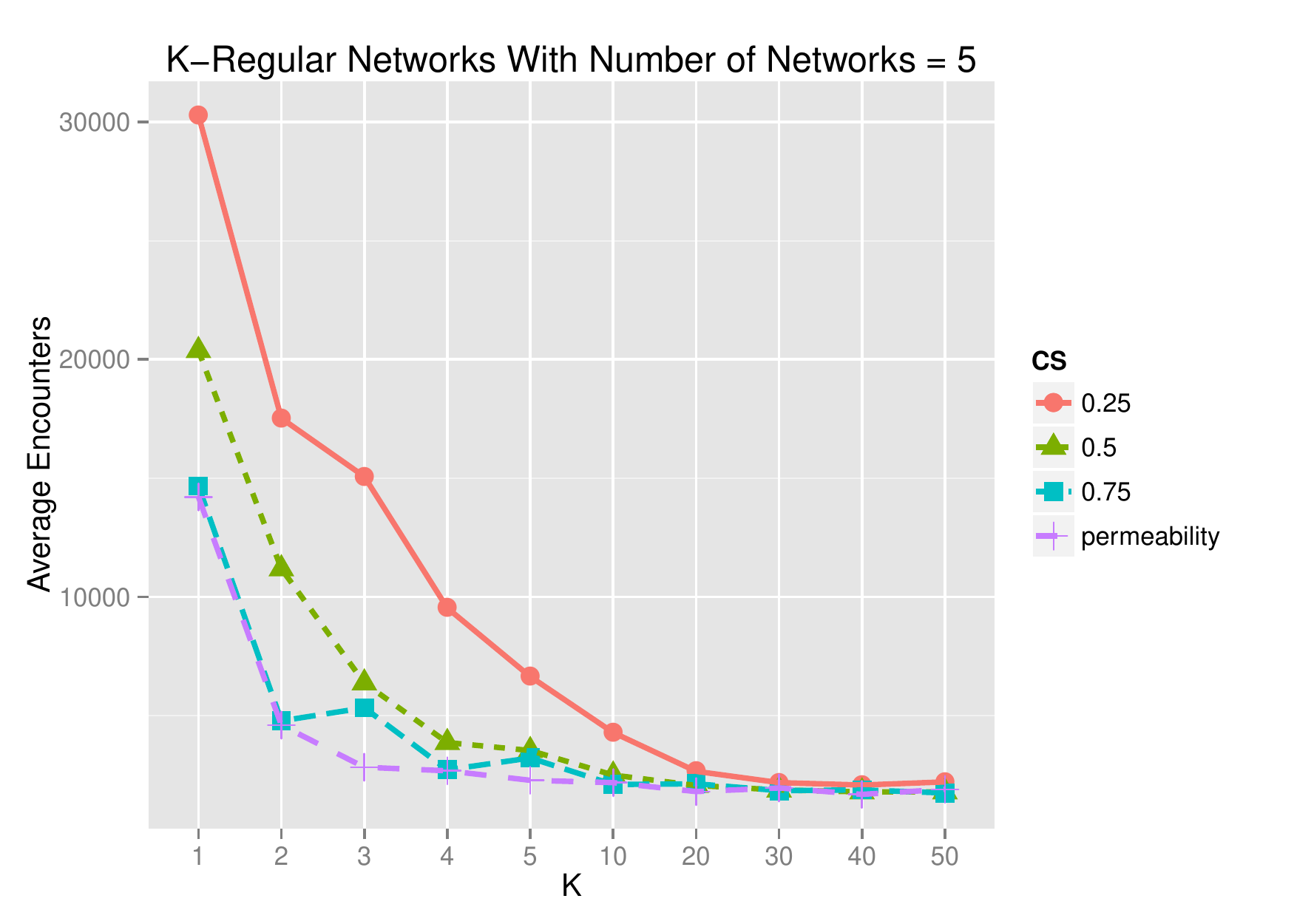}
          \caption{}
          \label{fig:ctx_switching_comp_kreg_5}
	\end{subfigure}
	\begin{minipage}{0.9\textwidth}
          \vspace{0.2cm}
          \caption{Average number of encounters during a simulation for 100 independent runs with
            \textit{k-regular} topologies in all the network layers. We compare the results of
            context switching with context permeability with a $number of networks =
            \{2,3,4,5\}$. The switching probability is the same in all the networks. }
		\label{fig:ctx_switching_comp_kreg}
	\end{minipage}
      \end{figure}

      Figure \ref{fig:ctx_switching_comp_kreg} shows that for 2 networks the results in terms of
      number of encounters for the context switching approximate those for context
      permeability. When we increase the number of networks ($n\ge3$), the configurations with
      higher switching probability lead the results to be closer to what happens in context
      permeability. This is no surprise since more switching makes agents switch more often between
      networks and consequently allows them too choose more often from different
      neighbourhoods. This is almost the same as having a larger neighbourhood to choose from, which
      is what happened in the context permeability model.

      The surprise was that for values of $k \ge 20$, the results where practically the same for
      both models independently of the switching probability -- even when we increased the number of
      networks, the number of encounters remained around $2000$ on average for all the models. Above
      a certain level of neighbourhood size (and overlapping, which causes the average path length
      to drop), the switching probability becomes less influential. It is still important that we
      get comparable results because we are introducing a temporal component that was never explored
      before in these types of opinion dynamics models: the fact that the agents can become active
      in different networks at different points in time.

      \begin{figure}[H]
	\centering
	\begin{subfigure}{.49\linewidth}
          \centering
          \includegraphics[width=1\linewidth]{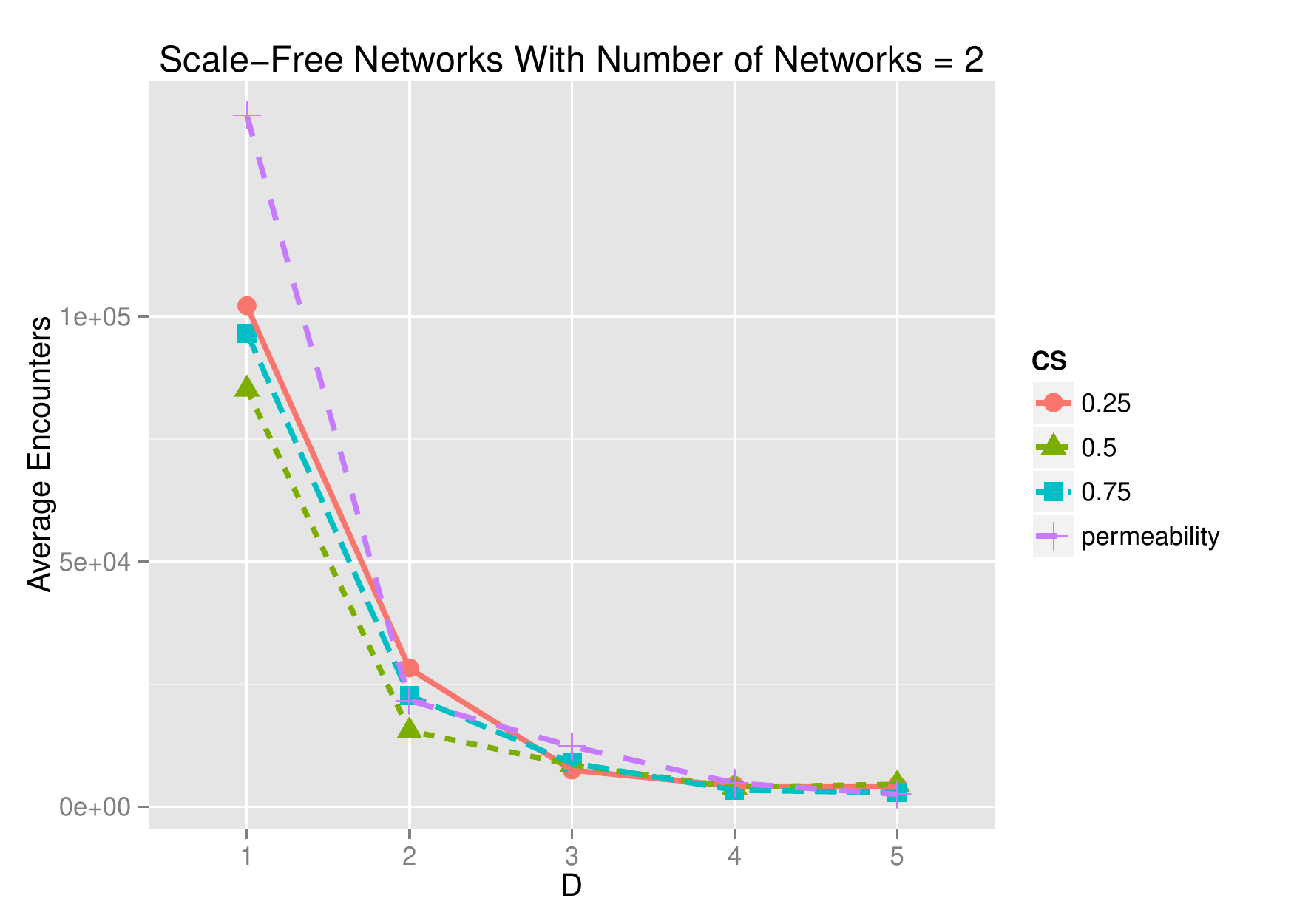}
          \caption{}
          \label{fig:ctx_switching_comp_sf_2}
	\end{subfigure}%
	\begin{subfigure}{.5\linewidth}
          \centering
          \includegraphics[width=1\linewidth]{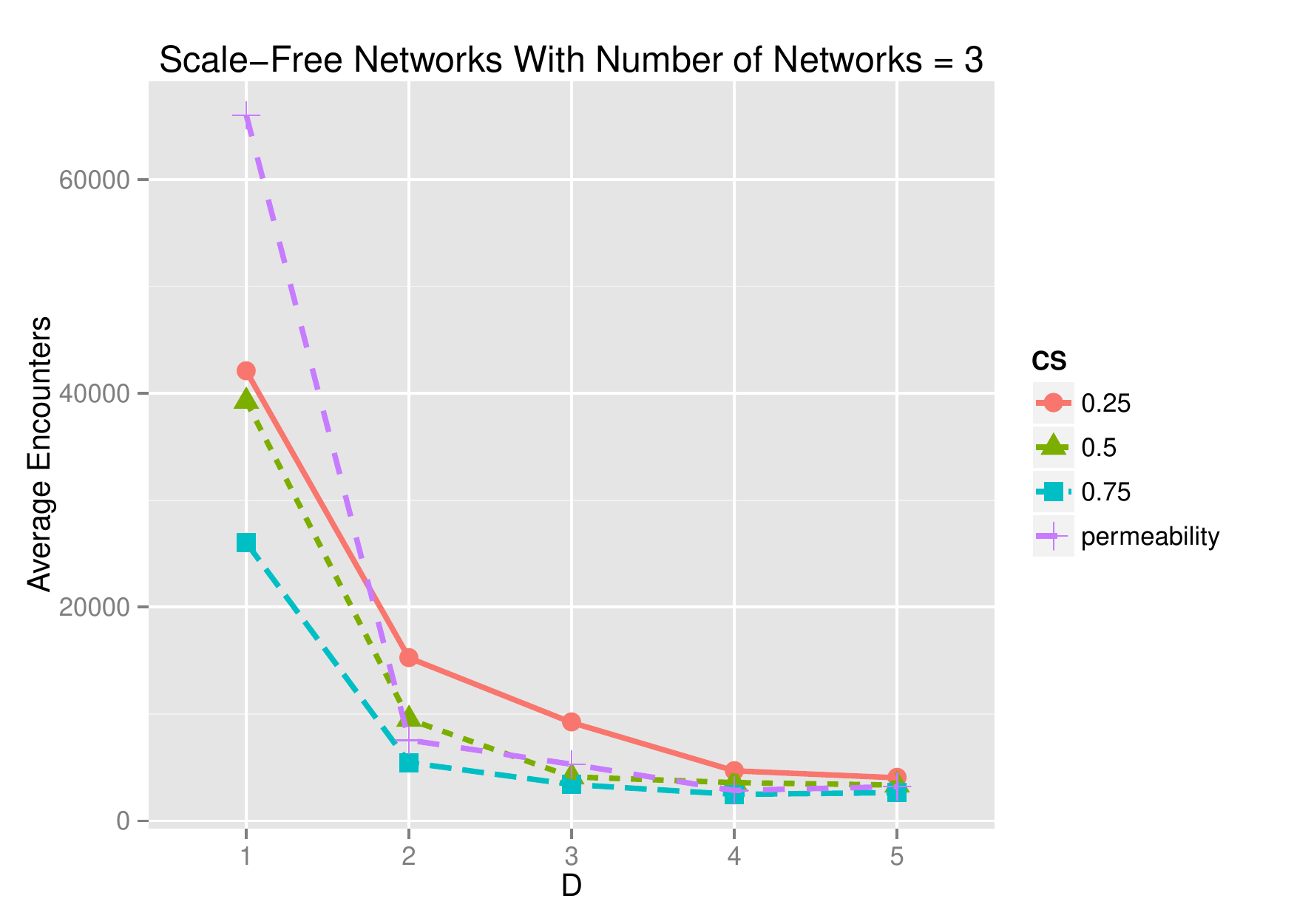}
          \caption{}
          \label{fig:ctx_switching_comp_sf_3}
	\end{subfigure}\\
	\begin{subfigure}{.49\linewidth}
          \centering
          \includegraphics[width=1\linewidth]{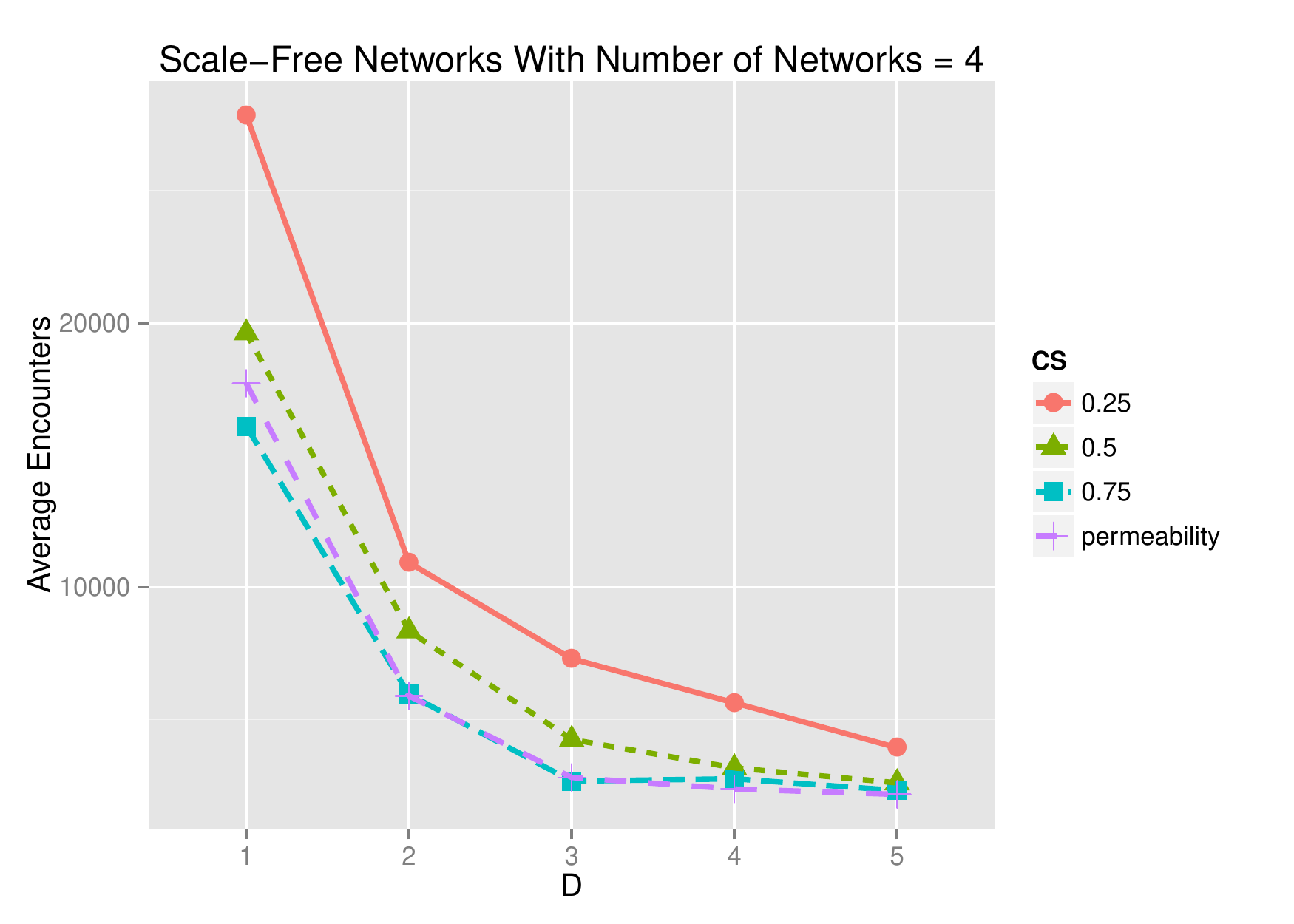}
          \caption{}
          \label{fig:ctx_switching_comp_sf_4}
	\end{subfigure}
	\begin{subfigure}{.5\linewidth}
          \centering
          \includegraphics[width=1\linewidth]{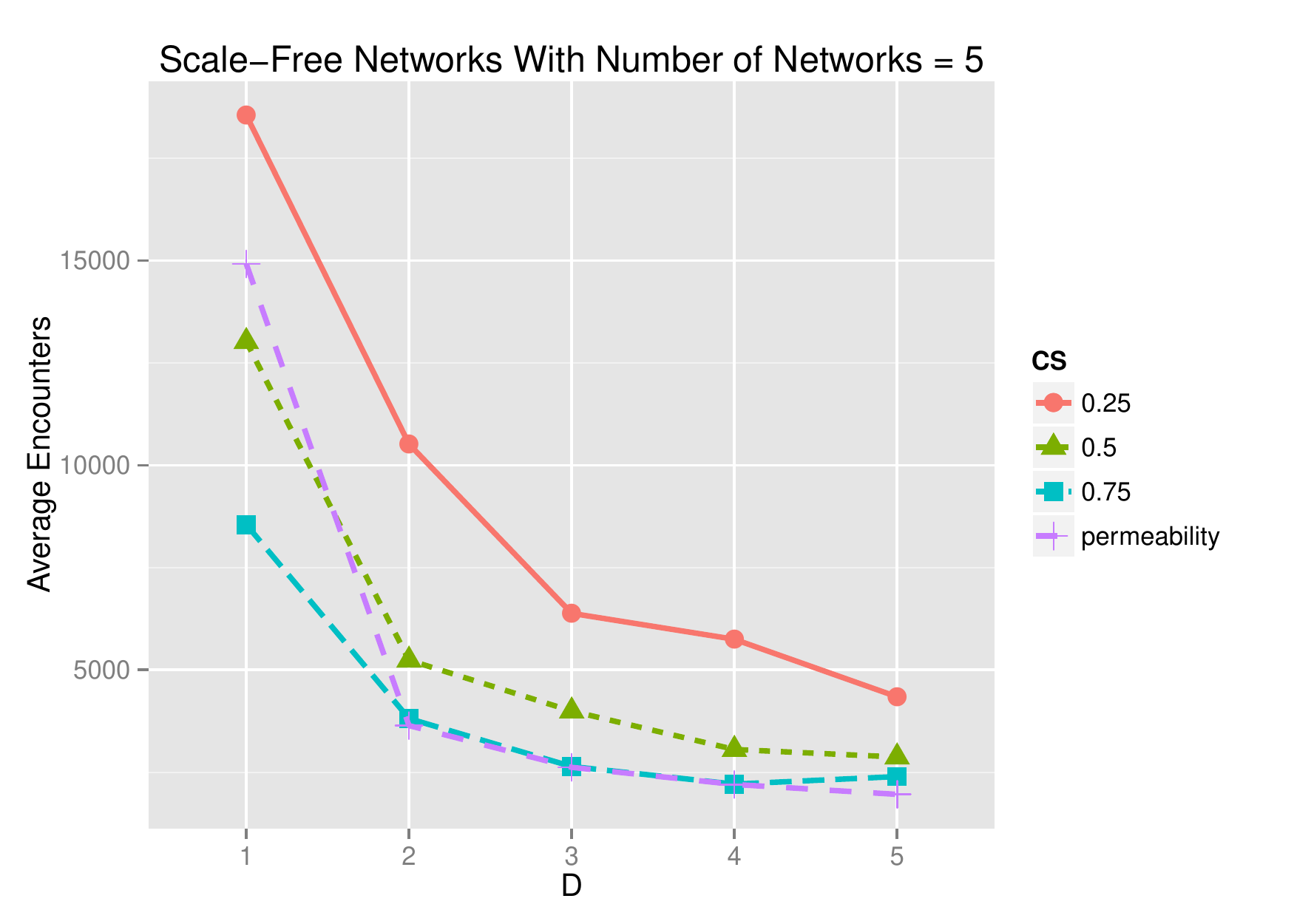}
          \caption{}
          \label{fig:ctx_switching_comp_sf_5}
	\end{subfigure}
	\begin{minipage}{0.9\textwidth}
          \vspace{0.2cm}
          \caption{Average number of encounters during a simulation for 100 independent runs with
            \textit{scale-free} topologies in all the network layers. We compare the results of
            context switching with context permeability with a $number of networks =
            \{2,3,4,5\}$. The switching probability is the same in all the networks.}
          \label{fig:ctx_switching_comp_sf}
	\end{minipage}
      \end{figure}

      For \textit{scale-free} networks, figure \ref{fig:ctx_switching_comp_sf} shows us similar
      results. Switching with a higher probability displays results similar with those of context
      permeability. In this case, adding more networks makes so that lowering switching
      probabilities displays slightly worse results. This influence comes from the neighbourhoods
      being considerably smaller in \textit{scale-free} networks -- even with values of $d=5$ -- so
      the impact of low switching was bound to be noticed.

      An increasing number of networks seems to always reduce the number of encounters to achieve
      consensus in both models. This happens due to what we called \textit{permeability}. Having
      multiple points of dissemination that permeate between different social networks enhances the
      overall convergence to consensus. With permeability there are more previously isolated nodes
      being reached in sparsely connected networks such as scale-free networks.

      \section{Conclusions}
      \label{sec:conclusion-future}
      \noindent In this article, we analysed and discussed our modelling framework for
      multi-relational models of social spaces. These results show that not only contexts play an
      important role in the dissemination of consensus in artificial agent societies, but also that
      simple mechanisms can generate a great deal of complexity (especially in those scenarios where
      conventions are being co-learned by multiple agents). Adding more networks to our opinion
      dynamics model leads to outcomes where consensus was achieved not only more often but also
      faster. In particular, our results show that achieving convergence (i.e. total consensus in a
      network) is not a matter of connectivity. With scale-free networks we managed to achieve
      consensus more often and quicker than with less clustered networks. This conclusion is
      especially important for application (or models) dealing with real-world social phenomena,
      many times represented with such networks.

      The second innovation presented in this article was the fact that our model describes an
      abstract way to represent the time spent on each network using the switching probability. This
      temporal component introduces a new dynamic to the study of opinion dynamics. Considering that
      in agent neighbourhoods, agents might not be available at all times -- and spend different
      amounts of time in different neighbourhoods--, this is an important factor to be included in
      simulation models (either representing real-world systems with different levels of
      abstraction, or artificial agent societies for particular applications).

      Future work includes the analysis of different complex network models. While
      \textit{k-regular} and \textit{scale-free} networks are two of the most pervasive network
      examples in social simulation, there has been a growing number of models with different
      properties inspired by real-world phenomena that can be considered. We will also analyse the
      networks according to different properties such as betweeness centrality of certain
      nodes. Moreover, we plan to apply what we learned with our consensus games to cooperation
      problems. Constructing agents that can learn both behaviour and coordination procedures is a
      difficult task (especially when working with large distributed artificial
      societies). Consensus games such as this one can be used to reduce the search space in
      problems where agents need to cooperate or coordinate in a decentralised fashion.

      While the convergence and number of encounters seem to be correlated with the average path
      length, this might not be the only property that plays a role in faster convergence to
      consensus. Correlation is just an informal tool that helps us assert if some monotonic
      relationship exists. More research is needed to find out, for instance, if self-reinforced
      structures --like the ones reported by \cite{Villatoro2013}-- exist in our multi-network
      structures. In particular, we are interested in finding out if permeability mitigates the
      effects of such phenomenon.

      We have seen that context permeability by context overlapping or switching can provide a
      fairly straightforward modelling methodology. This can be used to compose simulation scenarios
      for more complex social spaces. What remains to be done is to identify more precise contextual
      structures in real-world networks. While capturing the structure of real social network can be
      a daunting task, we do have pervasive records of networking activity between social actors:
      online social networks. Analysing these can lead us to insights about complex structural
      properties that emerge from user activity, and contexts that can be identified within those
      networks. A difficult --but not impossible-- prospect for further research would be to track
      similar contexts between different real online social networks.

      \section*{Acknowledgements}
      \noindent Work supported by the Funda\c{c}\~ao para a Ci\^encia e a Tecnologia under the grant
      SFRH/BD/86034/2012 and by centre grant (to BioISI, Centre Reference:
      UID/MULTI/04046/2013), from FCT/MCTES/PIDDAC, Portugal

      \newpage
      \begin{appendices}
	\include{appendix}

      \end{appendices}

      \bibliographystyle{elsarticle-num} 
      \bibliography{main}

\end{document}

%% file: appendix.tex
\renewcommand\thefigure{\thesection.\arabic{figure}}    
\setcounter{figure}{0}   

\renewcommand\thetable{\thesection.\arabic{table}}    
\setcounter{table}{0}     

\section{Overlapping Network Properties}

\subsection{Homogeneous Network Topology Configurations}
\begin{figure}[H]
\centering
\begin{subfigure}{.49\linewidth}
  \centering
  \includegraphics[width=1\linewidth]{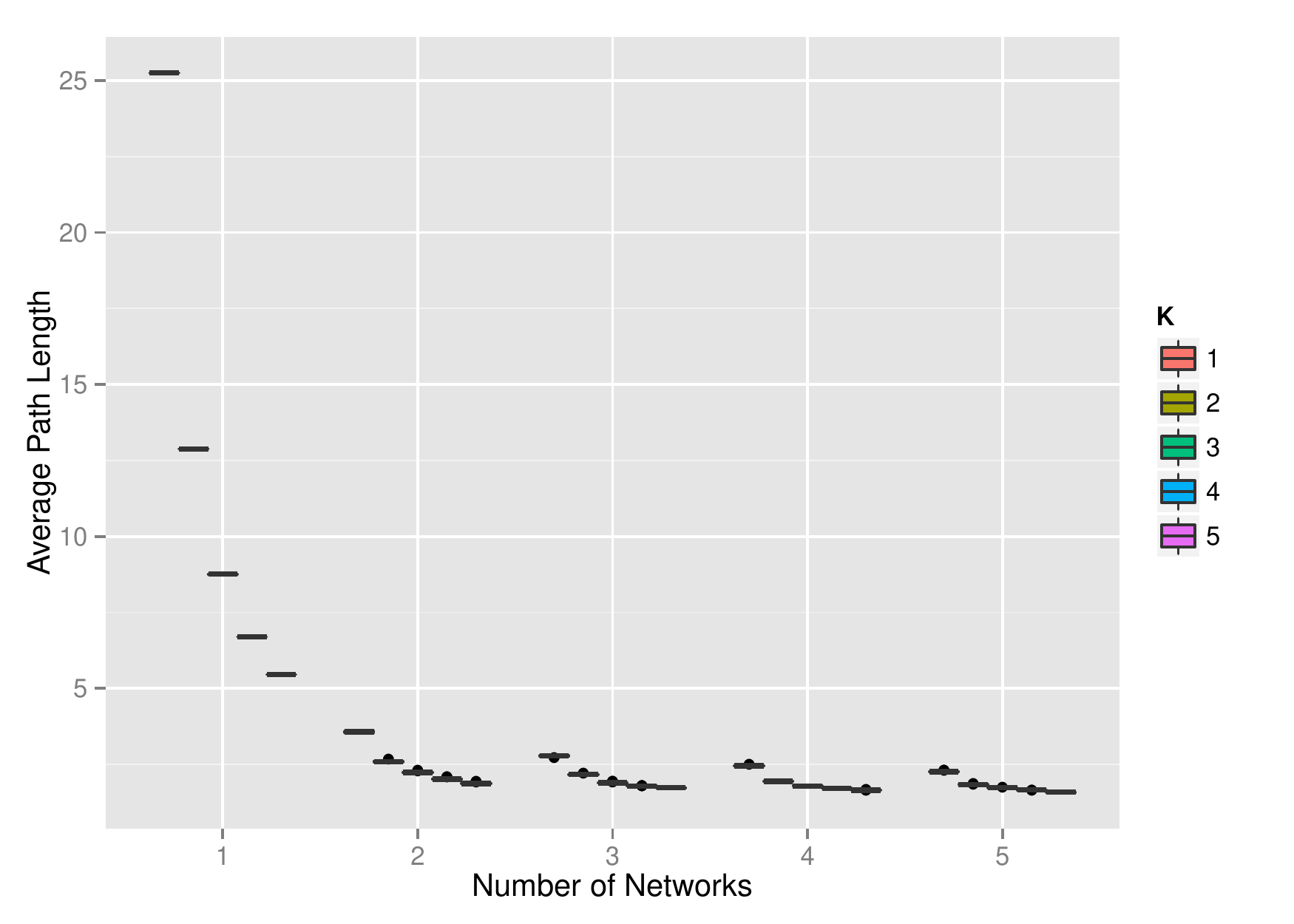}
 \caption{}
 \label{append_fig:network_properties_apl_kreg_12345}
\end{subfigure}%
\begin{subfigure}{.49\linewidth}
  \centering
 \includegraphics[width=1\linewidth]{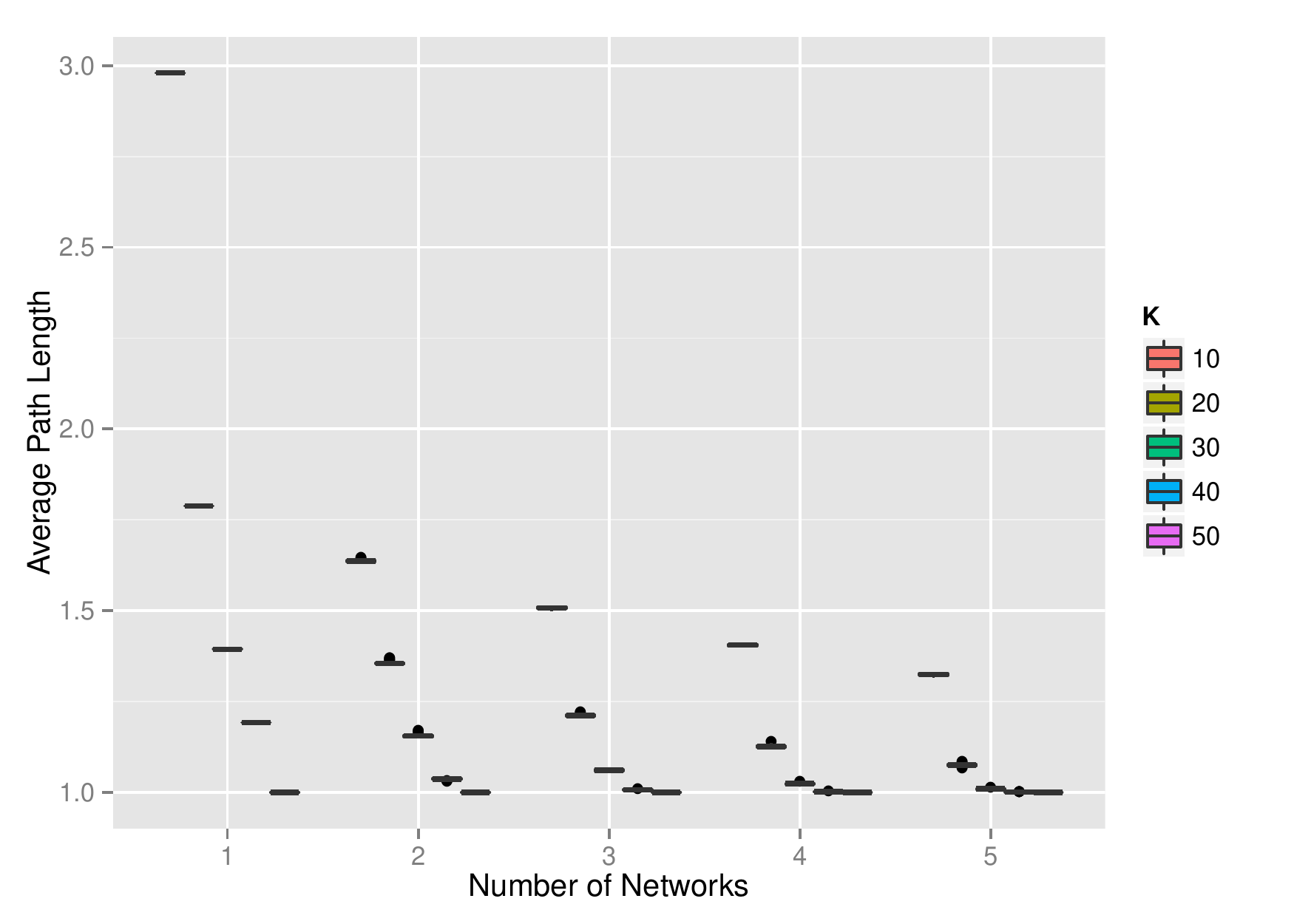}
 \caption{}
 \label{append_fig:network_properties_apl_kreg_1020304050}
\end{subfigure}
\begin{minipage}{0.9\linewidth}
\vspace{0.2cm}
\caption{\textit{Average path length} for 100 instances of overlapping \textit{k-regular} networks(containing 100 nodes each) with ~$k=\{1,2,3,4,5\}$~(\ref{append_fig:network_properties_apl_kreg_12345}), and  $k= \{10,20,30,40,50\}$ (\ref{append_fig:network_properties_apl_kreg_1020304050}). Since there is barely any variation in the property values for each configuration, the colours cannot be seen correctly, they are presented in the same order as the legend nonetheless.}
\label{append_fig:network_properties_apl_kreg}
\end{minipage}

\end{figure}

\begin{figure}[H]
\centering
\begin{subfigure}{.5\linewidth}
  \centering
 \includegraphics[width=1\linewidth]{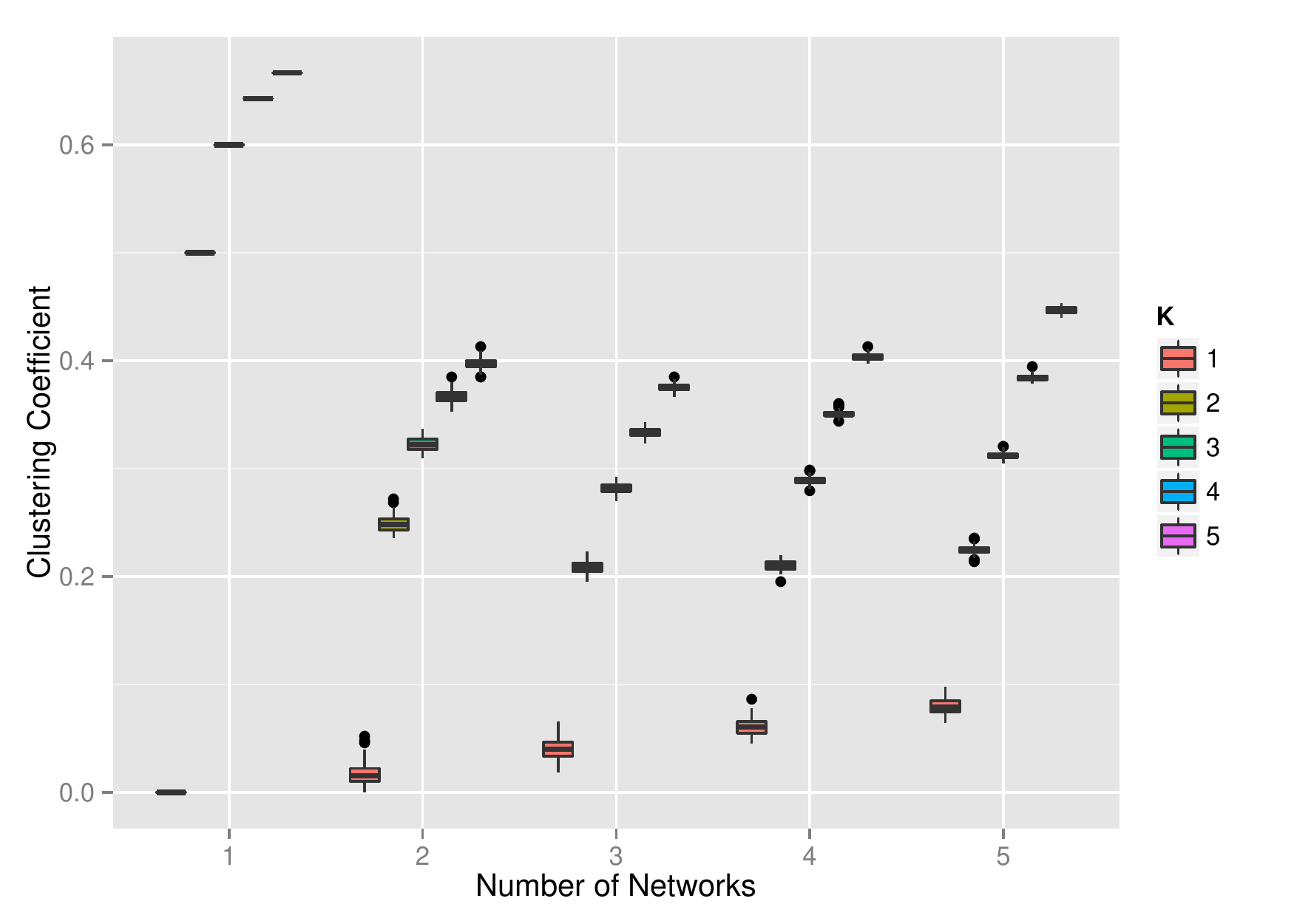}
 \caption{}
 \label{append_fig:network_properties_cc_kreg_12345}
\end{subfigure}%
\begin{subfigure}{.5\linewidth}
  \centering
 \includegraphics[width=1\linewidth]{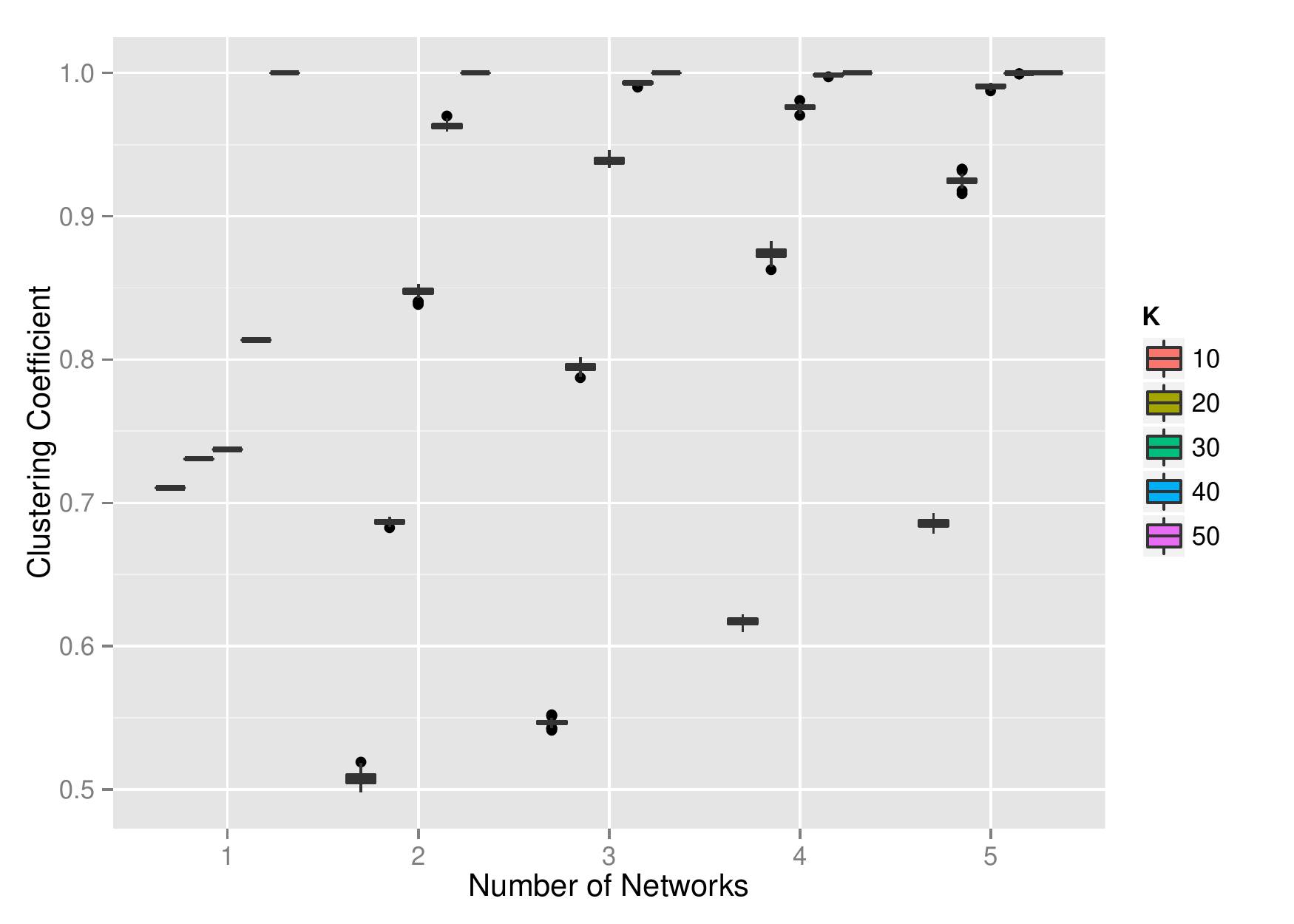}
 \caption{}
 \label{append_fig:network_properties_cc_kreg_1020304050}
\end{subfigure}
\begin{minipage}{0.9\linewidth}
\vspace{0.2cm}
\caption{\textit{Clustering coefficient} for 100 instances of overlapping \textit{k-regular} networks (containing 100 nodes each) with ~$k=\{1,2,3,4,5\}$~(\ref{append_fig:network_properties_cc_kreg_12345}) and  $k= \{10,20,30,40,50\}$ (\ref{append_fig:network_properties_cc_kreg_1020304050}). Note that the variation is bigger for networks with less connectivity. Also, for $k=50$, the network is fully connected.}
\label{append_fig:network_properties_cc_kreg}
\end{minipage}
\end{figure}

\begin{figure}[H]
\centering
\begin{subfigure}{.5\linewidth}
  \centering
 \includegraphics[width=1\linewidth]{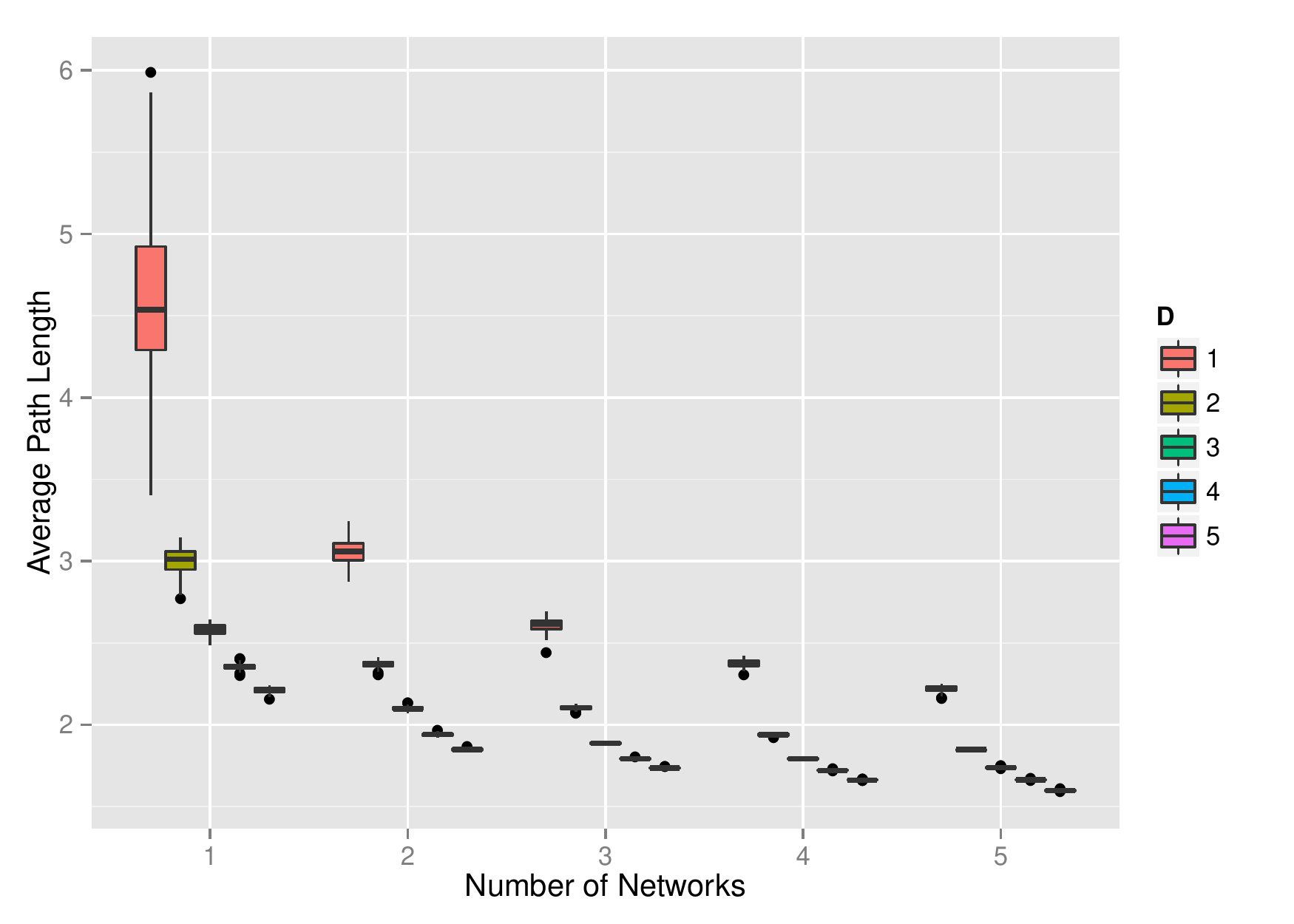}
 \caption{Average path length}
 \label{append_fig:network_properties_apl_sf_12345}
\end{subfigure}%
\begin{subfigure}{.5\linewidth}
  \centering
 \includegraphics[width=1\linewidth]{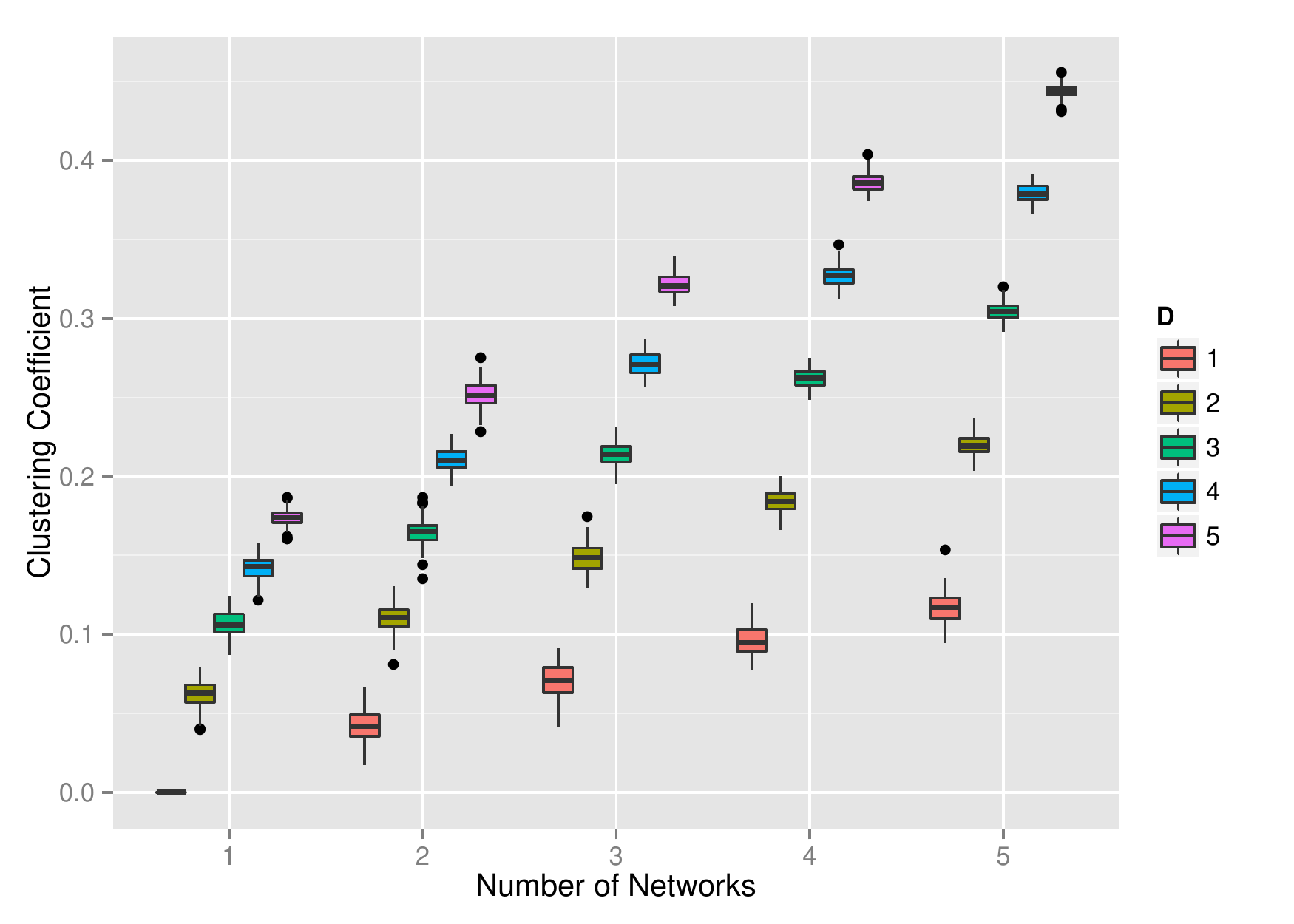}
 \caption{Clustering Coefficient}
 \label{append_fig:network_properties_cc_sf_12345}
\end{subfigure}
\begin{minipage}{0.9\linewidth}
	\vspace{0.2cm}
	\caption{\textit{Average path length} and \textit{clustering coefficient} for 100 instances of overlapping \textit{scale-free} networks. D is a parameter used to construct the network. It dictates how many edges are added by preferential attachment each time a new node is added to the networks. For more details refer to \cite{Barabasi1999}.}
	\label{append_fig:network_properties_sf}
\end{minipage}
\end{figure}

\subsection{Heterogeneous Network Topology Configurations}
\begin{figure}[H]
\centering
\begin{subfigure}{.49\linewidth}
  \centering
 \includegraphics[width=1\linewidth]{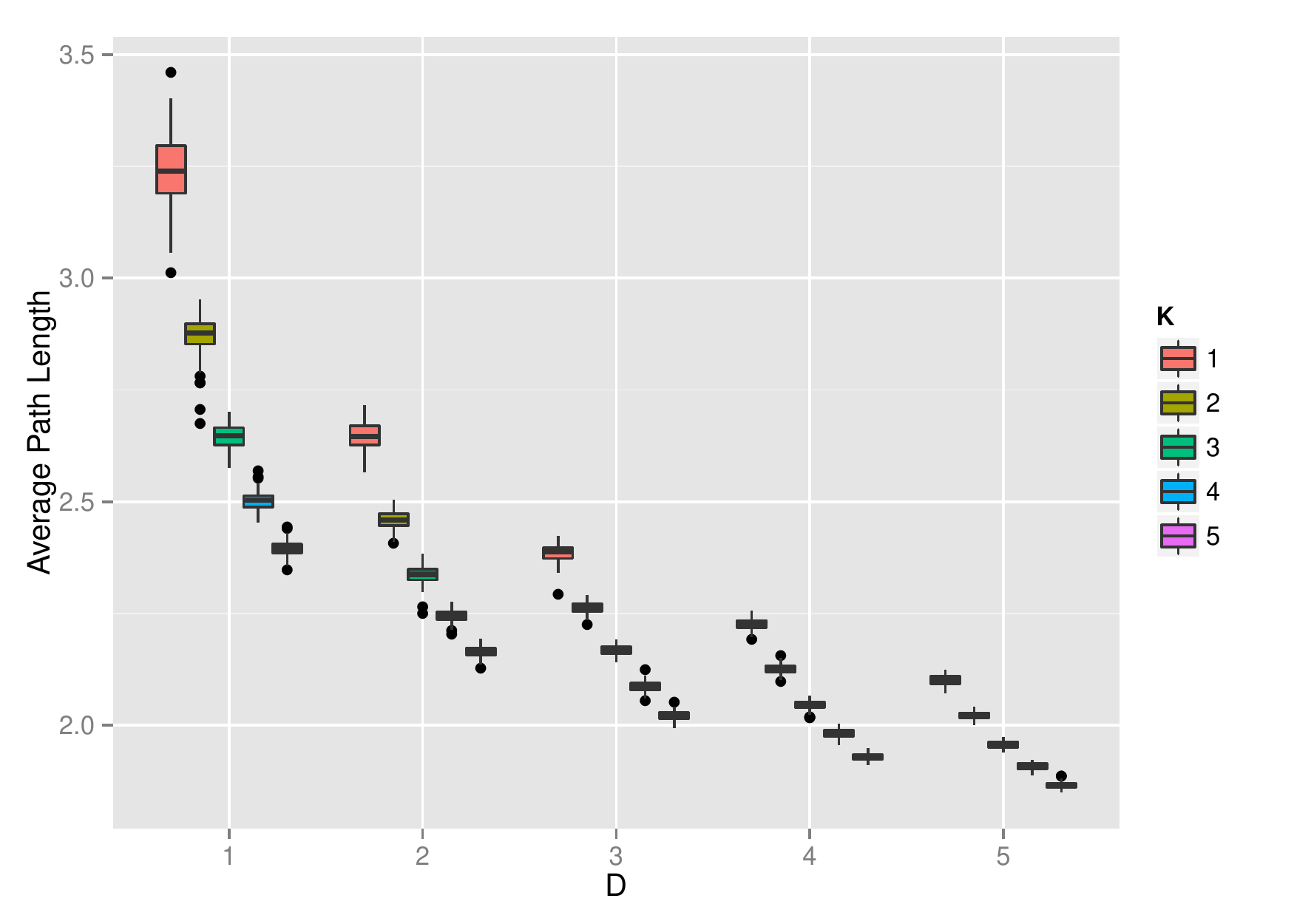}
 \caption{}
 \label{append_fig:network_properties_apl_sf_kreg_12345}
\end{subfigure}%
\begin{subfigure}{.49\linewidth}
  \centering
 \includegraphics[width=1\linewidth]{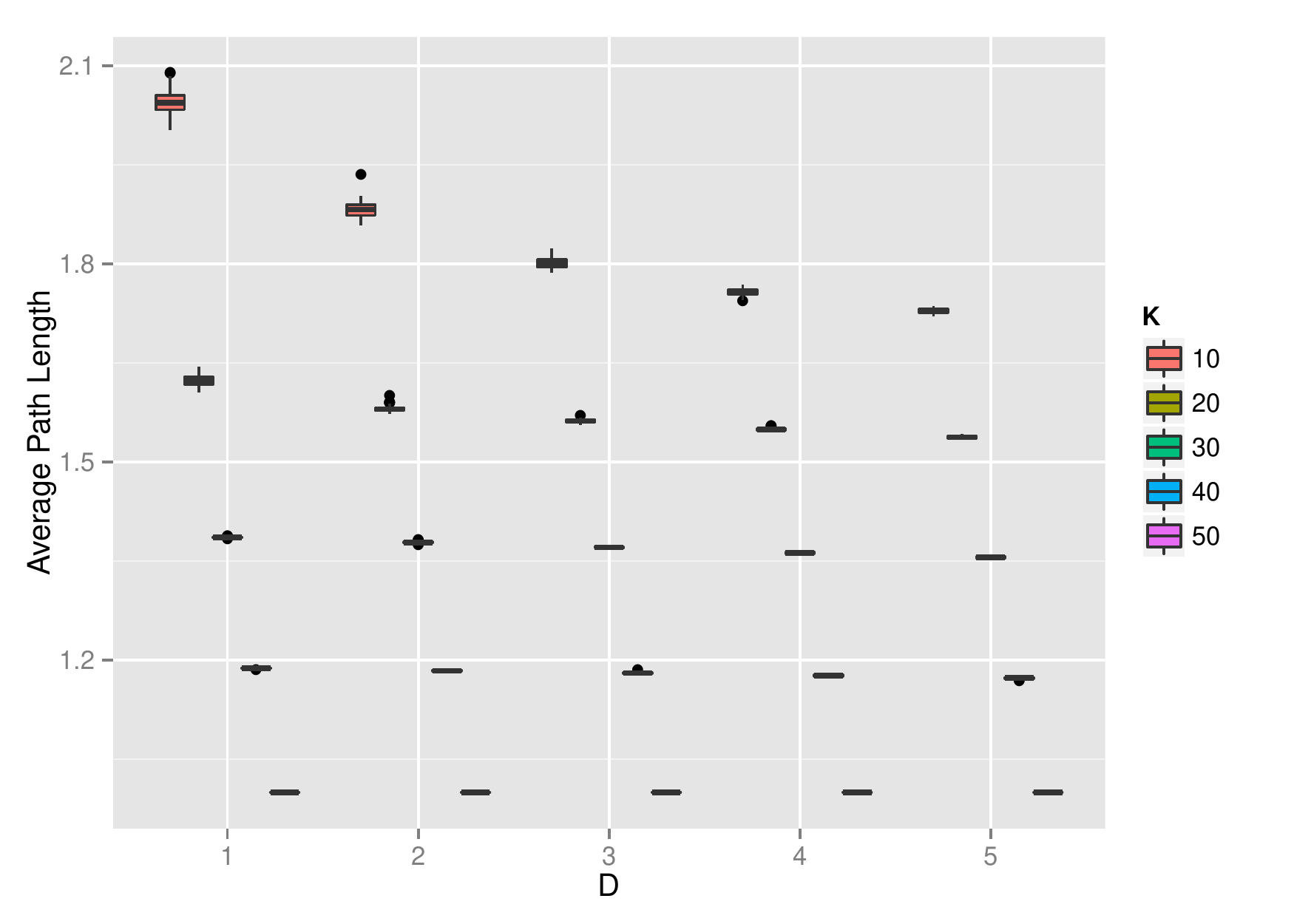}
 \caption{}
 \label{append_fig:network_properties_apl_sf_kreg_1020304050}
\end{subfigure}

\begin{minipage}{0.9\linewidth}
\vspace{0.2cm}
\caption{\textit{Average path length} for 100 instances of two overlapping networks: 1 \textit{scale-free} with $d=\{1,2,3,4,5\}$ and 1 \textit{k-regular} network with ~$k=\{1,2,3,4,5\}$~(\ref{append_fig:network_properties_apl_sf_kreg_12345}), and  $k=\{10,20,30,40,50\}$ (\ref{append_fig:network_properties_apl_sf_kreg_1020304050}). Since there is barely any variation in the property values for each configuration, the colours cannot be seen correctly in some cases, they are presented in the same order as the legend nonetheless.}
\label{append_fig:network_properties_apl_sf_kreg}
\end{minipage}

\end{figure}

\begin{figure}[H]
\centering
\begin{subfigure}{.49\linewidth}
  \centering
 \includegraphics[width=1\linewidth]{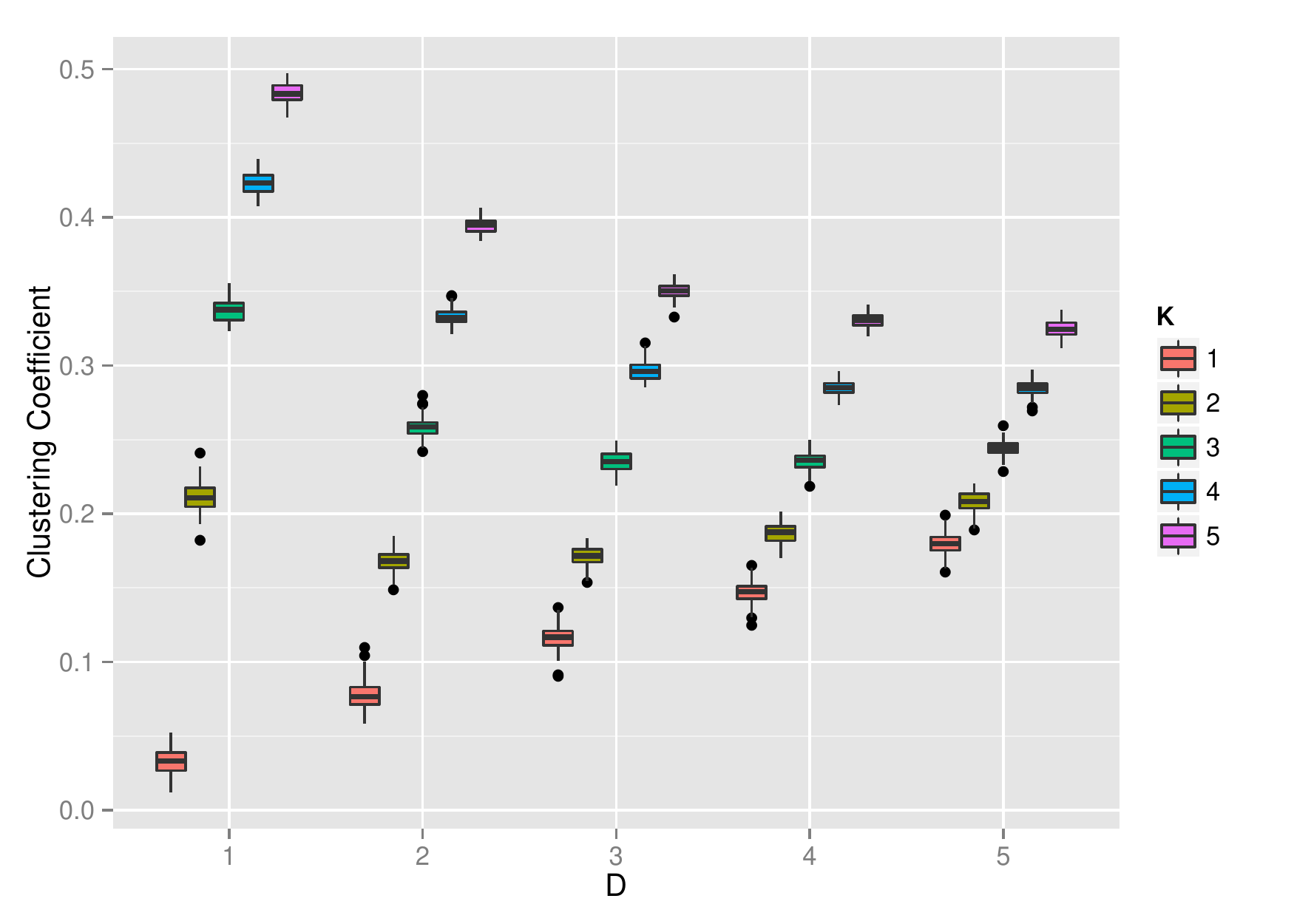}
 \caption{}
 \label{append_fig:network_properties_cc_sf_kreg_12345}
\end{subfigure}%
\begin{subfigure}{.49\linewidth}
  \centering
 \includegraphics[width=1\linewidth]{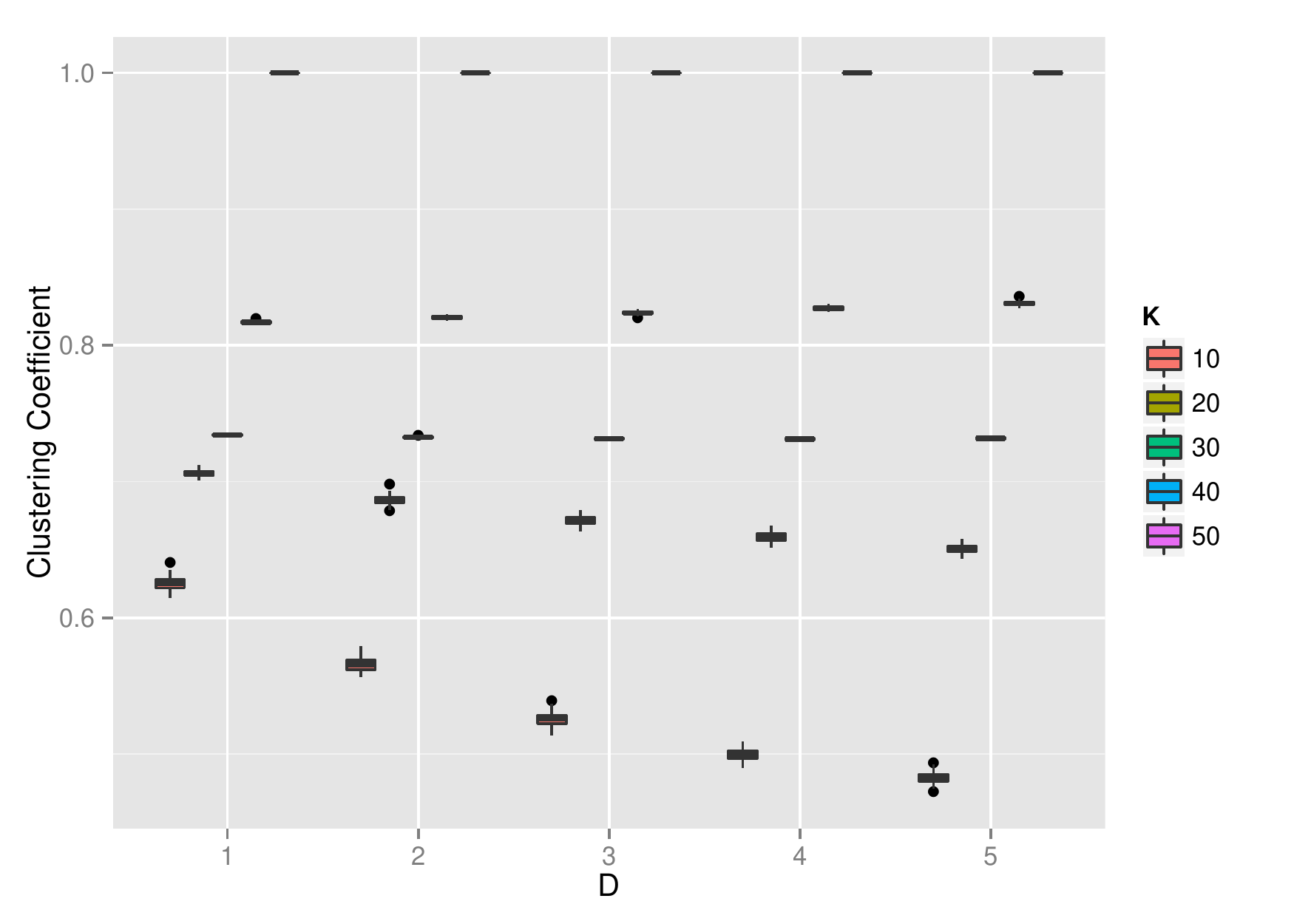}
 \caption{}
 \label{append_fig:network_properties_cc_sf_kreg_1020304050}
\end{subfigure}

\begin{minipage}{0.9\linewidth}
\vspace{0.2cm}
\caption{\textit{Clustering coefficient} for 100 instances of two overlapping networks: 1 \textit{scale-free} with $d=\{1,2,3,4,5\}$ and 1 \textit{k-regular} network with ~$k=\{1,2,3,4,5\}$~(\ref{append_fig:network_properties_apl_sf_kreg_12345}), and  $k=\{10,20,30,40,50\}$ (\ref{append_fig:network_properties_apl_sf_kreg_1020304050}). Since there is barely any variation in the property values for each configuration, the colours cannot be seen correctly in some cases, they are presented in the same order as the legend nonetheless.}
\label{append_fig:network_properties_cc_sf_kreg}
\end{minipage}

\end{figure}
\clearpage

\renewcommand\thefigure{\thesection.\arabic{figure}}    
\setcounter{figure}{0}   

\renewcommand\thetable{\thesection.\arabic{table}}    
\setcounter{table}{0}

\begin{landscape}
	\section{Context Permeability Analysis}
	\label{append_ctx_permeability}
	\vspace{-0.1cm}
\begin{table}[H]
	\centering
\begin{minipage}{0.9\linewidth}	
	\caption{Average number of meetings during a simulation run for 100 agents and a maximum of 2000 simulation cycles for different configuration of \textit{k-regular} networks.}
	\label{append_tab:enc_ctx_permeability_k}
\end{minipage}

\renewcommand{\arraystretch}{0.85}
\begin{tabular}{lcccccccccc}
	\toprule
	& \multicolumn{10}{c}{Number of Networks} \\ 
	& \multicolumn{2}{c}{1} & \multicolumn{2}{c}{2} & \multicolumn{2}{c}{3} & \multicolumn{2}{c}{4} & \multicolumn{2}{c}{5} \\ 
	
	k  & avg. & sd & avg. & sd & avg. & sd & avg. & sd & avg. & \multicolumn{1}{c}{sd} \\ 
\midrule
	1  & $200000$ & $\phantom{0000}0.0$ & $191237$ & $35780.6$ & $97770$ & $86648.0$ & $44827$ & $71714.8$ & $14210$ & $31633.6$ \\
	2  & $200000$ & $\phantom{0000}0.0$ & $\phantom{0}99744$ & $89096.9$ & $22690$ & $50813.5$ & $\phantom{0}7514$ & $21352.6$ & $\phantom{0}4602$ & $\phantom{0}6116.7$ \\
	3  & $198033$ & $19670.0$ & $\phantom{0}68555$ & $82808.5$ & $10217$ & $33963.1$ & $\phantom{0}6539$ & $22227.3$ & $\phantom{0}2828$ & $\phantom{0}2283.6$ \\
	4  & $198093$ & $19070.0$ & $\phantom{0}38359$ & $69305.4$ & $\phantom{0}5028$ & $10680.0$ & $\phantom{0}3710$ & $\phantom{0}4618.7$ & $\phantom{0}2678$ & $\phantom{0}2127.4$ \\
	5  & $196042$ & $27845.6$ & $\phantom{0}21287$ & $48141.1$ & $\phantom{0}3444$ & $\phantom{0}5144.0$ & $\phantom{0}2834$ & $\phantom{0}4334.5$ & $\phantom{0}2278$ & $\phantom{0}1921.5$ \\
	10  & $178207$ & $62301.7$ & $\phantom{00}8066$ & $28298.4$ & $\phantom{0}2894$ & $\phantom{0}6424.5$ & $\phantom{0}1800$ & $\phantom{0}1148.2$ & $\phantom{0}2180$ & $\phantom{0}2078.3$ \\
	20  & $\phantom{0}75541$ & $95871.3$ & $\phantom{00}2338$ & $\phantom{0}2739.9$ & $\phantom{0}2178$ & $\phantom{0}1588.3$ & $\phantom{0}1936$ & $\phantom{00}989.4$ & $\phantom{0}1802$ & $\phantom{00}823.6$ \\
	30  & $\phantom{0}11703$ & $43425.0$ & $\phantom{00}1952$ & $\phantom{0}1087.9$ & $\phantom{0}1726$ & $\phantom{00}936.6$ & $\phantom{0}1930$ & $\phantom{0}1096.3$ & $\phantom{0}1956$ & $\phantom{0}1273.5$ \\
	40  & $\phantom{00}1694$ & $\phantom{00}954.5$ & $\phantom{00}1646$ & $\phantom{00}821.9$ & $\phantom{0}1862$ & $\phantom{00}872.3$ & $\phantom{0}1806$ & $\phantom{00}884.6$ & $\phantom{0}1686$ & $\phantom{00}647.1$ \\
	50  & $\phantom{00}1788$ & $\phantom{00}786.9$ & $\phantom{00}1724$ & $\phantom{00}744.7$ & $\phantom{0}1776$ & $\phantom{00}827.7$ & $\phantom{0}1842$ & $\phantom{00}896.8$ & $\phantom{0}1896$ & $\phantom{00}911.4$ \\
\bottomrule
\end{tabular}

\end{table}

\begin{table}[H]
	\centering
	\begin{minipage}{0.9\linewidth}
		\caption{Average number of meetings during a simulation run for 100 agents and a maximum of 2000 simulation cycles for different configuration of \textit{scale-free} networks.}
		\label{append_tab:enc_ctx_permeability_d}
	\end{minipage}
	
	\renewcommand{\arraystretch}{0.85}
	\begin{tabular}{lcccccccccc}
		\toprule
		& \multicolumn{10}{c}{Number of Networks} \\ 
		& \multicolumn{2}{c}{1} & \multicolumn{2}{c}{2} & \multicolumn{2}{c}{3} & \multicolumn{2}{c}{4} & \multicolumn{2}{c}{5} \\ 
		
		d  & avg. & sd & avg. & sd & avg. & sd & avg. & sd & avg. & \multicolumn{1}{c}{sd} \\ 
		\midrule
		1  & $200000$ & $\phantom{0000}0$ & $141085$ & $83720$ & $66020$ & $82601$ & $17712$ & $42444$ & $14917$ & $38343.6$ \\
		2  & $150632$ & $78710$ & $\phantom{0}21654$ & $41341$ & $\phantom{0}7555$ & $21407$ & $\phantom{0}5890$ & $10889$ & $\phantom{0}3638$ & $\phantom{0}5127.0$ \\
		3  & $\phantom{0}37824$ & $64073$ & $\phantom{0}12309$ & $36441$ & $\phantom{0}5299$ & $20035$ & $\phantom{0}2804$ & $\phantom{0}3791$ & $\phantom{0}2618$ & $\phantom{0}2445.4$ \\
		4  & $\phantom{0}13751$ & $33295$ & $\phantom{00}4846$ & $\phantom{0}7921$ & $\phantom{0}2852$ & $\phantom{0}2766$ & $\phantom{0}2370$ & $\phantom{0}1441$ & $\phantom{0}2200$ & $\phantom{0}1433.4$ \\
		5  & $\phantom{00}8802$ & $18754$ & $\phantom{00}2536$ & $\phantom{0}1880$ & $\phantom{0}3204$ & $11008$ & $\phantom{0}2166$ & $\phantom{0}1269$ & $\phantom{0}1958$ & $\phantom{00}971.6$ \\
	\bottomrule
	\end{tabular}
	
\end{table}

\begin{table}[H]
\centering
\begin{minipage}{0.9\linewidth}	
	\caption{Average number of meetings during a simulation run for 100 agents and a maximum of 2000 simulation cycles for different configuration of 1 \textit{k-regular} and 1 \textit{scale-free} networks.}
	\label{append_tab:ctx_perm_kreg_sf_encounters}
\end{minipage}

\renewcommand{\arraystretch}{0.85}
\begin{tabular}{lcccccccccc}
\toprule
 & \multicolumn{10}{c}{d} \\ 
 & \multicolumn{2}{c}{1} & \multicolumn{2}{c}{2} & \multicolumn{2}{c}{3} & \multicolumn{2}{c}{4} & \multicolumn{2}{c}{5} \\ 
k  & avg. & sd & avg. & sd & avg. & sd & avg. & sd & avg. & \multicolumn{1}{c}{sd} \\ 
\midrule
1  & $173312$ & $61519$ & $100175$ & $88406$ & $63191$ & $81071$ & $73715$ & $84549$ & $51313$ & $67704$ \\
2  & $118591$ & $90107$ & $\phantom{0}72054$ & $86077$ & $43308$ & $70318$ & $31756$ & $58760$ & $43281$ & $72182$ \\
3  & $109155$ & $88722$ & $\phantom{0}38270$ & $66714$ & $34899$ & $65533$ & $26412$ & $55589$ & $23436$ & $49978$ \\
4  & $\phantom{0}80035$ & $88719$ & $\phantom{0}33339$ & $64075$ & $21289$ & $45725$ & $15575$ & $39516$ & $24034$ & $52407$ \\
5  & $\phantom{0}89179$ & $91823$ & $\phantom{0}31044$ & $62485$ & $13908$ & $36791$ & $14112$ & $39490$ & $\phantom{0}9252$ & $28404$ \\
10  & $\phantom{0}62416$ & $81036$ & $\phantom{0}11017$ & $35071$ & $\phantom{0}9095$ & $27266$ & $12182$ & $39798$ & $\phantom{0}5559$ & $20448$ \\
20  & $\phantom{0}61371$ & $81409$ & $\phantom{00}6680$ & $17900$ & $\phantom{0}6944$ & $27826$ & $\phantom{0}4721$ & $19832$ & $\phantom{0}2390$ & $\phantom{0}2313$ \\
30  & $\phantom{0}51559$ & $77468$ & $\phantom{00}4980$ & $12740$ & $\phantom{0}5095$ & $20028$ & $\phantom{0}2268$ & $\phantom{0}1355$ & $\phantom{0}2046$ & $\phantom{0}1225$ \\
40  & $\phantom{0}58625$ & $83049$ & $\phantom{00}5483$ & $20169$ & $\phantom{0}2726$ & $\phantom{0}3735$ & $\phantom{0}2346$ & $\phantom{0}1503$ & $\phantom{0}2412$ & $\phantom{0}1874$ \\
50  & $\phantom{0}48702$ & $74718$ & $\phantom{00}6568$ & $22611$ & $\phantom{0}2566$ & $\phantom{0}1716$ & $\phantom{0}2362$ & $\phantom{0}1619$ & $\phantom{0}1980$ & $\phantom{0}1081$ \\
\bottomrule
\end{tabular}
\end{table}

\end{landscape}
\clearpage

\renewcommand\thefigure{\thesection.\arabic{figure}}    
\setcounter{figure}{0}   

\renewcommand\thetable{\thesection.\arabic{table}}    
\setcounter{table}{0}

\section{Context Switching Analysis}
\label{append_ctx_switching}

\begin{figure}[H]
	\centering
	\includegraphics[width=1\linewidth]{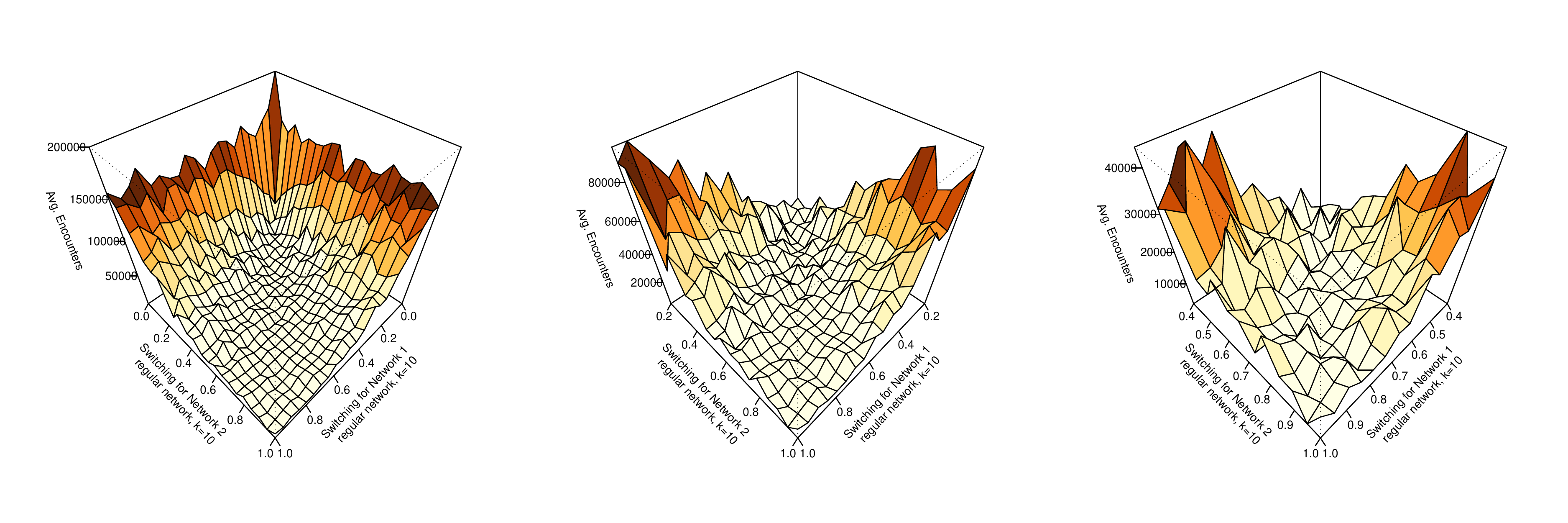}
	\begin{minipage}{0.9\textwidth}
		\caption{Perpective plot for the average number of meets during a simulation for 100 independent runs: 2 \textit{10-regular} networks ($k = 30$).}
		\label{append_fig:ctx_cs_2_10kreg}
	\end{minipage}
\end{figure}

\begin{figure}[H]
	\centering
	\includegraphics[width=1\linewidth]{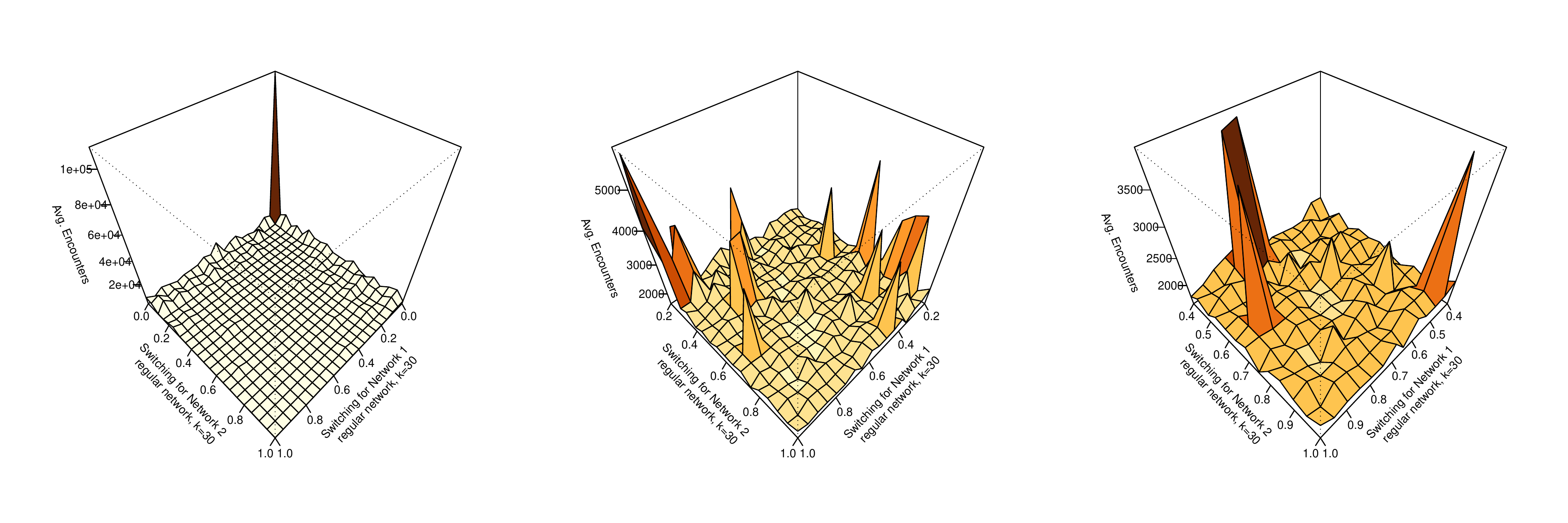}
	\begin{minipage}{0.9\textwidth}
		\caption{Perpective plot for the average number of meets during a simulation for 100 independent runs: 2 \textit{30-regular} networks ($k = 30$).}
		\label{append_fig:ctx_cs_2_30kreg}
	\end{minipage}
\end{figure}

\begin{figure}[H]
	\centering
	\includegraphics[width=1\linewidth]{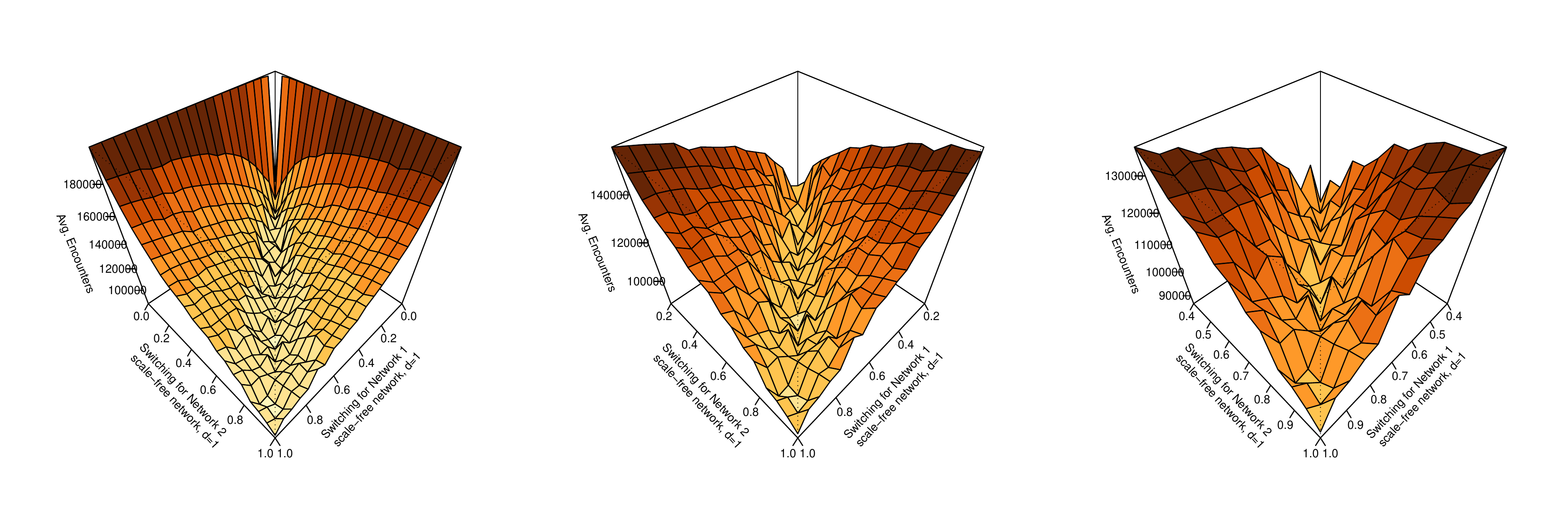}
	\begin{minipage}{0.9\textwidth}
		\caption{Perpective plot for the average number of meets during a simulation for 100 independent runs: 2 \textit{scale-free} networks with $d=1$.}
		\label{append_fig:ctx_cs_2_sd_d1}
	\end{minipage}
\end{figure}

\begin{figure}[H]
	\centering
R	\includegraphics[width=1\linewidth]{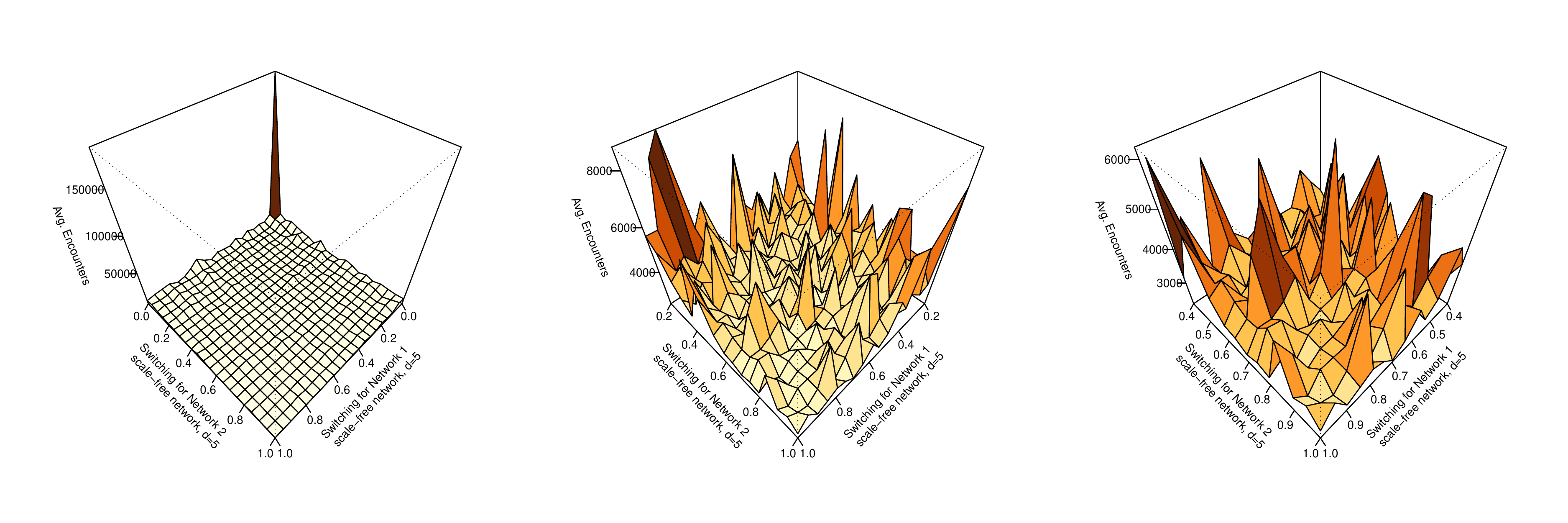}
	\begin{minipage}{0.9\textwidth}
		\caption{Perpective plot for the average number of meets during a simulation for 100 independent runs: 2 \textit{scale-free} networks with $d=5$.}
		\label{append_fig:ctx_cs_2_sd_d5}
	\end{minipage}
\end{figure}

\begin{figure}[H]
	\centering
	\includegraphics[width=1\linewidth]{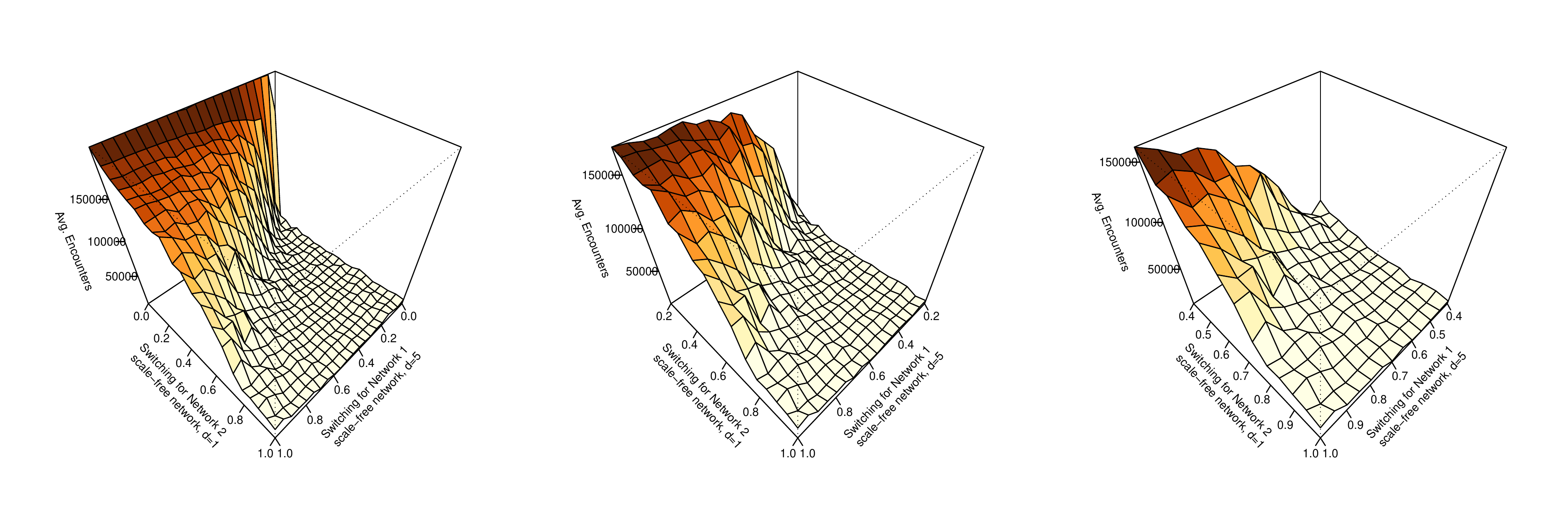}
	\begin{minipage}{0.9\textwidth}
		\caption{Perspective plot for the average number of meets during a simulation for 100 independent runs: 2 \textit{scale-free} networks, the first with $d=5$ and the second with $d=1$.}
		\label{append_fig:ctx_cs_sf_d5_d1}
	\end{minipage}
\end{figure}

\begin{figure}[H]
	\centering
	\includegraphics[width=1\linewidth]{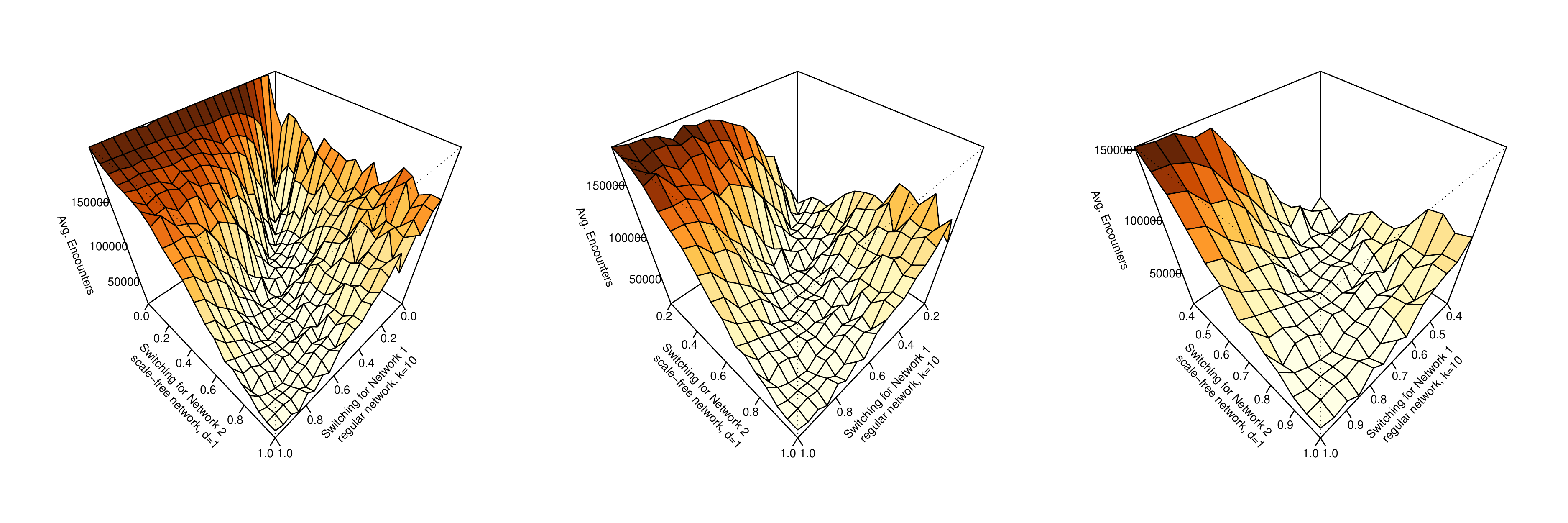}
	\begin{minipage}{0.9\textwidth}
		\caption{Perspective plot for the average number of meets during a simulation for 100 independent runs: 1 \textit{10-regular} network ($k=10$) and a scale-free network with $d=1$.
		}
		\label{append_fig:ctx_cs_kreg10_sfd1}
	\end{minipage}
\end{figure}

%% file: main.bbl
\begin{thebibliography}{10}
\expandafter\ifx\csname url\endcsname\relax
  \def\url#1{\texttt{#1}}\fi
\expandafter\ifx\csname urlprefix\endcsname\relax\def\urlprefix{URL }\fi
\expandafter\ifx\csname href\endcsname\relax
  \def\href#1#2{#2} \def\path#1{#1}\fi

\bibitem{Emmeche1997}
C.~Emmeche, Aspects of complexity in life and science, PHILOSOPHICA-GENT-
  (1997) 41--68.

\bibitem{Simon1954}
H.~A. Simon, Bandwagon and underdog effects and the possibility of election
  predictions, The Public Opinion Quarterly 18~(3) (1954) pp. 245--253.

\bibitem{Axelrod1997}
R.~Axelrod, {The Dissemination of Culture}, Journal of Conflict Resolution
  41~(2) (1997) 203--226.

\bibitem{Weisbuch2004}
G.~Weisbuch, Bounded confidence and social networks, The European Physical
  Journal B - Condensed Matter and Complex Systems 38 (2004) 339--343.

\bibitem{Albin1975}
P.~S. Albin, The analysis of complex socioeconomic systems, Lexington, Mass. :
  Lexington Books, 1975.

\bibitem{Roccas2002}
S.~Roccas, M.~B. Brewer, Social identity complexity, Personality and Social
  Psychology Review 6~(2) (2002) 88--106.

\bibitem{Ellemers2002}
N.~Ellemers, R.~Spears, B.~Doosje, Self and social identity*, Annual Review of
  Psychology 53~(1) (2002) 161--186.

\bibitem{French1956}
J.~French, A formal theory of social power, Psychological Review 63 (1956)
  181--194.

\bibitem{Degroot74}
M.~H. DeGroot, {Reaching a Consensus}, Journal of the American Statistical
  Association 69~(345) (1974) 118--121.

\bibitem{Lehrer1975}
K.~Lehrer, Social consensus and rational agnoiology, Synthese 31 (1975)
  141--160.

\bibitem{Galam1997}
S.~Galam, Rational group decision making: A random field ising model t = 0,
  Physica A: Statistical Mechanics and its Applications 238~(1) (1997) 66--80.

\bibitem{Antunes2009}
L.~Antunes, D.~Nunes, H.~Coelho, J.~a. Balsa, P.~Urbano, {Context Switching
  versus Context Permeability in Multiple Social Networks}, in: Progress in
  Artificial Intelligence, Vol. 5816 of Lecture Notes in Computer Science,
  2009, pp. 547--559.

\bibitem{Deffuant2000}
G.~Deffuant, D.~Neau, F.~Amblard, G.~Weisbuch, Mixing beliefs among interacting
  agents, Advances in Complex Systems 3~(01n04) (2000) 87--98.

\bibitem{Deffuant2002}
G.~Deffuant, F.~Amblard, G.~Weisbuch, T.~Faure, {How can extremism prevail? A
  study based on the relative agreement interaction model}, Journal of
  Artificial Societies and Social Simulation 5~(4) (2002) 1.

\bibitem{Hegselmann2002}
R.~Hegselmann, U.~Krause, Opinion dynamics and bounded confidence: models,
  analysis and simulation, Journal of Artificial Societies and Social
  Simulation 5~(3) (2002) 2.

\bibitem{Lewis1969}
D.~K. Lewis, {Convention: a philosophical study}, Harvard University Press
  Cambridge, 1969.

\bibitem{Delgado2002}
J.~Delgado, Emergence of social conventions in complex networks, Artificial
  intelligence 141~(1) (2002) 171--185.

\bibitem{Shoham1994}
Y.~Shoham, M.~Tennenholtz, Co-learning and the evolution of social activity,
  Tech. rep., DTIC Document (1994).

\bibitem{Schelling1971}
T.~C. Schelling, Dynamic models of segregation†, Journal of mathematical
  sociology 1~(2) (1971) 143--186.

\bibitem{Sakoda1971}
J.~M. Sakoda, The checkerboard model of social interaction, The Journal of
  Mathematical Sociology 1~(1) (1971) 119--132.

\bibitem{FlacheHegselmann2001}
A.~Flache, R.~Hegselmann, Do irregular grids make a difference? relaxing the
  spatial regularity assumption in cellular models of social dynamics., Journal
  of Artificial Societies and Social Simulation 4~(4).

\bibitem{Sullivan2000}
D.~B. O’Sullivan, Graph-based cellular automaton models of urban spatial
  processes, Ph.D. thesis, University of London (2000).

\bibitem{Erdos1959}
P.~Erdős, A.~Rényi, On random graphs, Publ. Math. Debrecen 6 (1959) 290--297.

\bibitem{Barabasi1999}
A.~Barab\'{a}si, R.~Albert, {Emergence of scaling in random networks}, science
  286~(5439) (1999) 509.

\bibitem{Watts1998}
D.~J. Watts, S.~H. Strogatz, {Collective dynamics of 'small-world' networks},
  Nature 393~(6684) (1998) 440--442.

\bibitem{Gaston2005}
M.~E. Gaston, M.~desJardins, Agent-organized networks for dynamic team
  formation, in: Proceedings of the fourth international joint conference on
  Autonomous agents and multiagent systems, ACM, 2005, pp. 230--237.

\bibitem{Nunes2012}
D.~Nunes, L.~Antunes, Consensus by segregation - the formation of local
  consensus within context switching dynamics, Proceedings of the 4th World
  Congress on Social Simulation.

\bibitem{Hayes1997}
P.~J. Hayes, Contexts in context, in: Context in knowledge representation and
  natural language, AAAI Fall Symposium, 1997.

\bibitem{Antunes2007}
L.~Antunes, J.~Balsa, P.~Urbano, H.~Coelho, The challenge of context
  permeability in social simulation, in: Proceedings of the Fourth European
  Social Simulation Association, 2007.

\bibitem{Antunes2010}
L.~Antunes, J.~Balsa, P.~Urbano, H.~Coelho, Exploring context permeability in
  multiple social networks, in: K.~Takadama, C.~Cioffi-Revilla, G.~Deffuant
  (Eds.), Simulating Interacting Agents and Social Phenomena, Vol.~7 of
  Agent-Based Social Systems, Springer Japan, 2010, pp. 77--87.

\bibitem{Luke2005}
S.~Luke, C.~Cioffi-Revilla, L.~Panait, K.~Sullivan, G.~Balan, {MASON: A
  Multiagent Simulation Environment}, Simulation 81~(7) (2005) 517--527.

\bibitem{Nunes:Software:11067}
D.~Nunes, L.~Antunes, \href{{http://dx.doi.org/10.5281/zenodo.11067}}{{Context
  Permeability Models}}\href {http://dx.doi.org/{10.5281/zenodo.11067}}
  {\path{doi:{10.5281/zenodo.11067}}}.
\newline\urlprefix\url{{http://dx.doi.org/10.5281/zenodo.11067}}

\bibitem{R2008}
R.~D.~C. Team, \href{http://www.R-project.org}{R: A Language and Environment
  for Statistical Computing}, R Foundation for Statistical Computing, Vienna,
  Austria, {ISBN} 3-900051-07-0 (2008).
\newline\urlprefix\url{http://www.R-project.org}

\bibitem{knitr2014}
Y.~Xie, \href{http://www.crcpress.com/product/isbn/9781466561595}{knitr: A
  comprehensive tool for reproducible research in {R}}, in: V.~Stodden,
  F.~Leisch, R.~D. Peng (Eds.), Implementing Reproducible Computational
  Research, Chapman and Hall/CRC, 2014.
\newline\urlprefix\url{http://www.crcpress.com/product/isbn/9781466561595}

\bibitem{NunesAntunes2014:Analysis:11898}
D.~Nunes, L.~Antunes, \href{{http://dx.doi.org/10.5281/zenodo.11898}}{{Context
  Permeability Analysis}}\href {http://dx.doi.org/{10.5281/zenodo.11898}}
  {\path{doi:{10.5281/zenodo.11898}}}.
\newline\urlprefix\url{{http://dx.doi.org/10.5281/zenodo.11898}}

\bibitem{Nunes:Software:11069}
D.~Nunes, L.~Antunes, \href{{http://dx.doi.org/10.5281/zenodo.11069}}{{B-Have
  Network Library}}\href {http://dx.doi.org/{10.5281/zenodo.11069}}
  {\path{doi:{10.5281/zenodo.11069}}}.
\newline\urlprefix\url{{http://dx.doi.org/10.5281/zenodo.11069}}

\bibitem{igraph2006}
G.~Csardi, T.~Nepusz, \href{http://igraph.org}{The igraph software package for
  complex network research}, InterJournal Complex Systems (2006) 1695.
\newline\urlprefix\url{http://igraph.org}

\bibitem{Kamada19897}
S.~Tomihisa, K.~Kawai, An algorithm for drawing general undirected graphs,
  Information Processing Letters 31~(1) (1989) 7 -- 15.
\newblock \href {http://dx.doi.org/10.1016/0020-0190(89)90102-6}
  {\path{doi:10.1016/0020-0190(89)90102-6}}.

\bibitem{Villatoro2013}
D.~Villatoro, J.~Sabater-Mir, S.~Sen, Robust convention emergence in social
  networks through self-reinforcing structures dissolution, ACM Trans. Auton.
  Adapt. Syst. 8~(1) (2013) 2:1--2:21.
\newblock \href {http://dx.doi.org/10.1145/2451248.2451250}
  {\path{doi:10.1145/2451248.2451250}}.

\end{thebibliography}
